%% file: WienerHopfControl_arXiv.tex
\documentclass[10pt,a4paper]{article}

	\usepackage[numbers,sort&compress]{natbib}
\usepackage{pstool,geometry,import}

\usepackage{graphicx}
\usepackage{epstopdf, epsfig}
\usepackage{graphicx,subfig,multirow}
\usepackage{amsmath,amsfonts,amssymb,color}

\newcommand{\temporarySupression}[1]{{\color{red}}}
\newcommand{\modification}[1]{#1}

\newcommand{\Matrix}[1]{\left[\begin{matrix}#1\end{matrix}\right]}

\def\mathbi#1{\textbf{\em #1}}
\newcommand{\M}[1]{\mathbi{#1}} 
\newcommand{\Mh}[1]{\hat{\M{#1}}} 
\newcommand{\V}[1]{\boldsymbol{#1}} 
\newcommand{\Vh}[1]{\hat{\V{#1}}} 

\newcommand{\Real}{\mathbb{R}} 

\newcommand{\ii}{\mathrm{i}}
\newcommand{\e}{\mathrm{e}}
\newcommand{\GAMMA}{\boldsymbol{\Gamma}}
\newcommand{\LAMBDA}{\boldsymbol{\Lambda}}

\newcommand{\Tr}[1]{Tr\left(#1\right)}

\newcommand{\x}{\V{u}}
\newcommand{\z}{\V{z}}
\newcommand{\f}{\V{f}}
\newcommand{\y}{\V{y}}
\newcommand{\n}{\V{n}}

\newcommand{\HH}{\M{H}_l}
\newcommand{\Hh}{\M{H}_r}
\newcommand{\GG}{\M{G}_l}
\newcommand{\Gg}{\M{G}_r}

\newcommand{\Tx}{\T_{\x}}
\newcommand{\A}{\M{A}}

\newcommand{\Bf}{{\M{B}_{\f}}}
\newcommand{\Ba}{{\M{B}_{\V{a}}}}
\newcommand{\Cy}{{\M{C}_{\y}}}
\newcommand{\Cz}{{\M{C}_{\z}}}
\newcommand{\R}{\M{R}}
\newcommand{\I}{\M{I}}
\newcommand{\F}{\M{F}}
\newcommand{\N}{\M{N}}

\newcommand{\Y}{\M{Y}}
\newcommand{\T}{\M{T}}

\newcommand{\Rfy}{\M{R}_{\y\f}}
\newcommand{\Rfz}{\M{R}_{\z\f}}
\newcommand{\Ray}{\M{R}_{\y\V{a}}}
\newcommand{\Raz}{\M{R}_{\z\V{a}}}

\title{Resolvent-based tools for optimal estimation and control via the Wiener-Hopf formalism}
\date{}
\author{Eduardo~Martini$^{1,2}$\thanks{Correspondence address: eduardo.martini@univ-poitiers.fr},
	Junoh~Jung$^3$,	
	Andr\'e~V.~G.~Cavalieri$^1$,\\
	 Peter~Jordan$^2$, 
	 Aaron~Towne$^3$ \\
\\
	$^1$ Instituto Tecnol\' ogico de Aeron\'{a}utica,   S\~ao Jos\' e dos Campos/SP - Brazil \\ 
	 $^2$  D\'epartement Fluides, Thermique et Combustion, Institut Pprime, CNRS, \\
	 Universit\' e de Poitiers, ENSMA, 86000 Poitiers, France\\
	 $^3$  University of Michigan, Ann Arbor, MI 48109, USA
}

\begin{document}

\maketitle

\begin{abstract}
	The application of control tools to complex flows frequently requires approximations, such as reduced-order models and/or simplified forcing assumptions, where these may be considered low-rank or defined in terms of simplified statistics (e.g. white noise). In this work, we propose a resolvent-based control methodology with causality imposed via a Wiener-Hopf formalism. Linear optimal causal estimation and control laws are obtained directly from full-rank, globally stable systems with arbitrary disturbance statistics, circumventing many drawbacks of alternative methods.  We use efficient, matrix-free methods to construct the matrix Wiener-Hopf problem, and we implement a tailored method to solve the problem numerically. The approach naturally handles forcing terms with space-time colour; it allows inexpensive parametric investigation of sensor/actuator placement in scenarios where disturbances/targets are low rank; it is directly applicable to complex flows disturbed by high-rank forcing; it has lower cost in comparison to standard methods; it can be used in scenarios where an adjoint solver is not available; or it can be based exclusively on experimental data. The method is particularly well-suited for the control of amplifier flows, for which optimal control approaches are typically robust. Validation of the approach is performed using the linearized Ginzburg-Landau equation. Flow over a backward-facing step perturbed by high-rank forcing is then considered. Sensor and actuator placement are investigated for this case, and we show that  while the flow response downstream of the step is dominated by the Kelvin-Helmholtz mechanism, it has a complex, high-rank receptivity to incoming upstream perturbations, requiring multiple sensors for control. 
\end{abstract}

	\section{Introduction}\label{sec:intro}

	Flow control is a challenging problem both from the academic perspective and in terms of concrete applications, where problems such as laminar-to-turbulent transition, drag reduction, and flow-induced vibration are of practical concern \citep{rowley2006linear,kim2007linear,bagheri2011transition,brunton2015closedloop,luhar2014opposition,jin2020feedback}. \modification{The difficulties arise from the non-linearity of flow dynamics, which is often avoided by considering a linearized system. Nevertheless, it is not clear how to model the neglected non-linear terms, and effective methods to determine the optimal location of sensors and actuators are still a topic of research. } 
	Among the linear control strategies presently available, inverse feed-forward, wave-cancellation,  optimal and robust control are popular choices.  	\modification{Throughout this work, by optimal and robust control we mean the corresponding linear control approaches.} 
	
	In flows dominated by convection, wave-cancellation has been proposed as a simple-but-effective control strategy. Waves are identified by upstream sensors, and actuators situated between the sensors and targets act to minimize perturbations at the target location \citep{sasaki2016closedloop}.   
	Although the approach is not guaranteed to be causal, in so far as computation of the actuation signal may require future sensor readings, causality can be imposed by ignoring the non-causal part of the control kernel. This has been shown to closely reproduce optimal control when there is sufficient distance between sensors, actuators, and targets \citep{sasaki2018wavecancelling}. While this can significantly reduce the  effectiveness of the controller~\citep{brito2021experimental}, the approach has been successfully used in numerous studies~\citep{hanson2014feedback,sasaki2018closedloop,maia2021realtime}. A similar approach was used by \cite{luhar2014opposition}, where opposition control based on the resolvent operator was performed. Adaptive control strategies are often similar to wave-cancellation approaches, but use additional downstream sensors to adapt the control law to changes in the flow~\citep{fabbiane2014adaptive,simon2016inflight}. 
	
	 Optimal control, on the other hand, minimizes a quadratic cost functional~\citep{bagheri2009amr}, frequently associated with the mean perturbation energy.  This approach provides a maximal reduction of perturbation energy and has been used to control flow instabilities \citep{bewley1998optimal} or to reduce the receptivity of a flow to disturbances \citep{barbagallo2009closed,semeraro2011feedback,juillet2013control,juillet2014experimental,morra2020realizable,sasaki2020role}, to delay transition to turbulence, for instance. 
	
	However, the robustness of optimal control may be hindered by feedback between actuators and sensors and/or by small errors in the flow model. In some cases, such issues may even cause the control law to further destabilize the system. While the model can be accurately known in some scenarios, for instance when simulating transitional flows subject to small disturbances, applications in off-design conditions, or in cases where  non-linear dynamics are important, will generally result in a reduction of the accuracy with which the linear model represents the physical system. 
 	Robust control allows a balance to be struck between the cost-functional reductions and the control robustness: at the cost of achieving lower energy reduction than optimal control, robust control can tolerate higher modeling errors. This kind of approach has been used in several recent studies~\citep{dahan2012feedback,jones2015modelling,jin2020feedback}. Robust control  is particularly important for the control of oscillator systems where actuators are frequently situated upstream of sensors \citep{bagheri2009amr}. For this kind of  configuration, optimal control is typically not robust~\citep{schmid2016linear}.
	
	There are, however,  many scenarios where optimal control is robust. \cite{schmid2016linear} showed that  this is the case for amplifier flows, where sensors are typically located upstream of actuators.  \cite{semeraro2011feedback} arrived at a similar conclusion for the control of boundary layer disturbances, where the optimal control was found to be robust to changes in Reynolds numbers and pressure gradients. Our study is focused on this scenario: the development of optimal control for amplifier flows.

	The usual approach to obtain optimal control laws involves solution of Riccati equations. The computational cost grows rapidly with the number of degrees of freedom (DOFs) of the system. As problems in fluid mechanics typically have many DOFs,  methods based on the solution of Riccati equations are impractical.
	A standard way of dealing with this issue is to base control design on a reduced-order models (ROMs) \citep{bagheri2009amr}. 
	Several bases have been used to obtain ROMs, such as proper orthogonal decomposition (POD) modes \citep{noack1993theoretical}, eigenmodes \citep{aakervik2007optimal}, and balanced modes \citep{bagheri2009amr,barbagallo2009closed}.  The
	eigen-system realization (ERA) algorithm~\citep{juang1985eigensystem} was shown to be equivalent to a ROM based on balanced modes, with only a fraction of the costs when external disturbances are low-rank \citep{ma2011reduced}. 
	A drawback of such techniques is that control laws obtained from these ROMs are not guaranteed to be optimal  when applied to the full system.  Model reduction with balanced modes has upper-error bounds  for modelling open-loop systems, but even when the open-loop system is accurately represented by the ROM, a control law based on the ROM can be ineffective when applied to the original problem \citep[page 349]{aastrom2010feedback}. 
	
	Several approaches exist for obtaining estimation and control laws for the full system.  Optimal estimation and control gains for the full system can be obtained iteratively 
	\citep{semeraro2013riccatiless,luchini2014adjoint}. This, however, requires integration of a large auxiliary system for real-time application; this adds significant cost when used in a numerical simulation and is likely unfeasible for experimental implementation. 
	Alternatively, the control law can be reduced a posteriori (design-then-reduce), instead of being designed based on a ROM (reduce-then-design).   \modification{Another approach applicable to the full system is to use estimation strategies using an ensemble Kalman filter~\citep{colburn2011state}. However, this requires the integration of multiple realizations of the full system, which adds significant computational cost when applied to numerical solutions.}
	
	Control strategies are highly dependent on actuator and sensor placement, and the optimal choice is often unclear.   As ROMs are frequently derived for a specific set of sensors and actuators, studies of the role of their placement for control often rely on the derivation of multiple ROMs. An example is the study of \cite{belson2013feedback}, in which an ERA ROM was required for each sensor position. This highlights the role of reduced-order models in studies of sensor and actuator placement and how this can be costly. As sensor and actuator positioning substantially impacts the control strategy~\citep{ilak2008modeling, illingworth2011feedback,belson2013feedback,freire2020actuator}, it is of interest to be able to obtain control laws without having to rely on ROMs for each possible choice of sensor and actuator.
 	
 	Most approaches used for flow control typically require simplified forcing assumptions, frequently modelled as white-in-time noise. Although complex spatial-temporal forcing colour can be used in the Kalman-filter framework, the approach requires use of an expanded system that filters a white noise input to create a coloured noise, with the extra assumption that the forcing CSD is a rational function of the frequency  \citep{aastrom2013computer}. Simplified forcing-colour models are thus typically used to avoid this complexity. However, it has been shown that the use of realistic spatiotemporal forcing colour is crucial for accurate estimation of complex flows \citep{chevalier2006state,martini2020resolventbased,amaral2021resolventbased}. This is an indication that control can be considerably enhanced if realistic forcing models are used.
	
		
	Another approach to estimation and control is possible using the Wiener-Hopf formalism. An optimal non-causal estimation method, known as a Wiener filter \citep{wiener1942extrapolation,barrett1987unified}, and optimal control strategies, known as Wiener regulators \citep{youla1976modern,grimble1979solution,moir1989wiener}, can be obtained based on cross-spectral densities (CSDs) between sensors, actuators, and flow states.  Although well described in the control literature, their potential for flow control has not been appropriately explored.
	To the best of the authors' knowledge,  \cite{martinelli2009feedback} is the only work in which the formalism has been used for flow control. Difficulties in solving the associated Wiener-Hopf problems limited that study to the use of one sensor and one actuator, and the significant cost of converging the CSDs probably hindered further application of the method.
	
	Wiener-Hopf problems appear when causality constraints are imposed on the estimation and control kernels. The Wiener-Hopf method has been used in the fluid mechanics community to obtain solutions to linear problems with spatial discontinuities, such as acoustic scattering by edges~\citep{noble1959methods,peake2004unsteady}.  Solutions to this class of problem are typically based on a factorization of the Wiener-Hopf kernel into components that are regular on the upper/lower halves of the complex frequency plane.  Such factorization can be achieved analytically only for scalar problems \citep{crighton1970scattering,peake2004unsteady} and  for some special classes of matrices \citep{daniele1978factorization,rawlins1981matrix}. Lacking general analytical solutions, several numerical approaches can be used \citep{tuel1968computer,daniele2007fredholm,atkinson2008algorithm,kisil2016approximate}. The potential of the method for estimation and control comes from the fact that the size of the matrices to be factorized scales with the number of sensors and actuators, in contrast to  control strategies based on solutions of algebraic Riccati equations that scale with the system's size.
	
	While the use of a linear control strategy for non-linear systems has clear limitations, i.e., it is expected to work only when the disturbances around a reference flow can be reasonably approximated by a linear model, there are many examples in the literature (several of which were cited above) that show how the approach is useful in a broad variety of flows. 
	
	\modification{The objective of our study is to obtain a method for real-time estimation and control that avoids some of the drawbacks of previous approaches. To achieve this, we strategically combine three pre-existing tools: the Wiener-Hopf regulator for flow control \citep{martinelli2009feedback}; a numerical method to solve matrix Wiener-Hopf problems~ \citep{daniele2007fredholm}; and matrix-free methods for obtaining the  resolvent operator of large systems \citep{martini2020resolventbased,martini2021efficient,farghadan2021randomized_aaron}. Combining these tools provides a novel method that can, for the first time, simultaneously handle complex forcing colour and be applied directly to large systems without the need for a priori model reduction,  allowing parametric investigation of sensor/actuator placement at  low cost when forcing/targets are low rank. To the best of the authors' knowledge, this is also the first time that the Wiener-Hopf regulator is constructed from first principles, i.e., from the linearized equations of motion and a model of the forcing, as in the LQG framework. This construction provides physical insights on the structures that can be estimated and controlled. The approach can also be used to improve wave-canceling strategies \citep{li2006active,sasaki2018wavecancelling,fabbiane2014adaptive,maia2021realtime,brito2021experimental},  minimizing the effect of kernel truncation.} 
	
	
	The method can be viewed as an extension of the resolvent-based estimation methods recently developed by \cite{towne2020resolvent}  and \cite{martini2020resolventbased}. In the former study, a resolvent-based approach was developed to estimate space-time flow statistics from limited measurements. The central idea is to use the measurements to approximate the non-linear terms that act as a forcing of the linearized Navier-Stokes  equations, which in turn provide an estimate of the flow state upon application of the resolvent operator in the frequency domain. The latter study extends this resolvent-based methodology to obtain optimal estimates of the time-varying flow state and forcing. The estimator is derived in terms of transfer functions between the measurements and the forcing terms and can be written in terms of the resolvent operator and the CSD of the forcing.  However, these methods are nominally non-causal, making them appropriate for flow reconstruction but limiting their applicability for flow control.   
	
	The method developed in the current paper follows that of \cite{martini2020resolventbased}, but with additional constraints to enforce causality. It is these constraints that lead to the aforementioned Wiener-Hopf problem.  The causality of the new resolvent-based estimator makes it applicable for real-time estimation, and we use a similar approach to develop an optimal resolvent-based controller. The resulting estimation and control methods are thus obtained directly for the full-rank flow system, without requiring ROMs, and they make use of the spatiotemporal forcing statistics, thus avoiding the simplified forcing assumptions that can lead to significant reductions in performance.
	
	The paper is structured as follows. The derivation of optimal estimation and control kernels based on the  Wiener-Hopf formalism are constructed in \S~\ref{sec:estimation_and_control}, and solutions are compared to those obtained from algebraic Riccati equation using a linearized Ginzburg-Landau problem.  Implementation of the method using numerical integration of the linearized system and using experimental data are described in \S~\ref{sec:inplementation}. An application to flow over a backward-facing step is presented in \S~\ref{sec:bfs}. Final  conclusions are drawn in  \S~\ref{sec:conclusions}. 
	An introduction to Wiener-Hopf problems (which appear in \S~\ref{sec:estimation_and_control}) and their solution is presented in Appendix  \ref{app:WH}.

	\section{Estimation and control using Wiener-Hopf methods} \label{sec:estimation_and_control}

		In what follows, we define the linear system considered, followed by the derivation of  optimal estimation and control kernels based on the Wiener-Hopf approach.  Optimality here is defined in terms of quadratic cost functionals. An approach to recover estimation and control gains from the proposed approach is presented. The derivation for full-state control is presented in Appendix \ref{sec:WHcontrol}.

	\subsection{System definition}
	We consider  the  linear time-invariant system
	\begin{align}\label{eq:system}
		\begin{aligned}
		\dfrac{d\x}{dt}(t) =& \A   \x(t) +  \Bf \f(t) + \Ba \V{a}(t), \\
		\y(t) =& \Cy \x(t) + \V{n}(t),\\
		\z(t) =& \Cz \x(t),
		\end{aligned}
	\end{align}
	where $ \x \in  \mathbb{C}^{n_u}$  represents the flow state, $ \f \in  \mathbb{C}^{n_f}$ is an unknown stochastic forcing, which can represent external disturbances and/or non-linear interactions  \citep{mckeon2010criticallayer}, $ \V{a}\in  \mathbb{C}^{n_a} $ represents flow actuation used for control, $ \y \in  \mathbb{C}^{n_y} $ is a set of system observables, $ \n \in  \mathbb{C}^{n_y}$ is measurement noise, and $ \z \in  \mathbb{C}^{n_z}$ is a set of  targets for the control problem, for instance, perturbations at a given position or surface loads, to be minimized. The system evolution is described by the matrix $ \A \in  \mathbb{C}^{n_u\times n_u}$, representing the linearized Navier-Stokes operator. The matrices  $ \Bf \in  \mathbb{C}^{n_u\times n_f}$, $ \Ba \in  \mathbb{C}^{n_u\times n_a}$, $ \Cy \in  \mathbb{C}^{n_y\times n_u}$, and  $ \Cz \in  \mathbb{C}^{n_z\times n_u}$ determine the spatial support of external disturbances, actuators, sensors, and targets, as in \cite{bagheri2009input}.  Forcing and sensor noise are modelled as stochastic zero-mean processes with two-point space-time correlations given by
	\begin{align}\label{key}
		\M{F}(t-t') & = \langle \V{f}(t)\V{f}^\dagger(t') \rangle ,
		\\
		\M{N}(t-t') & = \langle \V{n}(t)\V{n}^\dagger (t')\rangle,
	\end{align} 
	with $\dagger$ representing the adjoint operator using a suitable inner product and $ \langle\cdot\rangle $ representing  the ensemble average.  We emphasize that these forcing and noise statistics are more general than those assumed in the derivation of the Kalman filter and LQG control, in which they must be uncorrelated in time.  A frequency-domain representation of the two-point correlation is given by the CSDs
	\begin{align}
		 \Mh{F}(\omega) &= \langle \hat{\f} \hat{\f}^\dagger \rangle, \\
		  \Mh{\N} (\omega) &= \langle \hat{\n} \hat{\n}^\dagger \rangle.
	\end{align}
	Note that this differs from the standard definition, i.e., $ \langle \Vh{f}\Vh{f}^H \rangle $, with $ H $ representing the conjugate transpose of the matrices; this modified definition simplifies the derivations that follow.  
	
	Throughout this work, we assume the system to be stable, i.e.,  all eigenvalues $\lambda$ of $\A$, i.e., $ \lambda $ for which $\A \hat{\x} = -\ii \lambda \hat{\x} $ has non trivial solutions, lie in the lower-half plane. As discussed in \S~\ref{sec:intro}, optimal control strategies are best suited for amplifier flows, which satisfy this stability requirement.
	
	It is useful to split \eqref{eq:system} into two systems, one of which is driven by the forcing and includes sensor noise,
	\begin{align}\label{eq:system_1}
	\begin{aligned}
	\dfrac{d\x_1}{dt}(t) =& \;\A \;  \x_1(t) +  \Bf \f(t), \\
	\y_1(t) =& \Cy \x_1(t) + \V{n}(t),\\
	\z_1(t) =& \Cz \x_1(t),
	\end{aligned}
	\end{align}
	and another noiseless system driven by actuation only,
	\begin{align}\label{eq:system_2}
	\begin{aligned}
	\dfrac{d\x_2}{dt}(t) =& \;\A\;  \x_2(t) +  \Ba \V{a}(t), \\
	\y_2(t) =& \Cy \x_2(t) ,\\
	\z_2(t) =& \Cz \x_2(t).
	\end{aligned}
	\end{align}
	
\begin{figure}
		\centering
		\def\svgwidth{.7\linewidth}
		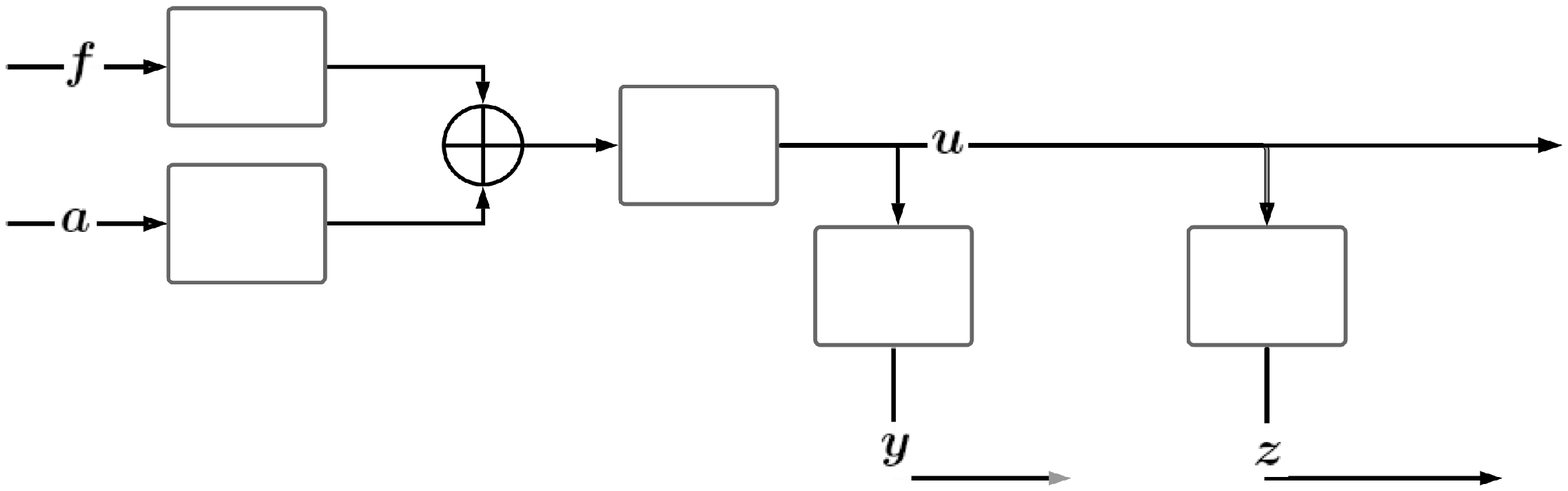
%
		\caption{Block diagram representing the system  \eqref{eq:system}.}
		\label{fig:system}
	\end{figure}

\begin{figure}
	\centering
	\def\svgwidth{.7\linewidth}
	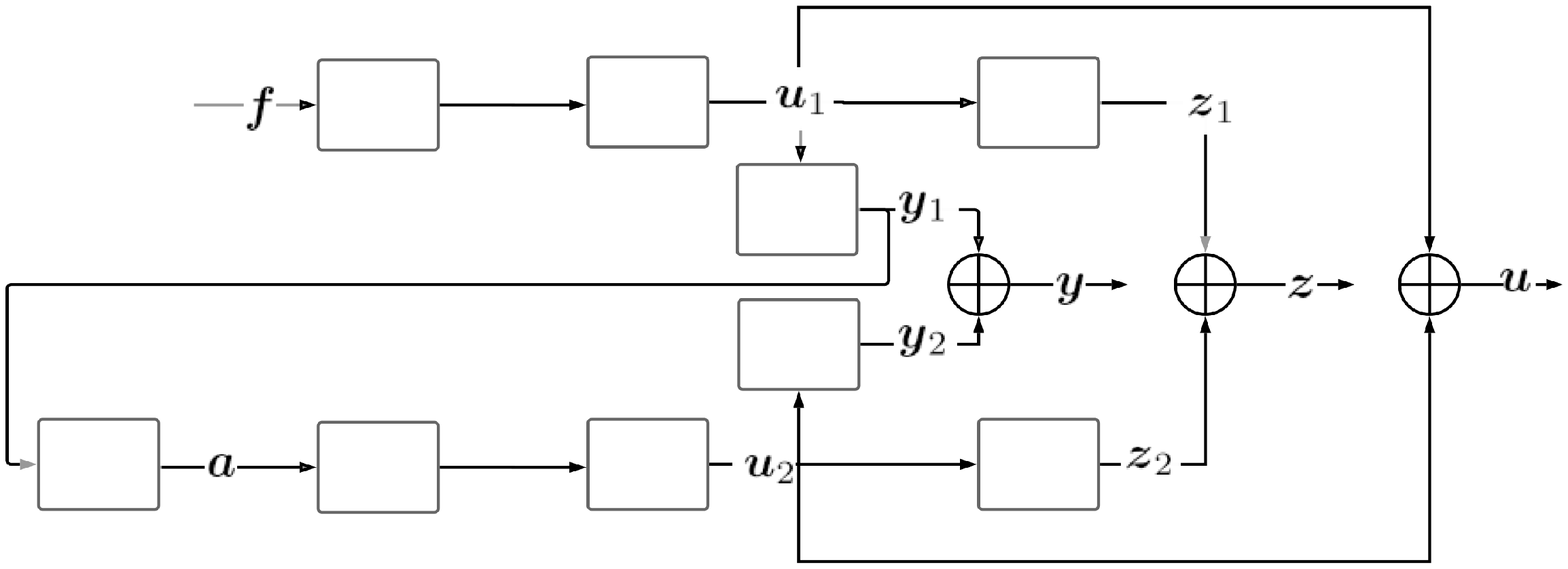

	\caption{Block diagram representing the system  \eqref{eq:system_1} and \eqref{eq:system_2}.}
	\label{fig:separatedsystem}
	
	\end{figure}

	The original system can be recovered adding the variables with subscripts ``1" and ``2", that is,
	\begin{align}
		\x &= \x_1 + \x_2, &
		\y &= \y_1 + \y_2, &
		\z &= \z_1 + \z_2.
	\end{align}
	Figures \ref{fig:system} and \ref{fig:separatedsystem} illustrate these systems.
	
	The derivations that follow make use of the frequency-domain form of these equations.  Taking a Fourier transform of \eqref{eq:system_1} and  \eqref{eq:system_2} yields the input-output relationships
	\begin{align} 
	\label{eq:ftoy_1}
		\hat{\y}_1(\omega) =& \Rfy(\omega) \hat{\f}(\omega) + \hat{\n}(\omega) , & \text{with }   & \Rfy(\omega) = {\Cy \R \Bf}, \\
	\label{eq:ftoz_1}
		\hat{\z}_1(\omega) =& \Rfz(\omega)\hat{\f}(\omega),  & \text{with }   & \Rfz(\omega) = {\Cz \R \Bf}, \\
	\label{eq:ftoy_2}
	\hat{\y}_2(\omega) =& \Ray(\omega) \Vh{a}(\omega),  & \text{with }   & \Ray(\omega) = {\Cy \R \Ba}, \\
	\label{eq:ftoz_2}
	\hat{\z}_2(\omega) =& \Raz(\omega)\Vh{a}(\omega),  & \text{with }   & \Raz(\omega) = {\Cz \R \Ba}, 
	\end{align} 
	 where 
	\begin{equation}\label{eq:resOp}
		\R = (-\ii\omega\I-\A)^{-1},
	\end{equation} 
	is the resolvent operator. Note that the above input-output relations are exact within the linearized model used here, and thus has the same limitation as any linear modelling of the original non-linear system, i.e., treating the non-linear interactions as exogenous stochastic noise. This framework is shared by all linear control laws that have been developed.

	\subsection{The estimation problem } \label{sec:estimation}
	
	Before deriving the optimal causal estimator that we seek, we first briefly review the derivation of the optimal non-causal estimator \citep{martini2020resolventbased}.  We will see later that the two cases are closely related but that imposing causality leads to the appearance of an additional term.  We focus on the uncontrolled estimation problem, i.e. $\V{a}(t) = 0$,   thus making  \eqref{eq:system} and \eqref{eq:system_1}  equivalent.
	
	The optimal estimator minimizes the cost functional
	\begin{align}\label{eq:estFunc}
		J = \int_{-\infty}^\infty \langle \V{e}^\dagger(t) \V{e}(t) \rangle dt = \dfrac{1}{2\pi}	\int_{-\infty}^\infty \langle \Vh{e}^\dagger(\omega) \Vh{e}(\omega) \rangle d\omega,
	\end{align}
	defined in terms of the estimation error
	\begin{align}\label{eq:est_error}
	\V{e}(t) = \z(t) - \tilde {\z}(t),
	\end{align}
	where $\z$ is the target, i.e., the quantity to be estimated, and $\tilde{\z}$ is its estimate.  Note that full-state estimation is recovered using $ \Cz=\M{I} $.
	 
	We seek an estimate of the targets in the form of a linear combination of sensor readings, in the form of 
	\begin{align}\label{eq:stateEst}
		\tilde \z(t) =& \int_{-\infty}^\infty  \T_{z,nc}(\tau) \y(t-\tau) d\tau,& \hat {\tilde \z} (\omega) =& \hat{\T}_{z,nc}	 \hat {\y}(\omega),
	\end{align}
	where the subscript $nc$ is a reminder that  the kernel, $ \T_{z,nc} \in  \mathbb{C}^{n_z\times n_y }$, is, in general, non-causal, and $ \hat{ \T}_{z,nc}$ is its Fourier transform. Expanding \eqref{eq:estFunc} leads to
		\begin{align}\label{eq:stateEst2}
			\begin{aligned}
		J  
		&= 
\dfrac{1}{2\pi}	\int_{-\infty}^\infty \left( \left( \Rfz - \hat{\T}_{z,nc}	\Rfy  \right) ^\dagger \hat{\F} \left(\Rfz - \hat{\T}_{z,nc}	\Rfy   \right) +	\Rfy^\dagger \hat{\T}_{z,nc}^\dagger \hat{\N}\hat{\T}_{z,nc}	\Rfy \right) d\omega.
			\end{aligned}
	\end{align}
	In the above equations, and henceforth, the frequency dependence is omitted for clarity.
	The  minimum is found by taking the derivative of the cost function \eqref{eq:stateEst2} with respect to $ \Mh{T} _{z,nc} ^\dagger(\omega) $ and setting it to zero.  The optimal estimation kernel is obtained as the solution of
	\begin{align}\label{eq:WFnoncausal}	 		 
	 \Mh{T} _{z,nc} \hat{\GG}=& \hat{\Gg} , 
	 \end{align}
	 where,
	 \begin{align}\label{eq:Gs}
	 \hat{\GG} =&	\Rfy \Mh{F} \Rfy^\dagger  + \Mh{\N} , \\
	 \hat{\Gg}= &\Rfz  \Mh{F} \Rfy ^\dagger.
	\end{align}
	 Note that the forcing CSD $ \Mh{F} $ appears explicitly in the equation, and is thus naturally handled by the approach.

	Causality of the estimation kernel can be enforced in \eqref{eq:estFunc} using Lagrange multipliers~\citep{martinelli2009feedback}, as
	\begin{align}\label{eq:estFuncCausal}
	\begin{aligned}
	J' &= J +  \int_{-\infty}^\infty \Tr{ \LAMBDA_-(t)  \T_{z,c}(t)+\LAMBDA^\dagger_- (t) \T^\dagger_{z,c}(t)}dt \\
	&
	 = J +   \int_{-\infty}^\infty \Tr{  \hat{\LAMBDA}_- \Mh{T} _{z,c}+ \hat{\LAMBDA}^\dagger_- \Mh{T} ^\dagger_{z,c} } d\omega,
	\end{aligned}
	\end{align}
	where the Lagrange multipliers $ \LAMBDA_-,{\LAMBDA}_+ \in  \mathbb{C}^{n_y\times n_z}$  are required to be zero for $t > 0$, thus enforcing the condition $ {\T_{z,c}(t<0)=0 }$. 
	The subscript $c$ is used to emphasize the causal nature of this kernel. 
	
	The functionals  $ J $ and $ J' $ differ only by linear terms in the estimation kernel, and thus it is straightforward to see that taking the derivative of $ J' $ and setting it to zero leads to
	\begin{align}\label{eq:WFcausal}
	\Mh{T} _{z,c} \hat{\GG} + \hat{\LAMBDA}_- =  \hat{\Gg} .
	\end{align}
	This apparently simple equation hides significant complexities. First, this is a single equation with two variables, $  {\hat{\T}_{z,c}} $ and  $ {\hat{\LAMBDA}}$. Nevertheless, it admits a unique solution due to the requirements that $  {\T_{z,c}(t<0)=0 } $ and  $ {\LAMBDA(t>0)=0 } $, which in the frequency domain impose restrictions on  $  {\hat{\T}_{z,c}} $ and  $ {\hat{\LAMBDA}}$, namely that these quantities are regular on the upper and lower complex planes, respectively. This restriction means that the values of $  {\hat{\T}_{z,c}} $ and  $ {\hat{\LAMBDA}}$ for different frequencies have a non-trivial relation between them, and thus cannot be chosen independently. Equation \eqref{eq:WFcausal}, with the regularity constrains, constitutes a Wiener-Hopf problem. As discussed in  \S~\ref{sec:intro}, analytical solutions are only known for special cases, and thus we resort to numerical methods.  An introduction to Wiener-Hopf problems and numerical methods to solve them is presented in Appendix~\ref{app:WH}. The solution of \eqref{eq:WFcausal}, once inverse Fourier transformed, is the optimal causal estimation kernel for the linear system at hand.
	
	In previous studies \citep{martini2020resolventbased,amaral2021resolventbased}, we have shown that using the spatiotemporal forcing statistics considerably improves the accuracy of the estimation of a turbulent channel flow.
	The estimation method presented here preserves the ability to handle these complex forcing models, while being applicable in real time  via the simple integration of
	\begin{align}
		\tilde \z(t) =& \int_{0}^\infty \T_{z,c}(\tau) \y(t-\tau) d\tau,
	\end{align}
	which is similar to \eqref{eq:stateEst}, but with integration restricted to positive $\tau$, implying that only present and past sensor measurements are used for estimation.
	
%

	\subsection{Partial-knowledge control }\label{sec:partialKownControl}
	Control can be divided in full-knowledge control, where the full state of the flow is assumed to be known, and partial-knowledge control, where the flow state needs to be estimated from limited, noisy, sensor readings. In this section we derive the optimal partial-knowledge control; the full-knowledge case is considered in Appendix~\ref{sec:WHcontrol} for completeness. 
	
	

	Analogous to the estimation problem, we seek a control law that constructs an actuation signal as a linear function of sensor readings, 
	\begin{align}
			\V{a}(t) &= \int_{-\infty}^\infty  \GAMMA'(\tau) {\y} (t-\tau) d\tau, &
			\Vh{a}   & = \hat{\GAMMA}' \hat{\y}, 
	\end{align}
	where $ \GAMMA'  \in  \mathbb{C}^{ n_a\times n_y }$ is the control kernel and $ \hat{\GAMMA}' $ is its Fourier transform. However, the derivation of the kernel is simplified if instead the actuation is expressed only in terms of $ \y_1 $, i.e., ignoring the influence of the actuator on the sensor, 
	\begin{align}
		\V{a}(t) &= \int_{-\infty}^\infty  \GAMMA(\tau) {\y}_1 (t-\tau) d\tau, &
		\Vh{a}   & = \hat{\GAMMA} \hat{\y}_1 \label{eq:Gamma},
	\end{align}
  	where again $ \GAMMA \in  \mathbb{C}^{ n_a\times n_y }$ and $ \hat{\GAMMA } $ its Fourier transform.
  	
    This implies no loss of generality:  as $ \y_2 $ is a function only of the previous actuation, which is known, it can be computed using the actuator impulse response and subtracted from $ \y $ to obtain $ \y_1 $.  The approach  is equivalent to the formalism of an internal model control \citep{morari1989robust}. 
	The control kernel $ \GAMMA $ is then chosen so as to  minimize a cost functional that trades off the expected value of targets and actuation, 
	\begin{equation}\label{eq:LQGfunc}
	J = \int_{-\infty}^{\infty} \langle \z^\dagger (t) \z(t) + \V{a}^\dagger(t)\M{P}\V{a}(t) \rangle dt 
	= \int_{-\infty}^{\infty} \langle \hat{\z}^\dagger  \hat{\z} + 
	\Vh{a}^\dagger\M{P}\Vh{a} \rangle d\omega,
	\end{equation}
	where $ \M{P} \in  \mathbb{C}^{ n_a\times n_a }$ is a positive-definite matrix containing actuation penalties. For simplicity, we do not include a state cost matrix. This does not imply any loss of generality, as any state cost matrix $ \M{Q}  \in \mathbb{C}^{ n_u\times n_u }$, which must be positive definite, can be absorbed into the definition of the targets as $ \Cz'= \Cz \M{Q}_c $, where $ \M{Q} = \M{Q}_c  \M{Q}_c^{H} $ is a Cholesky decomposition of $ \M{Q} $.

	Using the identity $ Tr(\boldsymbol{\Phi}\boldsymbol{\Psi})=Tr(\boldsymbol{\Psi}\boldsymbol{\Phi} ) $, valid for any matrices $\boldsymbol{\Psi}$ and $\boldsymbol{\Phi} $ with suitable sizes, the functional is re-written as
	\begin{equation}\label{eq:LQGfunc_Tr}
	J =
	\int_{-\infty}^{\infty} \langle Tr( \hat{\z}(\omega)\hat{\z}^\dagger (\omega))  + 
	Tr(\M{P}\Vh{a}(\omega)\Vh{a}^\dagger(\omega))  \rangle d\omega.
	\end{equation}
	The functional is expanded using
	\begin{align}
	\langle 	\hat{\z} \hat{\z}^\dagger \rangle =&
				\langle  \hat{\z}_1 \hat{\z}_1^\dagger  \rangle+ 
				\langle  \hat{\z}_2 \hat{\z}_1^\dagger  \rangle+
				\langle  \hat{\z}_1 \hat{\z}_2^\dagger  \rangle +
				\langle  \hat{\z}_2 \hat{\z}_2^\dagger \rangle ,\\
		\label{eq:espaa}
		\langle \Vh{a} \Vh{a} ^\dagger \rangle =&\hat{\GAMMA} \langle \hat{\y}_1 \hat{\y}_1^\dagger \rangle\hat{\GAMMA}^\dagger.  
	\end{align}
	 Before further expanding these terms, we define
	\begin{align} \label{eq:Hh_def}
		\hat{\HH}(\omega) =& \Raz^\dagger(\omega)  \Raz(\omega)  +\M{P} , &
		\hat{\Hh}(\omega) =& -\Raz^\dagger(\omega) .
	\end{align}
	As shown in Appendix \ref{sec:WHcontrol}, these terms are the counterparts of $ \hat{\GG} $ and  $ \hat{\Gg} $ for the full-knowledge control problem.
	It is now straight-forward to show that
	\begin{align}
	\label{eq:espy1y1}
	\langle \hat{\y}_1 \hat{\y}_1 ^\dagger \rangle =&
	\hat{\GG},\\
	\label{eq:espz1z1}
	\langle \hat{\z}_1 \hat{\z}_1 ^\dagger \rangle =& \Rfz \Mh{F}\Rfz^\dagger,  \\
	\label{eq:espz2z2}
	\langle \hat{\z}_2 \hat{\z}_2 ^\dagger \rangle =& 
	\hat{\Hh}^\dagger\hat{\GAMMA}  \hat{\GG}\hat{\GAMMA}^\dagger \hat{\Hh},\\
	\label{eq:espz2z1}
	\langle \hat{\z}_2 \hat{\z}_1 ^\dagger \rangle =& 
	-\hat{\Hh}^\dagger \hat{\GAMMA} \hat{\Gg}^\dagger.
	\end{align} 
	The cost functional can then be expressed as
	\begin{align}
	\begin{aligned}
	J = &
	\int_{-\infty}^{\infty} 
	Tr\left( 
	  \Rfz \Mh{F}\Rfz^\dagger + \hat{\Hh}^\dagger\hat{\GAMMA}_{nc}  \hat{\GG}\hat{\GAMMA}_{nc}^\dagger \hat{\Hh} -
	  \hat{\Hh}^\dagger \hat{\GAMMA}_{nc} \hat{\Gg}^\dagger \Cz^\dagger -
	  \Cz \hat{\Gg} \hat{\GAMMA}_{nc}^\dagger  \hat{\Hh} \ \right)d\omega
	\\+ &		\int_{-\infty}^{\infty} 
	Tr\left( \M{P}
	\hat{\GAMMA}_{nc} \hat{\GG} \hat{\GAMMA}_{nc}^\dagger   \right)d\omega,
	\end{aligned}
	\end{align}
	where the subscript ``nc" was added to emphasize the non-causal nature of the control that will be obtained. 
	Using  the cyclic property of the trace and isolating terms with $ \hat{\GAMMA}_{nc}^\dagger $ , we obtain
	\begin{align}
	\begin{aligned}\label{eq:JcontrolExpanded}
	J = &
	\int_{-\infty}^{\infty} 
	\left( 
	Tr \left(  \Rfz \Mh{F}\Rfz^\dagger 
	- \hat{\Hh}^\dagger \hat{\GAMMA}_{nc} \hat{\Gg}^\dagger \Cz^\dagger 
	\right) 
	+Tr\left( 
	\left(  \hat{\HH}
	 \hat{\GAMMA}_{nc} \hat{\GG}  -\hat{\Hh} \Cz \hat{\Gg}  \right) \hat{\GAMMA}_{nc}^\dagger   \right)
	 \right) d\omega,
	\end{aligned}
	\end{align}
	where $ \hat{\HH} = \hat{\Hh}\hat{\Hh}^\dagger  + \M{P} $ was used. The minimum of $J$ is found by differentiating \eqref{eq:JcontrolExpanded} with respect to~$ \hat{\GAMMA}_{nc}^\dagger $, leading to
	\begin{align} \label{eq:gamma}
	\hat{\HH} \hat{\GAMMA}_{nc} \hat{\GG}   =  \hat{\Hh}  \hat{\Gg},
	\end{align}
	which can be solved for the optimal control kernel as 
	\begin{equation}
	\label{eq:Gamma_nc}
	\hat{\GAMMA}_{nc} =	\hat{\HH}^{-1}\hat{\Hh}  \hat{\Gg} \hat{\GG}^{-1} . 
	\end{equation}

	From \eqref{eq:Gamma_nc},  it can be seen that the optimal non-causal control kernel is a combination of the optimal non-causal estimation of targets ($ \hat{\Gg} \hat{\GG}^{-1}  $, solution of \eqref{eq:WFnoncausal}) with the optimal full-knowledge, non-causal control, described in Appendix \ref{sec:WHcontrol},  ($\hat{\HH}^{-1}\hat{\Hh} $, solution of \eqref{eq:non-causalLQR}), acting to minimize these targets with actuation inputs. 
	
 	As this control kernel is, in general, not causal, it cannot be used in real-time applications. It does, however, provide upper bounds for the effectiveness of causal control, and it is the basis for the control method proposed by \cite{sasaki2018closedloop}, where the kernel is truncated to its causal part. This approach was also successfully applied to experiments \citep{maia2021realtime,brito2021experimental}. However,  truncating the non-causal control kernel is, in general, sub-optimal.
	
	To obtain the optimal causal control law,  causality is again enforced via Lagrange multipliers. The modified cost functional reads 
	\begin{equation}\label{eq:LQGfunc_Tr_causal}
		\begin{aligned}
	J'& = J + 
	\int_{-\infty}^{\infty} \Tr{
    \LAMBDA_- (t) \GAMMA_c(t) + \LAMBDA_-^\dagger (t) \GAMMA^\dagger_c(t) } dt \\
    & = 
	J +	\int_{-\infty}^{\infty} \Tr{
	\hat{\LAMBDA}_-  \hat{\GAMMA}_c + \hat{\LAMBDA}_-^\dagger  \hat{\GAMMA}^\dagger_c } d\omega,
		\end{aligned}
	\end{equation}
	where the subscript $c$ emphasizes the causal nature of the control that will be obtained.
	Just as in the estimation problem, the linear terms added to $ J $, contribute to an additional term when taking the derivative of $ J' $. The control kernel is now given by the solution of 
	\begin{align}\label{eq:gammaplus}
	\hat{\HH} \hat{\GAMMA}_c \hat{\GG}   +\hat{\LAMBDA}_- =    \hat{\Hh} \Cz \hat{\Gg}. 
	\end{align}
	Again, as in the estimation problem, the requirements that ${\GAMMA(t<0)=0}$ and ${\LAMBDA_-(t>0)=0}$ makes this a well-posed problem, and specifically another type of Wiener-Hopf problem.
	The procedure to solve this problem is equivalent to the one used for the estimation problem, and is presented in Appendix~\ref{app:WH}.
	

	The control kernel $ \GAMMA' $ can be now recovered from $ \GAMMA $. The expression for the actuation 
	\begin{align}
	\Vh{a} = \hat{\GAMMA} \y_1 = \hat{\GAMMA} (\hat{\y} -\hat{\y}_2)  =\hat{\GAMMA} (\hat{\y} -\Ray \Vh{a}) , 
	\end{align}
	can be re-written as
	\begin{align} \label{eq:gammap}
	\Vh{a} = \underbrace{ \left(\I + \hat{\GAMMA} \Ray \right)^{-1} \hat{\GAMMA}}_{\hat{\GAMMA}'} \hat{\y},
	\end{align}
	thus recovering $ \GAMMA' $.
	
\begin{figure}
		\centering
				\def\svgwidth{.7\linewidth}
				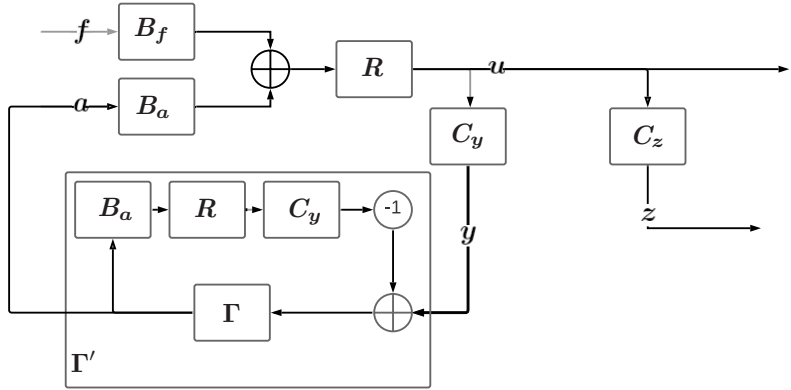
		\caption{Closed-loop control diagram. }
		\label{fig:controlschemeclosed}
	\end{figure}

	The closed-loop control diagram is illustrated in figure \ref{fig:controlschemeclosed}, where the relation between $ \hat{\GAMMA}' $ and $ \hat{\GAMMA} $ is shown. As $ \y_2 $ is a function only of the previous actuation, it can be computed in real-time and  subtracted from $ \y $ to obtain $ \y_1 $. This procedure is by construction included in    $ \GAMMA'$.  
	
	Note that it is the process of removing the actuator's response from the sensor readings that can lead to instabilities if the feedback is not accurately modelled \citep{belson2013feedback}. For amplifier flows, the actuators are typically located downstream of the sensors, the feedback tends to be small, and  the optimal control is thus  robust~\citep{schmid2016linear}. This explains the successful use of optimal control in many studies~\citep{semeraro2011feedback,barbagallo2012closedloop,morra2020realizable,sasaki2020role,maia2021realtime}. As our method targets amplifier flows, the issue of robustness is not further addressed. 
	

\subsection{ Recovering Kalman and LQR Control Gains} \label{sec:gains}

We now demonstrate that gain matrices for Kalman-filter estimation ($\M{L}$), and LQR control ($\M{K}$) can be recovered from the Wiener-Hopf formalism. As both methods have the same optimality properties, they amount to different approaches for obtaining the same result if applied to a system satisfying the common assumptions in the derivation of Kalman filters and LQR control, such as white-in-time disturbances., i.e., $\Mh{F}(\omega) = \M{F}_0$.


From the Kalman-filter estimation equation \citep{aastrom2013computer},
\begin{align}
	\dfrac{d\tilde{\x}}{dt} = (\A -\M{L} \Cy)\tilde {\x} + \M{L} \y,
\end{align}
the impulse response of the i-th sensor, $ y_i(t) =\delta_{j,i}\delta(t) $, is such that $ \tilde{\x}(0^+) = \M{L}_i $, where $ \M{L}_i $ is the i-th row of $ \M{L} $. Thus, from \eqref{eq:stateEst}, 
\begin{align}
	\M{L} =\Tx(0).
\end{align}

The LQR control gains can be recovered by emulating an initial condition $ \x_0 $ at $ t=0 $, which can be accomplished  by setting $ \Bf=\I $ and using $ \f(t) = \x_0 \delta(t) $, or equivalently $ \hat{\z}_1 = \Cz\R \x_0 $. From the LQR framework, 
\begin{equation}\label{eq:optimalAct1}
	\V{a}(t) = \M{K}\x(t).
\end{equation}
Since $ \x(t<0)=0 $, it follows that $ \V{a}(t<0) =0$. Comparing with the solution of \eqref{eq:WHcausalsol_H}, \eqref{eq:WHlqr} and \eqref{eq:ftoy_1}, we have
\begin{align}\label{eq:optimalAct2}
	\begin{aligned}
		\Vh{a}(\omega) & =   \hat{\HH}_+^{-1}(\omega) \left(\hat{\HH}_-^{-1} (\omega) \hat{\Hh} (\omega) \hat{\z}(\omega) \right)_+   \\
		& =   \underbrace{\hat{\HH}_+^{-1}(\omega) \left(\hat{\HH}_-^{-1} (\omega) \hat{\Hh} (\omega) \Cz \R(\omega)  \right)_+}_{\hat{\boldsymbol{\Pi}}_c(\omega)} \x_0 
	\end{aligned}.
\end{align}
Comparing \eqref{eq:optimalAct1} and \eqref{eq:optimalAct2}, we have $ \M{K} = \boldsymbol{\Pi}_c(0) $.

Knowledge of LQR gains $\M{K}$ and the Kalman filter $\M{L}$ may be useful to understand regions of the flow that require accurate estimation to obtain effective control strategies, as discussed by  \cite{freire2020actuator}.


%

	\subsection{ Validation using a linearized Ginzburg-Landau problem} \label{sec:comparison_on_GL}

	We here compare the kernels and  gains computed with the present method to those obtained with the Kalman-filter and the LQR approaches. We use a linearized Ginzburg-Landau problem, for which  standard tools can be used for the computation of estimation and control gains without resorting to model reduction. Such models have been used in many previous studies \citep{bagheri2009amr,lesshafft2018artificial,cavalieri2019amr}.  The model reads
	\begin{align} 
	\frac{\partial \x(x,t)}{\partial t} =&  \A  \x(x,t)  + \f(x,t) \label{eq:GL_timeDomain}, &
	\A &= - U \frac{\partial }{\partial x} + \mu(x) + \gamma \frac{\partial^2 }{\partial x^2},
	\end{align}
	and we use the same parameters as in  \cite{martini2020resolventbased}, namely: $ U=6 $, $ \gamma=1-\ii $ and $ \mu(x)=\beta \mu_c(1-x/20) $, where $\beta=0.1$ was used and $ \mu_c=U^2 \Real (\gamma) / |\gamma|^2$ is the critical value for onset of absolute instability \citep{bagheri2009amr}.

\begin{figure}[t]
		\centering
		
		\psfragfig[scale=0.5]{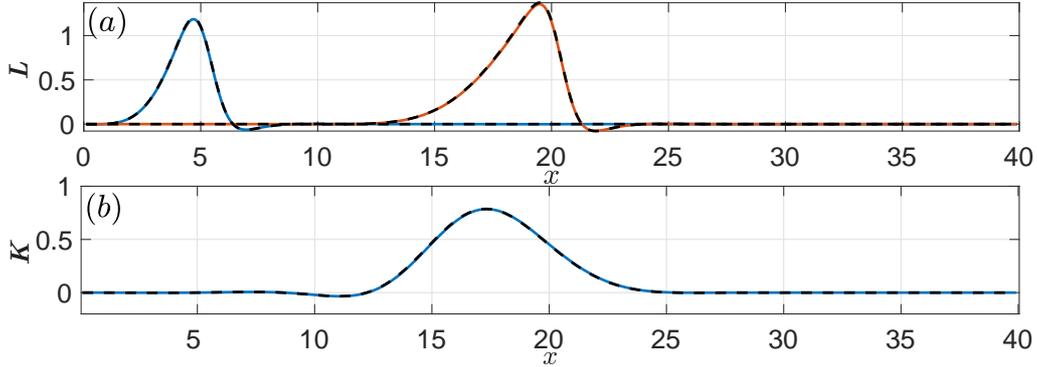} {		
			\psfrag{KK}{$\M K$}
			\psfrag{LL}{$\M L$}
		}	
		\caption{(a) Blue/red lines  show the estimation gains corresponding the first and second sensor obtained
			using proposed method. Dashed black lines show results obtained from the algebraic Riccati equation. (b) Same for control gains.  }
		\label{fig:GL_GAins}
	\end{figure}

\begin{figure}[th]
		\centering
		
		\psfragfig[scale=0.5]{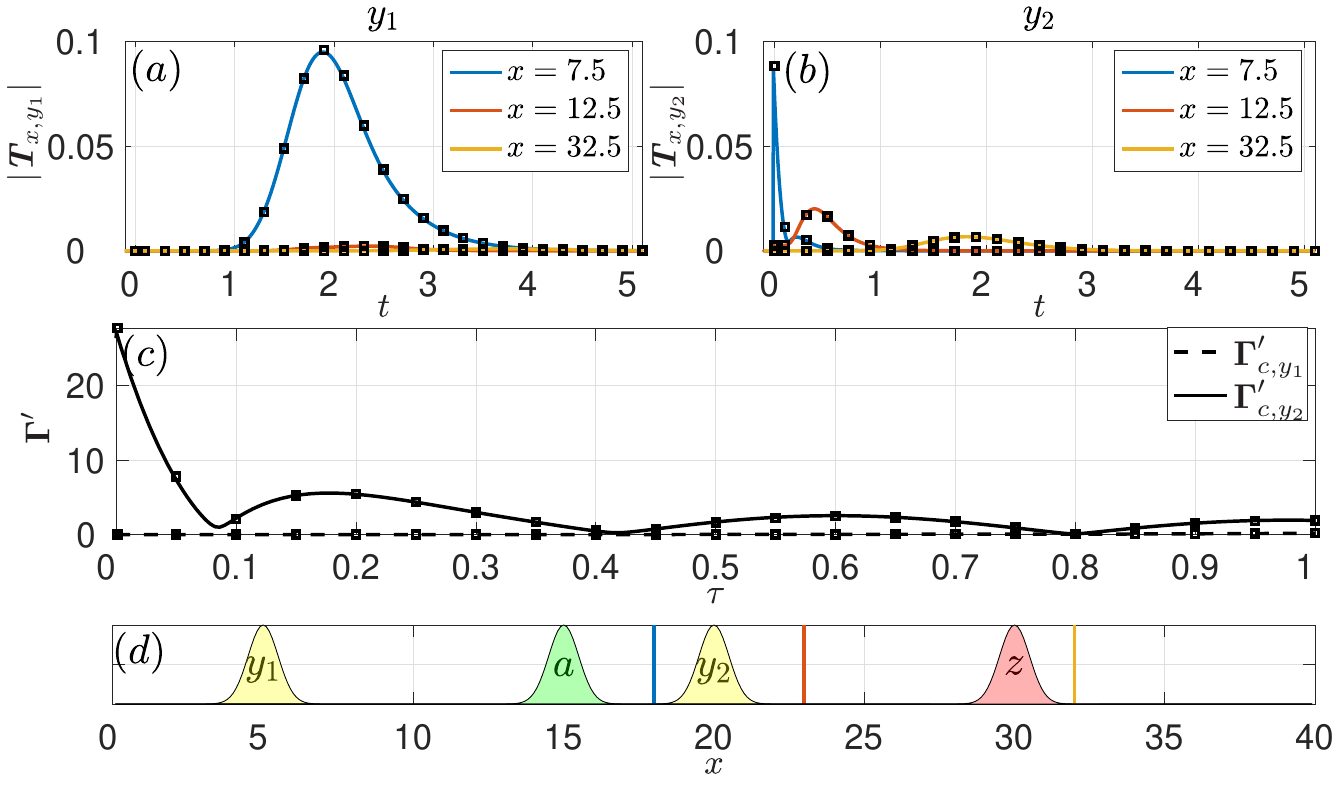}{
			\psfrag{Txy1}{$|\T_{x,y_1}|$}
			\psfrag{Txy2}{$|\T_{x,y_2}|$}
			\psfrag{GA1}{$\GAMMA_{c,y_1}'$}
			\psfrag{GA2}{$\GAMMA_{c,y_2}'$}
			\psfrag{GA}{$\GAMMA'$}
		}	
		
		\caption{(a-b) Kernels for the estimation of the system at different locations. Colour lines/black markers show results from the proposed method and from the algebraic Riccati equations. (c) Same for the control kernel. (d) Spatial support of the sensors, actuator, and target. Vertical lines show locations for which the estimation kernels  in (a-b) were computed.}
		\label{fig:glcausalestimationkernel}
	\end{figure}

	For this validation, we use $ \hat{\F}=\I $, two sensors located at $ x=5 $ and $ 20 $, an actuator at $ x=15 $, and a target at $ x=30 $.
	Figure \ref{fig:GL_GAins} compares gains obtained from the proposed approach and from the algebraic Riccati equations. Both approaches  produce identical gains, indicating their equivalence when white-noise forcing is assumed. Estimation and control kernels are shown in figure \ref{fig:glcausalestimationkernel}, again showing the equivalence between the two methods. The Kalman and LQG kernels are obtained as follows. From \eqref{eq:stateEst}, the state estimation is obtained as
	\begin{align}	
		\tilde \x(t) = \int_{-\infty}^\infty \T_{u,kal}(\tau) \y(t-\tau) d\tau,
	\end{align}
	with the estimation kernel given by,
	\begin{align}
		\T_{u,kal}(\tau) = \begin{cases}
		e^{ (\A- \M{L} \Cy)\tau} \M{L}&  \tau\ge  0  \\
		0&  \tau<0  
		\end{cases}.
	\end{align}
	The LQG kernel is obtained from
	\begin{align}
		\frac{d}{dt} \tilde{\x}(t) & = \A  \tilde \x(t)  + \Ba  \V{a}(t)+\M{L} \left( \y(t )- \M{L} \Cy  \tilde \x(t)  \right) , \\
	 	\V{a}(t) & = - \M{K}		 	\tilde \x(t),
	\end{align}
	with the solution reading
	\begin{align}	
		 \V{a}(t) = \int_{-\infty}^\infty \GAMMA'_{lqg}(\tau) \y(t-\tau) d\tau,
	\end{align}
	and where the control kernel is given by
	\begin{align}\label{eq:galmmalqg}
	\GAMMA'_{lqg}(\tau) = 
	\begin{cases}
		\M{K} e^{ \left(\A-\M{L}\Cy-\Ba\M{K}\right) \tau} \M{L} &  \tau\ge  0  \\
		0&  \tau<0  
	\end{cases}.
	\end{align}
	Figures \ref{fig:GL_GAins} and \ref{fig:glcausalestimationkernel} show that the Riccati-based and  proposed approaches provide the same results, illustrating their equivalence.




	\section{Implementation}\label{sec:inplementation}

	The size of the Wiener-Hopf problems  \eqref{eq:WFcausal} and \eqref{eq:gammaplus} is independent of the size of the linear system $n_{u}$. 
	The dominant cost to solve these problems is the Wiener-Hopf factorization of $ \hat{\HH} $ and $ \hat{\GG} $, which are matrices that scale with $ n_y $ and $ n_a $, respectively. Accordingly, solution can be obtained with low cost for arbitrarily large systems, as long as the number of sensors and actuators remains reasonable. However, the coefficient matrices of the Wiener-Hopf problems are functions of the resolvent operator \eqref{eq:resOp}, and thus requires the inversion of matrices of size $ n_u $ to be constructed, which is unfeasible for large systems.
		
	
	In  \S~\ref{sec:timemarching}, we show a method to construct the coefficients of the Wiener-Hopf problems efficiently, with  approach to incorporate effective forcing models  presented in \S~\ref{sec:colours}. In  \S~\ref{sec:adjointless}, we show that the terms in the equations correspond to CSD matrices that can be obtained directly from numerical and physical experiments. The latter technique allows application of the tools developed in this paper when adjoint solvers are not available or in experimental setups.

	\subsection{Matrix-free implementation} \label{sec:timemarching}

	%
	%
	%
	%
	
	In this section we apply methods developed in previous works \citep{martini2020resolventbased,martini2021efficient} to construct the terms $ \hat{\HH} $, $ \hat{\Hh} $, $ \hat{\GG} $ and $ \hat{\Gg} $, circumventing the need to construct the resolvent operator, which would make their construction prohibitive in any practical scenario. The approach consists of using time-domain solutions of the linearized equations to obtain the action of the resolvent operator on a vector, which in turn can be used to reconstruct the operator. 
	
	As an example, assume that the action of $ \Rfy $ on a vector $\Vh{v}_d$ can be efficiently computed. The operator can be constructed row-wise by setting $ \Vh{v}_d=\V{e}_i $, where $ \V{e}_i $ is an element of the canonical basis for the forcing space:  the resulting vector $ \Rfy\V{e}_i $ provides  the i-th row of $ \Rfy $. If $ n_f $ is small, $ \Rfy $ can recovered by repeating the procedure for $ i=1,\dots, n_f $, and subsequently used for the construction of $ \hat{\HH} $. 
	

	To obtain the action of the resolvent operator on a vector for all resolved frequencies simultaneously, we use an approach based on the transient-response method (TRM), developed in \cite{martini2021efficient}.  The action of $ \Rfy $ and $ \Rfz $ on a vector $ \Vh{v}_d(\omega)$ is obtained using the following system
	\begin{align}
		\label{eq:dir}
		\dfrac{d\V{q}}{dt}(t) &= \A  \;              \V{q}(t) + \Bf \; \V{v}_d(t), & \V{r}_d(t) &= \Cy \V{q}(t),& \V{s}_d(t) &= \Cz \V{q}(t),
		\\	\label{eq:dirfreq}
		 \Vh{q} &= \R\Bf \Vh{v}_d, &
		\Vh{r}_d &= \underbrace{\Cy\R\Bf}_{\Rfy} \Vh{v}_d, & 
		\Vh{s}_d &= \underbrace{\Cz\R\Bf}_{\Rfz} \Vh{v}_d,
	\end{align}
	where \eqref{eq:dirfreq} is obtained form a Fourier transform of \eqref{eq:dir}.
	Here $ \Vh{v}_d $ is regarded as an input, with $ \Vh{r}_d $ and $ \Vh{s}_d $ as outputs of the system, with their dimensions implicit by the context. Using $ \V{v}_d(t)=\V{e}_i\delta(t) $ ensures that the canonical basis is used for all frequencies.  A time-domain solution of \eqref{eq:dir} will be referred to as a forcing direct run.  An actuator direct run is obtained by replacing $ \Bf $ with $ \Ba $, and provides the action of $ \Ray $ and $ \Raz $ on a given vector. 
	
	Similarly, the action of the operators $ \Rfy^\dagger $ and $ \Ray^\dagger $ on a vector is constructed by time-marching  the adjoint system,
	\begin{align}
		\label{eq:adj}
		-\dfrac{d\V{q}}{dt}(t) =& \A^\dagger  \;              \V{q}(t) + \Cy^\dagger \; \V{v}_a(t), & \V{r}_a(t) &= \Bf^\dagger \V{q}(t), & \V{s}_a(t) &= \Ba^\dagger \V{q}(t),
		\\	\label{eq:adjreq}
		\Vh{q} &= \R^\dagger\Cy^\dagger \Vh{v}_a, &
		\Vh{r}_a &= \underbrace{\Bf^\dagger\R^\dagger\Cy^\dagger}_{\Rfy^\dagger} \Vh{v}_a, & 
		\Vh{s}_a &= \underbrace{\Ba^\dagger\R^\dagger\Cy^\dagger}_{\Ray^\dagger} \Vh{v}_a,
	\end{align}
	referred here as a sensor adjoint run. A target adjoint run is obtained by replacing $ \Cy $ with $ \Cz $ and can be used to obtain the action of $ \Rfz^\dagger $ and $ \Raz^\dagger $.
	
\begin{figure}[t]
		\centering
	\def\svgwidth{.9\linewidth}
	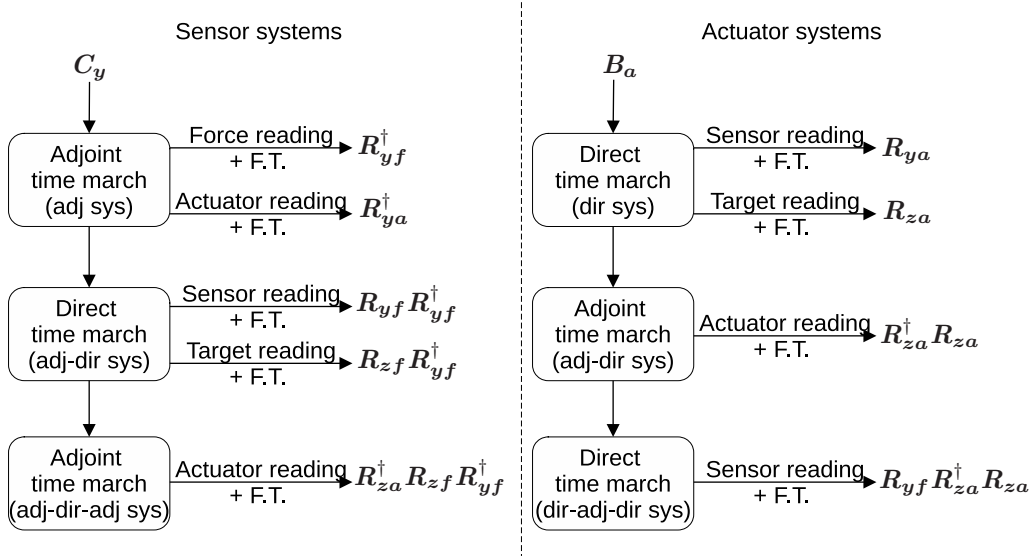
		\caption{Illustration of the terms obtained from different combinations of direct and adjoint time-domain solutions. The left and right flowcharts represent a sensor adjoint(-direct-adjoint) system and an actuator direct(-adjoint-direct) system. Readings refer to inner products of the state with sensors/forcing/targets/actuators spatial supports, and F.T. to a Fourier transform. Diagrams for the target and forcing systems are equivalent to the left(right) diagrams with $\Cy(\Ba)$ exchanged by $\Cz(\Bf)$, and  the $\V{y}(a)$ subscripts of the right-most terms in each quantity exchanged by $\V{z}(f)$.}
		\label{fig:schemeruns}
	\end{figure}

	The action of more complex terms can be obtained by solving \eqref{eq:dir} and \eqref{eq:adj} in succession. For example, if the output $ \V{r}_a $ of a sensor adjoint run as an input of an direct run, then the outputs of the latter will be given by ${ \Vh{r}_d = \Rfy\Rfy^\dagger\Vh{v}_a} $ and ${ \Vh{s}_d = \Rfz\Rfy^\dagger\Vh{v}_a }$. The resulting system is referred as a sensor adjoint-direct run. Following similar procedures, adjoint-direct, direct-adjoint-direct and adjoint-direct-adjoint  runs are constructed. The terms whose action are obtained from each of these runs are illustrated in figure \ref{fig:schemeruns}.
	
	In practice, to link together direct and adjoint runs as described above, checkpoints of the output of one time integration are saved to disk and subsequently read and interpolated in the following run to be used as forcing terms. Details of this procedure as well as the impact of different interpolation strategies are described by~\cite{martini2021efficient}, to which we refer the reader for details. Alternatively,  \cite{farghadan2021randomized_aaron} developed an approach that minimizes the data to be retained using streaming Fourier sums to obtain the action of the operator on a discrete set of frequencies.  
	
	The most effective approach for constructing the Wiener-Hopf problems depend on the rank of forcing and targets. In the following sections, we outline the best approach for four possible scenarios and discuss the associated computational cost and the types of parametric studies that can be easily conducted in each case.  
	
	\subsubsection{Low-rank $ \Bf $ and $ \Cz $}  \label{sec:lrBf_lrCz}
	
	 In this case scenario, the terms $\Rfy$, $ \Raz^\dagger $ and $ \Rfz^\dagger$  are small matrices and can be obtained using $ n_f $ direct and $ n_z $ adjoint runs. With those terms, $ \hat{\GG} $, $ \hat{\Gg} $, $ \hat{\HH} $ and $ \hat{\Hh} $ can be constructed.  
	 
	 Note that as $ \Rfy $ is obtained from state readings in $ \eqref{eq:dir} $, storing snapshots of $ \V{q}(t)$ allows for computing $ \Rfy $ for any $ \Cy $ from inexpensive data-post processing. Using the same strategy in \eqref{eq:adj} allows $ \Raz^\dagger $ to be computed for any $ \Ba $. It is thus possible to inexpensively compute the control kernels for any sensor and actuator when the forcing and targets are low rank.  
	
	\subsubsection{High-rank $ \Bf $ and low-rank $ \Cz $} \label{sec:hrBf_lrCz}
	
	In this scenario, $\Rfy$ and $ \Rfz $ are large matrices, and it is  impractical to construct and manipulate them. Instead, $\Rfy \Rfy^\dagger$ and $\Rfz \Rfy^\dagger$, which are still small matrices, are obtained from $ n_y $ sensors adjoint-direct runs, while $ \Ray $ can be constructed from $ n_z $ target adjoint runs, as in \S~\ref{sec:lrBf_lrCz}.
	
	These terms can then be used to construct $ \hat{\GG} $, $ \hat{\Gg} $, $ \hat{\HH} $ and $ \hat{\Hh} $, assuming that $ \F=\I $. A strategy to efficiently include complex forcing  models in this approach will be detailed in \S~\ref{sec:colours}. As the sensor position is an input of the sensor adjoint-direct run, it becomes expensive to perform parametric sensor studies in this scenario. However, actuator placement studies are still inexpensive.
	
	\subsubsection{Low-rank $ \Bf $ and high-rank $ \Cz $} \label{sec:lrBf_hrCz}
	This scenario is similar to the one in \S~\ref{sec:hrBf_lrCz}, with the role of sensors and actuators reversed.  Here $ \Rfy$ is constructed from $ n_f $ forcing direct runs, and $ \Raz $ is obtained from $ n_a $ actuator direct runs. Since $ \hat{\Hh} $ and $ \hat{\Gg} $ are both large matrices, the product $ \hat{\Hh}  \hat{\Gg} $  is constructed directly from $ \Raz^\dagger \Rfz$, which can be obtained from $ n_f $ forcing direct-adjoint runs, and $\Rfy^\dagger$, which is obtained directly from  $ \Rfy $.
	 
	In this scenario, parametric studies of actuator placement are costly, while such studies for the sensors are inexpensive.
	
	\subsubsection{High-rank $ \Bf $ and $ \Cz $}  \label{sec:hrBf_hrCz}
	Finally, in this scenario,  the product $ \hat{\Gg} \hat{\Hh} $ cannot be broken into the product of small matrices as before, and thus has to be constructed directly using $ n_y $ sensor adjoint-direct-adjoint runs or $ n_a $ actuator direct-adjoint-direct runs.  The term  $ \hat{\HH} $($ \hat{\Gg} $) is constructed from $ n_y $($ n_a $) sensor adjoint-direct(actuator direct-adjoint) runs. Again, we have assumed here that $ \F=\I $.
	
	In this scenario,  parametric studies of both sensors and actuators are expensive.

\subsubsection{Summary}
\modification{
A brief summary of the costs of each scenario  is presented in table \ref{tab:numberofruns}. Note that the descriptions above focused on the solution of the Wiener-Hopf problem, which provide $ \GAMMA $. In some scenarios,  supplementary runs are required to obtain $ \Ray $, which is required to construct $ \GAMMA' $ as in \eqref{eq:gammap}. This term can be obtained using sensor adjoint runs or actuator direct runs. We also report the extra runs required to obtain $ \Rfz^\dagger\Rfz$, which will be used to obtain offline estimates of the control performance in \S~\ref{sec:discussion_implementation}.

	\begin{table}
	\centering
	\begin{tabular}{c|c|c|c|c}
		\rule[-1ex]{0pt}{2.5ex} Scenario & Runs used &
		\multicolumn{3}{c}{Time-domain solutions required to} \\
		\rule[-1ex]{0pt}{2.5ex}  &   & construct $ \GAMMA $ & construct $ \GAMMA' $ & estimate performance\\
		\hline
		\multirow{2}{*}{\S~\ref{sec:lrBf_lrCz}}	& $ n_f $ force dir & \multirow{2}{*}{$n_f+n_z$ } & \multirow{2}{*}{$+\min(n_y,n_z)$ } & \multirow{2}{*}{$+0$}   \\
		& $ n_a $ target adj & & &  \\
		\hline
		\multirow{2}{*}{\S~\ref{sec:hrBf_lrCz}}	&  $ n_y $ sensor adj-dir  & \multirow{2}{*}{$2n_y+n_z$ } & \multirow{2}{*}{$+0$ } &  \multirow{2}{*}{$+n_z$}\\
		&  $ n_z $ target adj  & & & \\
		
		\hline
		\multirow{2}{*}{\S~\ref{sec:lrBf_hrCz}}	&  $ n_f $ force dir  & \multirow{2}{*}{$n_f+2n_z$ } &  \multirow{2}{*}{$+\min(n_y,n_a)$ } &  \multirow{2}{*}{$+0$}\\
		&  $ n_z $ target adj-dir  &  & & \\
		\hline
		\multirow{2}{*}{\S~\ref{sec:hrBf_hrCz}}	&  $ n_y $ sensor adj-dir(-dir)   & 
		{$2n_y+2n_a+$ }& \multirow{2}{*}{$+0$ } & \multirow{2}{*}{n.a. }  \\
		&  $ n_a $ actuator dir-adj(-dir)  &  $ \min(n_y,n_a) $ & & \\
		
	\end{tabular}
	\caption{Description and number of  time-domain solutions needed in each scenario to construct the $ \GAMMA $, and the extra time-domain solutions required to construct $ \GAMMA' $ and to perform an offline estimation of the control performance.}
	\label{tab:numberofruns}
\end{table}
}

\subsection{Using coloured forcing statistics} \label{sec:colours}
	Previous results show that incorporating accurate coloured forcing statistics in the model is important to obtain accurate estimates of the flow state~\citep{chevalier2006state,martini2020resolventbased,amaral2021resolventbased}. We now detail how to incorporate forcing colour for each of the scenarios described in the previous sections.
	
	In the scenarios described in \S~\ref{sec:lrBf_lrCz} and \S~\ref{sec:lrBf_hrCz}, since the term $\Rfy$ is a small matrix, forcing colour could be easily included a posteriori in the construction of $ \hat{\GG} $ and $ \hat{\Gg} $. In the scenarios \S~\ref{sec:hrBf_lrCz} and \S~\ref{sec:hrBf_hrCz}, where $\Rfy\Rfy^\dagger$ is directly obtained from a time-domain solution, the inclusion of forcing colour requires time-domain convolutions of the output of the sensor adjoint system with the forcing cross-correlation matrices, before its use as an input of the direct system. However the  convolutions of a large matrix ($\M{F}$) and a vector (output of the adjoint system) is typically unfeasible. To circumvent this limitation,  the approach proposed by \cite{martini2020resolventbased} can be used.  An estimated forcing CSD, obtained from low-rank flow measurements, is used to construct $ \hat{\GG} $ and $ \hat{\Gg} $. This can be done without explicit construction of $ \Mh{F} $, as will be shown next. 
	
	From a set of $ n_{y'} $ auxiliary sensors readings, $ \y' $, obtained as   $ \Cy'\x  $,  which we assume to contain the set of sensors that will be used for estimation and/or control ($ \y $), and their CSD, $ \M{Y}' $,  the estimated forcing colour is obtained as
	\begin{align}
		\Mh{F}' = \Mh{T} _f \Mh{\Y} ' \Mh{T} _f^\dagger, 
	\end{align}
	where
	\begin{align}
		\Mh{T} _f'  =\R^\dagger_{y'f}   \left( \R_{\y'\f} \R_{\y'\f}^\dagger + \Mh{\N} ' \right) ^{-1} 
	\end{align}
	is a transfer function that estimates forcing from measurements with a prior assumption of white-noise forcing~\citep{towne2020resolvent,martini2020resolventbased}.

  Estimates for $ \hat{\GG} $ and $ \hat{\Gg} $ using the estimated forcing CSD, $ \hat{\F}' $, referred to here as $ \hat{\GG}' $ and $ \hat{\Gg} ' $, read
\begin{align}
	\hat{\GG}' &= \Rfy \Mh{F}' \Rfy^\dagger + \Mh{\N}  
	= \Rfy \Mh{T} _f \Mh{\Y} ' \Mh{T} _f^\dagger   \Rfy^\dagger + \Mh{\N}  
\\
	\hat{\Gg} ' &= \Rfz \Mh{F}' \Rfy^\dagger \;\;\;\;\;\;\;\,
	= \Rfz \Mh{T} _f \Mh{\Y} ' \Mh{T} _f^\dagger   \Rfy^\dagger 
\end{align} 
Note that $ \Mh{T}' _f \in \mathbb{C}^{n_f \times n_{y'}}$, and that $ n_f $ was assumed to be large. To avoid operations with large matrices, the compound term
\begin{align}
	\Rfz \Mh{T}' _f=  \Rfz \R^\dagger_{\y'\f}   \left( \R_{\y'\f} \R_{\y'\f}^\dagger + \Mh{\N} '\right) ^{-1},
\end{align}
that has size $ n_y \times n_{y'} $,	can be constructed from $ { \Rfz \R^\dagger_{\y'\f} \in \mathbb{C}^{n_z\times n_{y'}}}$ and  $ \R_{\y'\f} \R_{\y'\f}^\dagger \in \mathbb{C}^{n_{y'}\times n_{y'}}$. As 
  the terms extra terms  $ \R_{\y'\f}  $ and $ \R_{\y'\f} \R_{\y'\f}^\dagger $ can be obtained from the matrix-free approach described, the time-domain convolution can be replaced by a few extra time-domain solutions of the linearized problems and inexpensive post-processing.

\temporarySupression{\color{red}
To construct Kalman and control gains, the solution of \eqref{eq:WFcausal}, given by \eqref{eq:WNsolutioncausal_G}, and \eqref{eq:optimalAct2} can be used. For large systems however, they apparently requires multiple convolutions to be made, of the need to Fourier transform large vectors, to represent $ \R $ and $ \R^\dagger $. A close inspection however shows that only one time integral in the time domain is needed, as presented next.

From the definition of $ \Pi $ in \eqref{eq:optimalAct2}, $ \Pi(0) $ can be written as 
\begin{align}
	\label{eq:Pi0_dev1}
	\Pi(0) &=  \int_{-\infty}^\infty \HH_+^{-1}(\tau_1)  \left(\HH_-^{-1} * \M{V}^\dagger \right)_+(-\tau_1) d\tau_1,
\end{align}
where $ \M{V}^\dagger = \Matrix{\V{v}_1^\dagger, \dots , \V{v}_i^\dagger} $.

Noting that both $ \HH^{-1}_+(\tau) $ and $  \left(\HH_-^{-1} * \M{V}^\dagger\right)_+(\tau) $ are zero for $ \tau<0 $, it is thus apparent that \eqref{eq:Pi0_dev1} implies that $\M{Y}K= \Pi(0)=0 $. This is of course not the case. In fact $ \HH^{-1}_\pm(\tau) $ have a delta-like behaviour at $ \tau=0 $ ( see examples for $ \HH_\pm $ in figure \ref{fig:convergence}). Splitting $ \HH_\pm^{-1} $ as 
\begin{align}
 \HH_\pm^{-1}(\tau) =   \HH_\pm^{s,-1}(\tau) + \HH_\pm^{d,-1} \delta (\tau) 
\end{align}
where $ \HH_\pm^{s,-1} $ contains its smooth component and $ \HH_\pm^{d,-1} $ its delta-like intensity,
\eqref{eq:Pi0_dev1} becomes
\begin{align}
\label{eq:Pi0}
\M{K}= \Pi(0) &= \HH_+^{d,-1} \left( \HH_{-}^{d,-1} \M{V}^\dagger (0) + \int_{0}^\infty
   \HH_{-}^{s,-1}(-\tau')  \M{V}^\dagger(\tau')d\tau'\right),
\end{align}

Similarly, $ \M{L} =  \T_u(0) $ is given by
\begin{align}
\label{eq:Tu0}
	\M{Y}	L = \T_u(0)   &= \left( \M{Q} (0) \GG_{-}^{d,-1}  + \int_{0}^{\infty}
\M{Y}Q(\tau') \GG_{-}^{s,-1}(-\tau')  d\tau'\right) \GG_+^{d,-1} ,
\end{align}
where $ \M{Q} = \Matrix{\V{q}_1, \dots , \V{q}_i} $.

Using \eqref{eq:Pi0} and \eqref{eq:Tu0}, Kalman and control gains can be computed using only simple integrations of time marching data.
}

\subsection{Adjointless method}\label{sec:adjointless}
	We now show that the terms needed to construct the estimation and control kernels can be obtained from experimental data alone. We first write the sensors and targets in terms of the external disturbances, 
	\begin{align}
	\Matrix{\hat \y_1\\\hat \z_1} = 		\Matrix{\Rfy\\\ \Rfz} \hat \f	+ \Matrix{1 \\0} \hat \n.
	\end{align}
	Their CSDs are then obtained from the forcing statistics \citep{towne2020resolvent}, as
	\begin{equation}\label{eq:yzCSDs}
	\begin{aligned}
	\Matrix{ {\M{S}}_{y_1,y_1} & {\M{S}}_{y_1,z_1} \\ {\M{S}}_{z_1,y_1} & {\M{S}}_{z_1,z_1} } & =
	\Matrix{\Rfy\\\ \Rfz}  \hat \F 	\Matrix{\Rfy ^\dagger &  \Rfz^\dagger} + \Matrix{\hat \N & 0 \\ 0 & 0} 
	= 
	\Matrix{ \hat{\GG}  & \hat{\Gg} \\ 
		\hat{\Gg} ^\dagger & \Rfz \hat \F \Rfz^\dagger }.
	\end{aligned}
	\end{equation}	
	
	 The non-causal estimation is obtained via the transfer function $ \M{S}_{y,z}  \M{S}_{y,y}^{-1}  $.
	 The terms $ \hat{\GG}$ and $ \Cz \hat{\Gg} $ can be constructed from measurements of the uncontrolled system, as they correspond to sensor CSDs ($ {\M{S}}_{y_1,y_1} $)  and the cross-spectra between sensors and targets ($ {\M{S}}_{y_1,z_1} $), respectively. As the terms  $ \hat{\HH} $ and $ \hat{\Hh} $ can be obtained from the actuators' impulse response, all the required terms can be obtained using physical or numerical experiments. This not only provides a simple framework to obtain optimal control based on a data-driven approach, but it also relaxes the requirement of  an adjoint solver to obtain these control strategies for large systems, either by using readings from the non-linear problem, as done by \cite{martinelli2009feedback}, or from the direct linearized problem excited by stochastic forcing.
	 
	\modification{The adjointless approach presented here is in fact the classical application of the Wiener regulator, which is constructed directly from sensor/target CSDs, with the derivation presented here showing that the two approaches, the classical and the resolvent-based, are equivalent when the exact force model is used. However, the statistical convergence of the CSDs is considerably more expensive, and typically less accurate, than its construction from first principles (\S~\ref{sec:timemarching}), and thus the latter is preferable to a numerical experiment if an adjoint solver is available. 
	For physical experiments, where obtaining long time series is typically inexpensive, the method presented here provides an efficient way to obtain a data-driven optimal control law.}
	
	\subsection{The role of sensors for the estimation problem}\label{sec:discussions}
		
		Insights into the problem of sensor placement can be obtained from further analysis of \eqref{eq:yzCSDs}, focusing the discussion on non-causal estimation and white-noise forcing for simplicity. Sensor and target sensitivities to the forcing terms  are given by  $  \Rfy^\dagger $ and $ \Rfz^\dagger $. We define two forcing subspaces, $\Mh{W}_y$ and $\Mh{W}_z$,  spanned by the the rows of  $  \Rfy^\dagger $ and $ \Rfz^\dagger $.  Defining $\Vh{w}_{i,y}$ and $\Vh{w}_{i,z}$ to be orthogonal bases for these subspaces, the forcing CSD can be decomposed into four different subspaces:
		\begin{itemize}
			\item $ \mathcal F_{y,z} $: spanned by components given by $ \Vh{w}_{i,y}\Vh{w}_{i,z}^\dagger + \Vh{w}_{i,z}\Vh{w}_{i,y}^\dagger $, corresponds  to forcing components correlated with responses in both $ \y $ and $ \z $,
			\item $ \mathcal F_{y} $: the subspace spanned by components $ \Vh{w}_{i,y}\Vh{w}_{i,y}^\dagger $ that are orthogonal to $  \mathcal F_{y,z} $, containing forcing components correlated with responses in $ \y $ but not in $ \z $,
			\item $ \mathcal F_{z} $: the subspace spanned by components $ \Vh{w}_{i,z}\Vh{w}_{i,z}^\dagger $ that are orthogonal to $ \mathcal F_{y,z} $, containing forcing components correlated with responses in  $ \z $ but not in $ \y $,
			\item $ \mathcal F^\perp $: the complement of the three above subspaces, containing forcing components that are  correlated  with neither  $ \y $ nor $ \z $.
		\end{itemize}
		Defining $ \Mh{F}_{y} $ as the projection of $ \Mh{F} $ into $ \mathcal F_y $, with similar definitions for the other subspaces, \eqref{eq:yzCSDs} can be re-written as
		
		\begin{equation}\label{eq:yzCSDs2}
			\begin{aligned}
				\Matrix{ {\M{S}}_{y_1,y_1} & {\M{S}}_{y_1,z_1} \\ {\M{S}}_{z_1,y_1} & {\M{S}}_{z_1,z_1} }  = &
				\Matrix{\Rfy \hat \F_y \Rfy^\dagger +\hat \N & 0\\ 
					0         & 0 } + \\&
				\Matrix{\Rfy \hat \F_{y,z} \Rfy^\dagger& \Rfy \hat \F_{y,z} \Rfz^\dagger\\ 
					\Rfz \hat \F_{y,z} \Rfy^\dagger         & \Rfz \hat \F_{y,z} \Rfz^\dagger } + 
				\Matrix{ 0 & 0 \\ 
					0   & \Rfz \hat \F_z \Rfz^\dagger }.
			\end{aligned}
		\end{equation}		
		
		\modification{Each of these sub-spaces plays a different role in the estimation. The subspace $ \mathcal{F}_y $ generates responses at the sensors but not at the target, and thus it does not provide any useful information that can be used to construct the estimates, instead appearing in the problem in a similar way as does the sensor noise.  On the other hand, the subspace $ \mathcal{F}_{y,z} $ generates responses at sensors and targets, and thus the response to this forcing measured by the sensors can be used to estimate the corresponding response of the flow at the target locations. This is the term that is effectively used for target estimation.  Finally, the subspace $ \mathcal{F}_{z} $ corresponds to forces that have responses at the target but not at the sensors. Accordingly, the responses associated with it cannot be estimated.}
		
		This analysis has a direct relation with the concept of observable forces discussed in previous works by \cite{towne2020resolvent}: only target  components that are excited by forcing components that also generate readings on the sensors can be estimated, and as a consequence, controlled.  \cite{martini2020resolventbased} discussed how the correlation between difference forcing components can allow the estimation of the responses to non-observable forcing components, and this was shown to considerably improve turbulent flow estimation.
		Note that, in general, non-causal estimation/control is needed for use of the full correlation, i.e.  estimation/control of  all the target components which are correlated with the sensors.  The causality constraint allows for a real-time application at the price of deteriorating the estimation/control.

		To optimize the non-causal estimation, one should seek to minimize $ \F_z $ and reduce the effect of $ \F_y $ on the estimation of $  \F_{y,z} $. Although causality imposes extra restrictions, which further complicates the problem, the above discussion can provide insights into the sensor placement problem, which is still  an active topic of research.

	\subsection{Discussion}\label{sec:discussion_implementation}
	
	As previously stated, the Wiener-Hopf formalism proposed here provides large computational savings when compared to tools based on algebraic Riccati equations for large systems. While the latter requires solutions of algebraic Riccati equations for matrices of size $ n_u $, the former only requires factorization of matrices of size $ n_y $ and $ n_a $, and can thus be solved for large systems. The largest cost of the approach is associated with construction of the terms that define the Wiener-Hopf problem. Our matrix-free methods allow the construction of these terms using direct and adjoint simulations or experimental data (as shown in \S~\ref{sec:adjointless}), making the proposed approach widely applicable.
	
	When the terms are obtained numerically, the construction of control laws for different sensor/actuator configurations is very cheap when forcing/targets are low-rank. Noting that a given control law, $\GAMMA $, e.g., the causal or non-causal optimal control,  can be analysed via the sensor/target CSDs allows for a fast offline evaluation of the control strategy, and thus for an efficient investigation of sensor/actuator placement.  The  target CSDs   ($ \M{S}_{z,z} $) of a system controlled with the control kernel $\GAMMA$ can be obtained from the uncontrolled target CSDs ($ \M{S}_{z_1,z_1} $) and the terms found in the Wiener-Hopf equations.  From  \eqref{eq:espy1y1}--\eqref{eq:espz2z1}, 
	\begin{align} \label{eq:Czz}
		\M{S}_{z,z}= \M {S}_{z_1,z_1} +
		\hat{\Hh}^\dagger   \hat{\GAMMA}  \hat{\GG} \hat{\GAMMA}^\dagger \hat{\Hh}
		-\hat{\Hh}^\dagger \hat{\GAMMA} \hat{\Gg}^\dagger \Cz^\dagger -  \Cz\hat{\Gg} \hat{\GAMMA}^\dagger \hat{\Hh}.
	\end{align}
	In the next section, this expression will be used to quickly estimate the control performance for several actuator locations.

	\modification{
		The cost of the method is dominated by the direct and adjoint runs used to construct the matrix coefficients in the Wiener-Hopf equations, and the cost of actually solving the Wiener-Hopf problem is comparatively negligible.  This is the case because the cost of the direct and adjoint runs scale (often linearly) with the problem dimension $n_{u}$, while the cost of the Wiener-Hopf problem is independent of the problem dimension and instead scales with the number of sensors and actuators, which are typically small.  As a point of reference, the direct and adjoint runs for the example problem in \S~\ref{sec:bfs} took around one hour, while solving the Wiener-Hopf took minutes on a laptop computer.  
		
		The costs of the current method are thus considerably smaller than the iterative method proposed by \cite{semeraro2013riccatiless}. For a case with a single sensor and actuator and low rank forcing and targets, they report approximately $40$ iterations to converge both estimation and control gains. With the proposed approach, only two runs are necessary, reducing the cost by a factor of $20$ while also providing  sensor and actuator parametric studies, an offline performance estimation of these configurations, and the possibility of handling complex force models.
	}


	\modification{The kernels that were obtained  in \S~\ref{sec:comparison_on_GL}, and as will be shown in \S~\ref{sec:bfs}, they are are smooth and decay rapidly for large values of $ \tau $. This means that the convolutions required for estimation and control can be efficiently computed with a finite number of points using standard numerical quadrature methods, and thus can be used in implementations with limited memory and computational power, as in experimental applications \citep{fabbiane2015role,brito2021experimental,maia2021realtime}.}
	
\section{Estimation and control of the flow over a backwards-facing step}
	\label{sec:bfs}
	
		\begin{figure}[t]
		\centering
		\includegraphics[width=.85\linewidth]{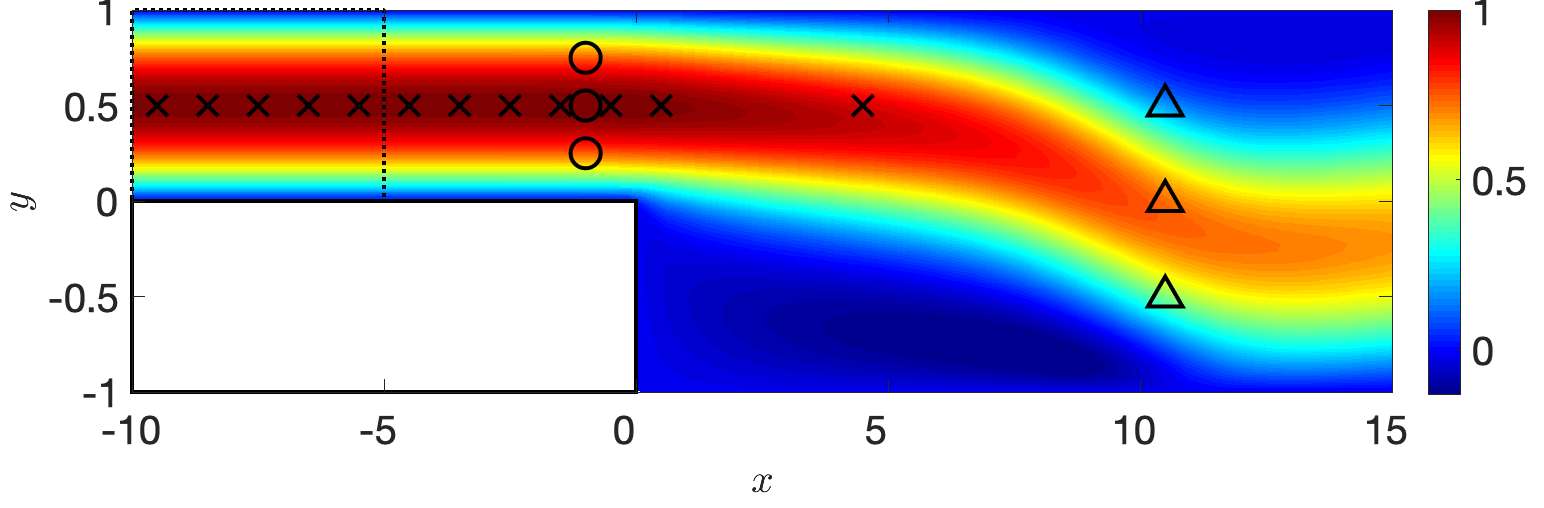}
		\caption{Baseflow for the backwards-facing step flow.  The dashed square indicates the region where the flow is disturbed. Sensor are located at $x=-1$ and $y=0.25,0.50$ and $ 0.75$, actuators at $y=0.5$ with a unit spacing between  $x=-9.5$ and $0.5$, and target at $x=10.5$ and $y=-0.50,0.00$ and $0.50$. They are indicated respectively by circles, crosses, and triangles. 
		}
		\label{fig:bfsbaseflow}
	\end{figure}

	In this section \modification{we illustrate the potential of the tools developed in this paper using an amplifier flow in which the level of non-linearity can be easily adjusted. This allows a smooth transition between linear (linearized Navier-Stokes equations, or small perturbation amplitude in the non-linear system) and non-linear (larger perturbation amplitude in the non-linear system) problems.}
	
		We study the two-dimensional flow over a backward-facing step with Reynolds number  $Re=500 $ based on the step height. In order to model disturbances coming from the upstream channel-like flow,  we consider high-rank disturbances localized upstream of the step, a case significantly more challenging than the similar problem studied by \cite{herve2012physics}, who considered rank-1 disturbances. The flow is illustrated in figure \ref{fig:bfsbaseflow}, where  the region to which the disturbances are applied is illustrated.   The inflow condition is laminar Poiseuille flow, and quantities are made non-dimensional with respect to the maximal inflow velocity and the step height. \modification{The system linearization is performed around the steady solution of the unforced non-linear problem. Note that for large disturbances, mean-flow distortion can arise, which can be partially accounted for by linearizing the system around the mean flow~\citep{hogberg2003relaminarization,chevalier2006state,sipp2007global,leclercq2019linear}.  However, choice of base flow is a generic issue for any linear method and not specific to our particular formulation, so we do not explore this further.    }
		
	The Navier-Stokes equations are solved using the spectral-element code \emph{Nek5000} \citep{fischer1989parallel,fischer1998projection}, which uses $n^{th}$-order Lagrangian interpolants within each element to solve a weak formulation of the incompressible Navier-Stokes equations, upon which the method proposed here was implemented. The resulting open-source code is available as a Git repository\footnote{ https://github.com/eduardomartini/Nek5000\_ResolventTools}.  The domain was represented by 600 elements, each discretized by $ 5^{th} $-order polynomials. Time integration was performed using a non-dimensional time step of $ 3\times 10^{-2}$. For the non-linear integration, used to obtain the base flow, an inflow with a Poiseuille profile was imposed at the leftmost boundary, an outflow condition imposed on the rightmost boundary, and  no-slip conditions imposed on all other boundaries. For the linear runs, Dirichlet boundary conditions for velocity fluctuations were used on all boundaries.
	The flow is globally stable, and thus the base flow is obtained by simple time marching of the Navier-Stokes equations until the time derivative is smaller than $ 10^{-10} $. When present, external forcing is obtained by a pseudo-random number generator, whose seed is initialized to the same value on each processing unit, making uncontrolled and controlled runs comparable as long as the same time step and number of cores are used. Although straightforward for the linear problem, CFL constraints rendered fixed time steps impractical when large perturbations are present, and thus equivalent time steps for different runs could not be obtained. For such cases, representative snapshots will be presented instead.
	
	The dynamics of the linearized system can be summarized as follows. Upstream of the step the flow is Poiseuille-like and only exhibits spatially decaying  waves. Once these waves reach the shear layer downstream of the separation, they excite Kelvin-Helmholtz instability waves, which undergo significant growth before the end of the recirculation bubble that forms in the wake of the step. 
	
	All sensors, actuator and targets considered here have Gaussian spatial support, given by $ \exp\left({-\frac{(x-x_c)^2}{2\sigma_x^2}-\frac{(y-y_c)^2}{2\sigma_y^2}}\right) $, with $ \sigma_x=0.2 $ and $ \sigma_y=0.1 $.  All sensors and actuators used  act on the streamwise direction only. Unless explicitly stated, $ \Mh{\N} $/$ \M{P} $ is taken as a constant identity matrix with diagonal entries corresponding to $ 10^{-2} $ of the maximum value of  $ \hat{\GG} $/$ \hat{\HH} $ without noise/penalty. \modification{The ability of the control law to suppress disturbances is typically a monotonic function of sensor noise/actuator penalty, asymptotically reaching a minimum value for small enough values  of these quantities. However, if very small values are used, numerical ill-conditioning can affect this trend. These effects are discussed by \cite{sasaki2018wavecancelling}. }  The values used here guarantee effective factorization, avoiding numerical ill-conditioning of the factorization problem (\S~\ref{sec:danielesMethod}). For most of the configurations studied, these values yield results that approach the zero cost/noise limit.
	
	In the following sections, control on the linearized problem is first investigated, where the role of the number and placement of sensors and actuators is  studied. As we have previously explored the non-causal estimation problem  \citep{martini2020resolventbased}, we focus on the causal control of disturbances, which depends on accurate causal estimation. Next, the same control strategies are obtained from a numerical experiment, as proposed in \S~\ref{sec:discussions}, showing that the method can be applied to experimental and adjoint-less scenarios. Finally, we illustrate the application of control laws obtained for the linearized system to the non-linear equations.
		
	\subsection{Control of the linearized problem} \label{sec:bfs_linear}

	We initially  focus on a linearized system disturbed by high-rank external forces, designing control strategies to minimize readings from low-rank targets. We thus perform a limited study on the sensor placement, but a large actuator placement study.  These results will be used to inform sensors and actuators to be used on a full-rank target scenario. 
	
	\subsubsection{Control with a single sensor}	
	We first consider the use of a single sensor (middle circle in figure \ref{fig:bfsbaseflow}) and investigate the placement of one actuator for control (locations indicated by crosses in figure \ref{fig:bfsbaseflow}). Three control approaches are explored: non-causal control (\emph{nc}), the truncated non-causal control (\emph{tnc}), and optimal-causal control (\emph{c}). Truncated non-causal control corresponds to a truncation of the optimal non-causal control to its causal part, corresponding to the approach used in previous works~\citep{sasaki2018closedloop,brito2021experimental}, which was seen, in some cases, to be a good approximation for optimal causal control, pointing to a  wave-cancelling nature of optimal control strategies in these scenarios~\citep{sasaki2018wavecancelling}.  Non-causal control cannot be used in real-time applications, but it provides upper bounds for the performance of any linear control strategy and thus can be useful to evaluate different sensor and actuator placements. 
	
	As we use high-rank forcing and a low-rank target, the configuration corresponds to scenario (ii) described in \S~\ref{sec:timemarching}. A total of three time-domain solutions were obtained, corresponding to one adjoint-direct system for the sensors, from which the estimation terms $ \GG $ and $ \Gg $ are obtained, and one adjoint-direct system for the targets, from which impulse responses from any actuator can be obtained. All the following results were obtained via inexpensive post-processing of the resulting data.
	
\begin{figure}[t!]
		\centering
		\subfloat[Actuator at $ x=-3.5 $]{\psfragfig[scale=0.5]{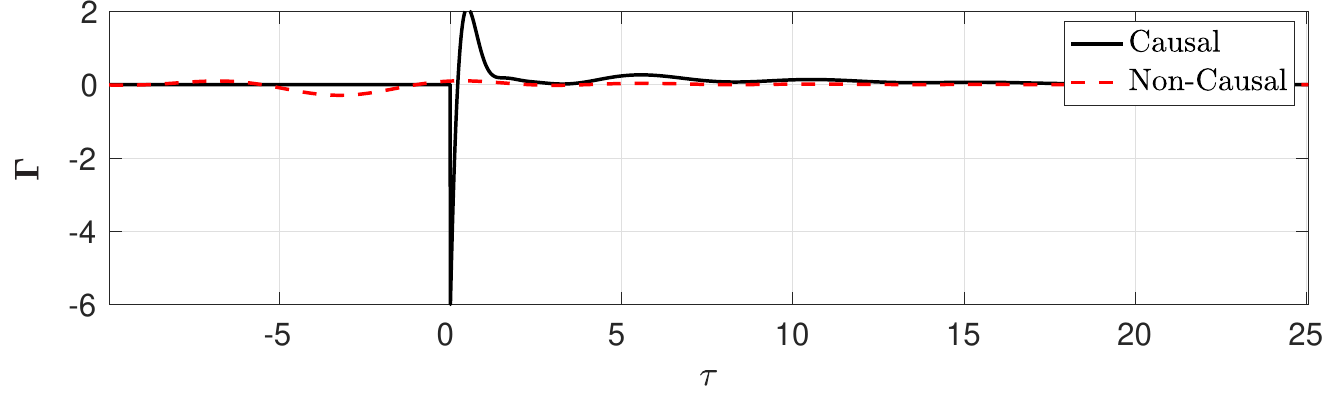}{		\psfrag{GA}{$\GAMMA$}}	}
		
		\subfloat[Actuator at $ x= 0.5 $\label{fig:sisosKernels_ax.5}]{\psfragfig[scale=0.5]{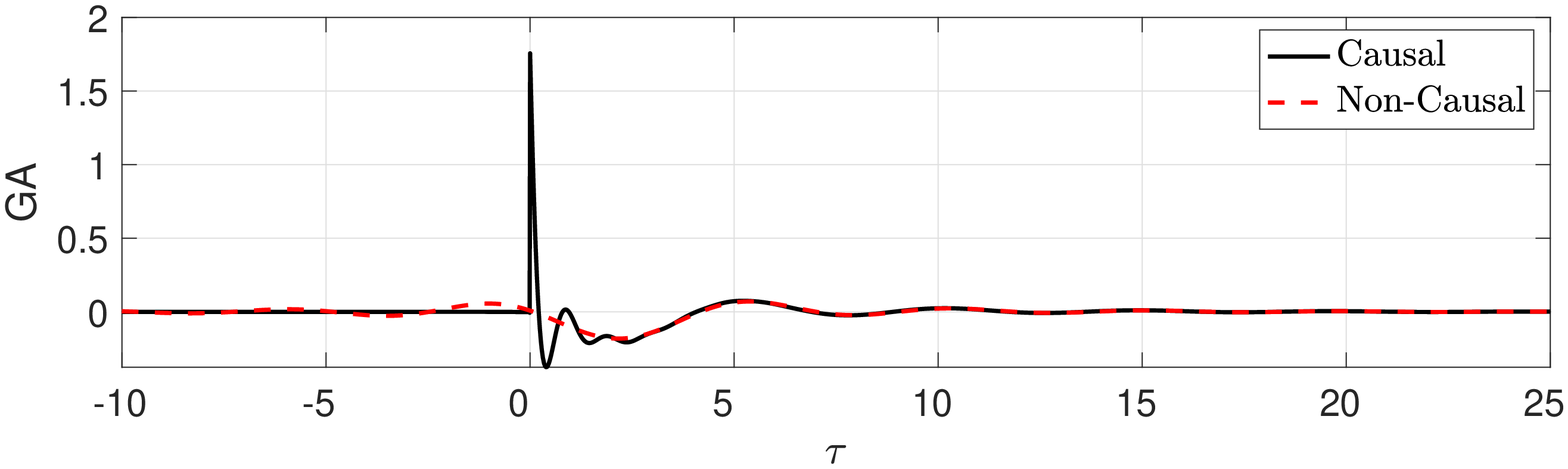}{		\psfrag{GA}{$\GAMMA$}}	}
		
		\subfloat[Actuator at $ x= 1.5 $]{\psfragfig[scale=0.5]{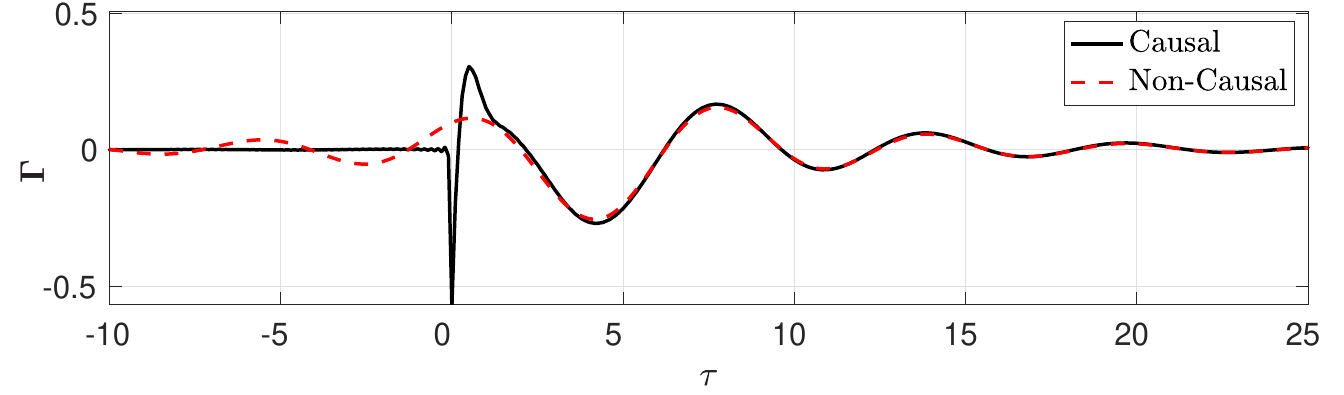}{		\psfrag{GA}{$\GAMMA$}}	}
		\caption{Causal and non-causal control for different actuator placements and a sensor at $ (x,y)=(-1,0) $.}
		\label{fig:sisoKernels}
	\end{figure}

	Representative control kernels for different actuator placements are shown in figure~\ref{fig:sisoKernels}. The causal kernel is seen to converge to the non-causal kernel for large $ \tau $ when the actuator is not far upstream of the sensor. When the non-causal control law uses non-causal information, the causal control law typically exhibits a spike at $\tau=0$. This behaviour can be understood as the control law using the most recent information to compensate for future information, which is not available.  Similar behaviour was reported by \cite{morra2020realizable}.

\begin{figure}[t]
		\centering
		\subfloat[Actuator at $ x=-9.5 $]{\includegraphics[scale=0.5]{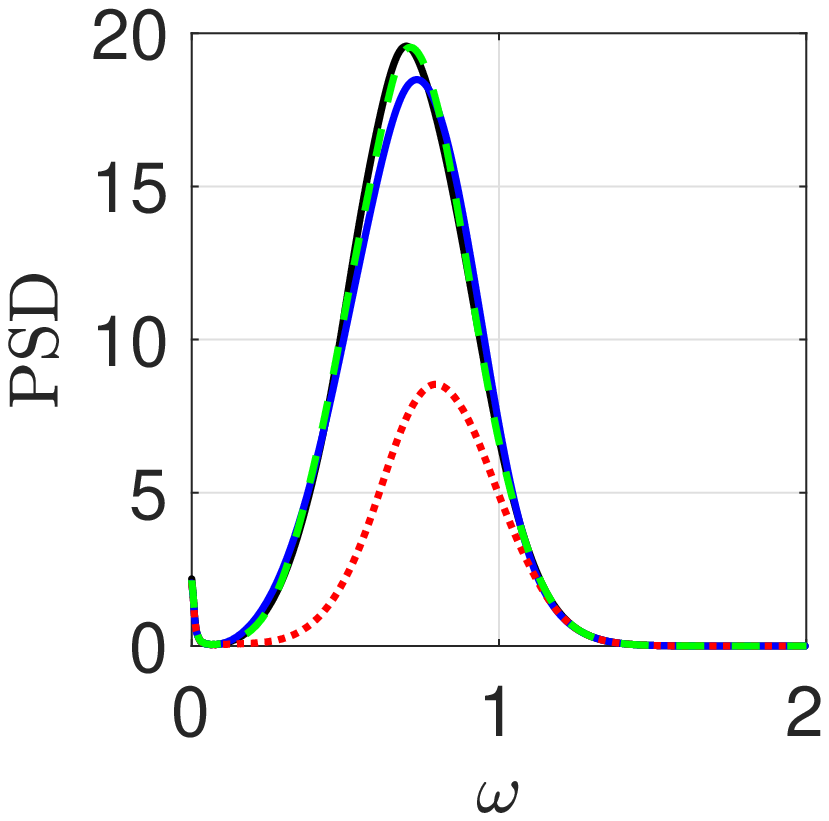}}
		\subfloat[\label{eq:SISO_b} Actuator at $ x=-3.5 $]{\includegraphics[scale=0.5]{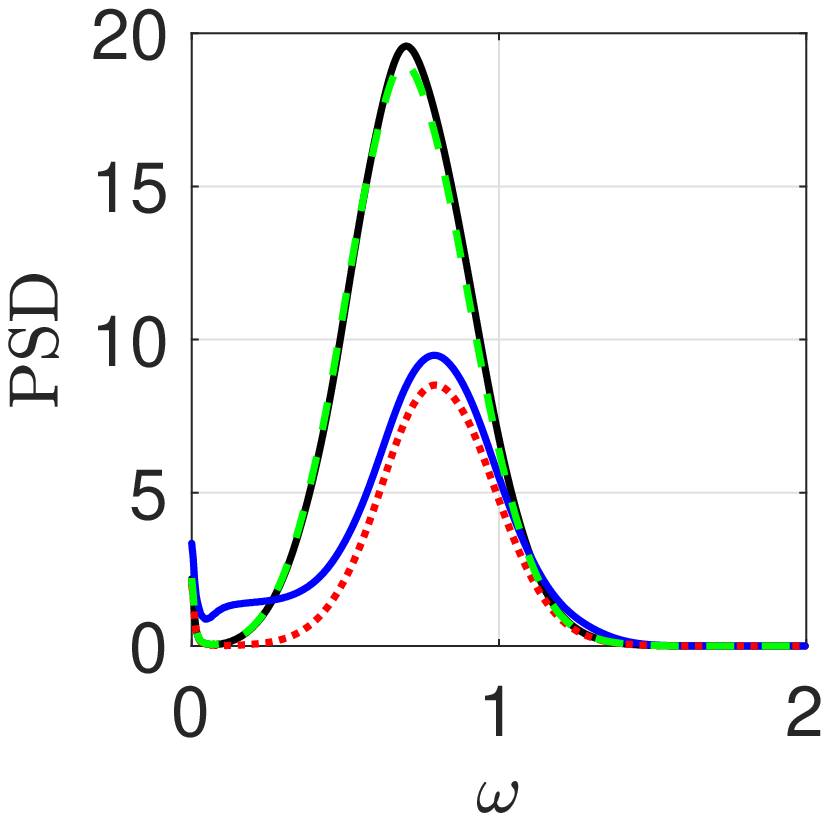}}
		\subfloat[Actuator at $ x= 0.5 $]{\includegraphics[scale=0.5]{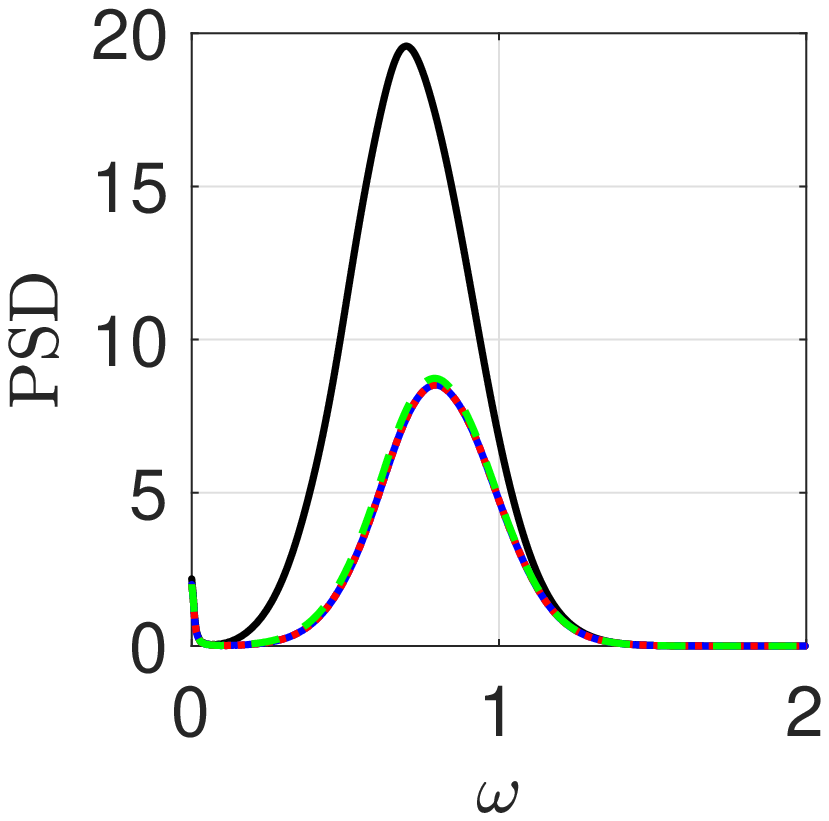}}
		
		\subfloat[Target energy]{ 
			\psfragfig[scale=0.5]{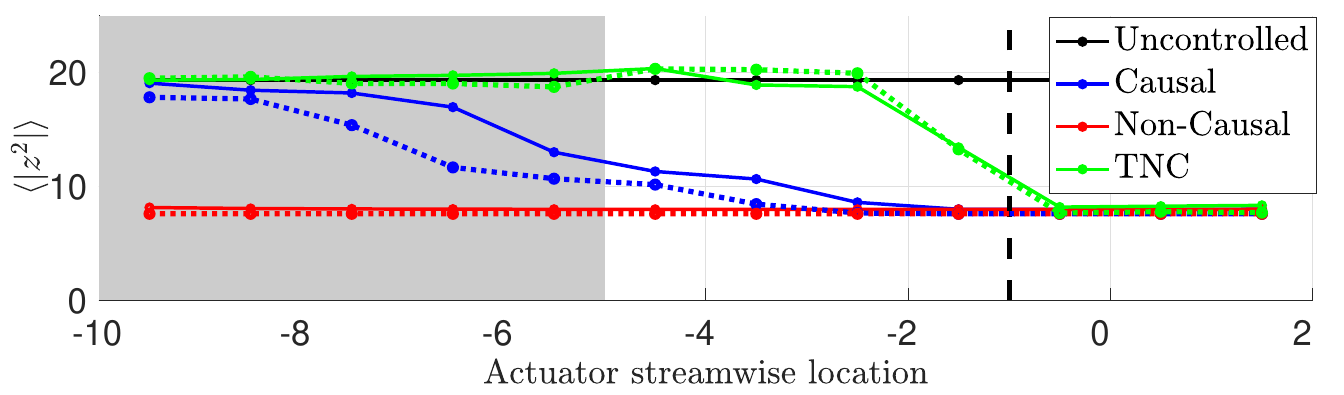}{
				\psfrag{meanz}{$\langle |\z^2|\rangle$}
			}
		}
		
		\caption{
			Target reading PSDs and energy for different  control strategies and actuator placement. $ \M{P}=10^{-2}\max(|\HH(\omega)|)$ was used in (a)-(c) and for the solid lines in (d). A value of $\M{P}=10^{-5}\max(|\HH(\omega)|) $ was used for dotted lines in (d).  
		}
		\label{fig:SISO_actPlaceent}
	\end{figure}

\begin{figure}[t]
		\centering
		\includegraphics[scale=0.5]{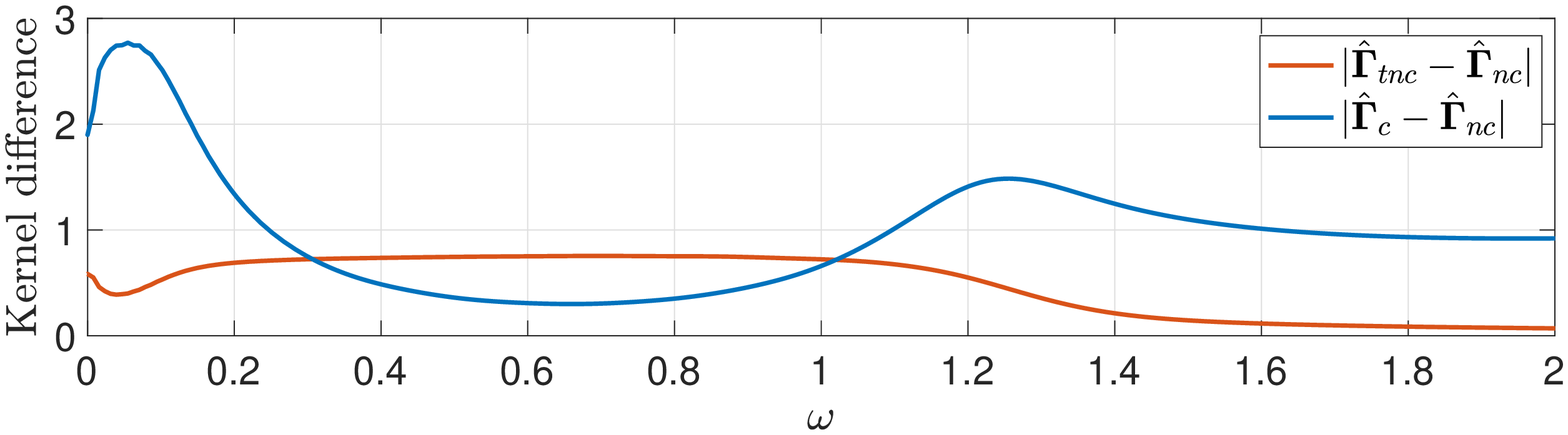}
		\caption{Difference between causal, and TNC control kernels with respect to the non-causal kernel  in the frequency domain. The configuration is the same as in figure \ref{eq:SISO_b}. As the quantities are complex, the absolute value of the difference is shown to measure how  the causal and TNC  approaches deviate from the non-causal kernel. }
		\label{fig:bfskernelsdiff}
	\end{figure}

	 To compare the different control strategies, \eqref{eq:Czz} is used to compute the target's PSD and expected energy, i.e. the integral of the  PSD, with the results presented in figure  \ref{fig:SISO_actPlaceent}.   All control strategies converge approximately to the same PSD reduction when the actuator is located downstream of the sensor.
	 Not surprisingly, non-causal control provides the highest PSD reduction for all configurations, with little dependence on the  actuator placement. As the non-causal control law can be solved independently for each frequency, the  PSDs obtained with this control are lower than that of  the uncontrolled case for all frequencies. Different trends are observed for the causal or the TNC control. Comparing the target PSDs for the causal and non-causal cases in figure \ref{eq:SISO_b}, the non-causal control provides the lowest PSD at all frequencies. The causal control increases target PSD at lower frequencies ($ |\omega|<0.2 $), which is however compensated by the  reduction for the other frequencies. Such a compromise between PSD reduction in different frequencies is unavoidable in causal control strategies, due to constraints imposed by causality.  
	 Figure~\ref{fig:bfskernelsdiff} compares the kernels in the frequency domain and shows that each approach better represents the non-causal kernel for different frequencies. A comparison of figures~\ref{eq:SISO_b} and  \ref{fig:bfskernelsdiff}  shows that the causal control better reproduces the non-causal control for the most relevant frequencies, i.e. the ones which provide a higher reduction in the target PSD, explaining its superior performance.

	Figure \ref{fig:SISO_actPlaceent}(d) shows that the three approaches show different trends as the actuator is moved upstream. While actuator placement has a negligible influence on the non-causal control, it significantly affects the TNC control, leading to an increase in the target energy for some configurations. Causal control, on the other hand, remains  efficient even when located upstream of the sensor, with the reduction in the PSD degraded by $ 50\% $  only when the actuator is located approximately 4.5 step heights upstream of  the sensor.  
	
	Two different effects can explain these observations.	Structures that emerge from the upstream disturbances can have significant coherence lengths, of the order of two convective time units, as later reported in figure \ref{fig:sensor-cross-correlation}. This coherence length allows the sensor to partially estimate their upstream components, which can thus be cancelled by the actuator. It is speculated that a similar mechanism is responsible for the control of a jet (a convectively unstable flow, as the present example) using downstream sensors obtained by \cite{bychkov2019plasmabased} and \cite{kopiev2019plasma}. An extreme example is when the flow is excited with a harmonic, rank-1, forcing: forces and flow response have infinite coherence lengths, and can thus be estimated and  controlled  based on only a few sensor readings, from which amplitudes and phases are extracted.
	
	Another possible mechanism is the exploration of different control approaches by the causal control. As the flow studied here is incompressible, the actuation has an instantaneous, although possibly small, effect throughout the flow. This response is typically negligible for the non-causal control, which favours more effective mechanisms, but may be the only one available for the causal control when the actuator is located far upstream and  exploited if  low actuation penalties are used. This interpretation is supported by the dotted lines in figure \ref{fig:SISO_actPlaceent}, where the actuation penalty was drastically reduced.   For upstream actuators, a significant reduction of the targets PSDs is observed for the causal control, but a small effect is observed for the non-causal control.    Actuator placements upstream of the sensors tend to be less robust to unmodelled dynamics~\citep{belson2013feedback}, and it is a configuration that is typically not used for amplifier flows~\citep{schmid2016linear}. Thus, upstream actuators are not further investigated in this work.

	\subsubsection{Control with multiple sensors} \label{sec:control_mimo}
	
	\cite{herve2012physics} reported higher perturbation reductions for a similar control configuration than those obtained here using a single sensor and actuator. However, the cited work considered only rank-1 disturbances. The present work deals with a more complex forcing scenario, with several waves exciting the Kelvin-Helmholtz instability. The identification of such multiple waves invariably requires multiple sensors; we thus explore this scenario. Four additional time-domain solutions were necessary for the results presented next, corresponding to adjoint-direct systems for the added sensors, the top and bottom circles in figure \ref{fig:bfsbaseflow}.
	
\begin{figure}[t]
		\centering
		\psfragfig[scale=0.5]{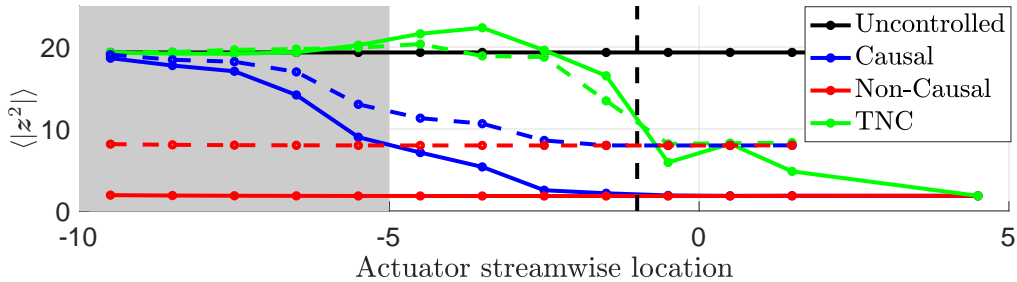}{
			\psfrag{meanz}{$\langle |\z^2|\rangle$}
		}
		
		\caption{Same as figure \ref{fig:SISO_actPlaceent}, but using three sensors. Results for SISO control are reproduced in the dashed lines for reference. }
		\label{fig:MIMO_actPlaceent}
	\end{figure}

\begin{figure}[t!]
		\centering
		\subfloat[Control kernels using 1 sensor]{\psfragfig[scale=0.5]{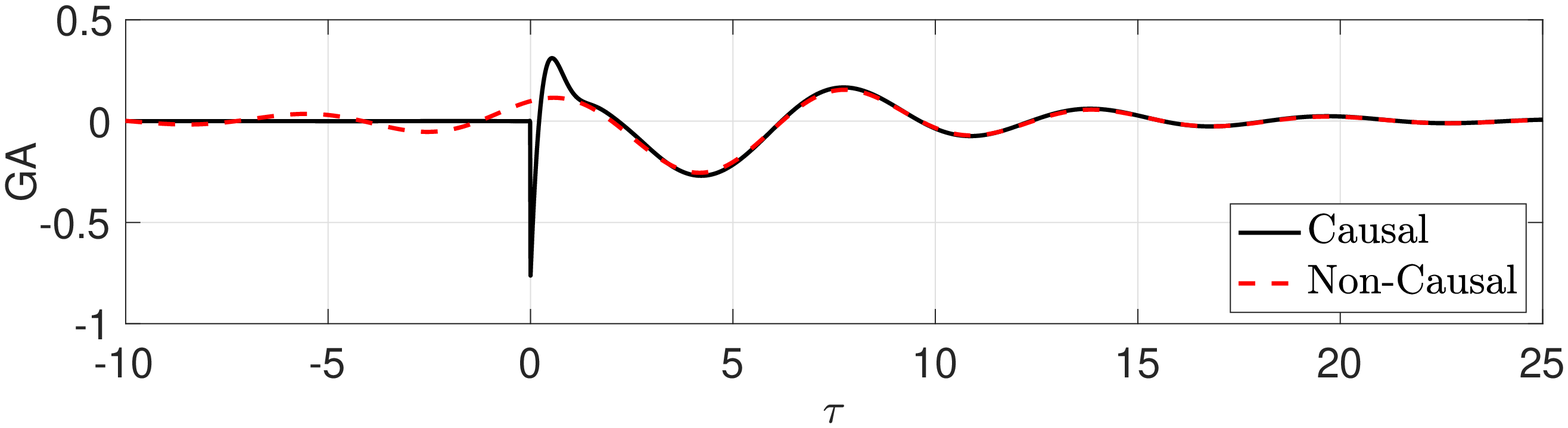}{		\psfrag{GA}{$\GAMMA$}}	}
		
		\subfloat[Control kernels using 3 sensors]{\psfragfig[scale=0.5]{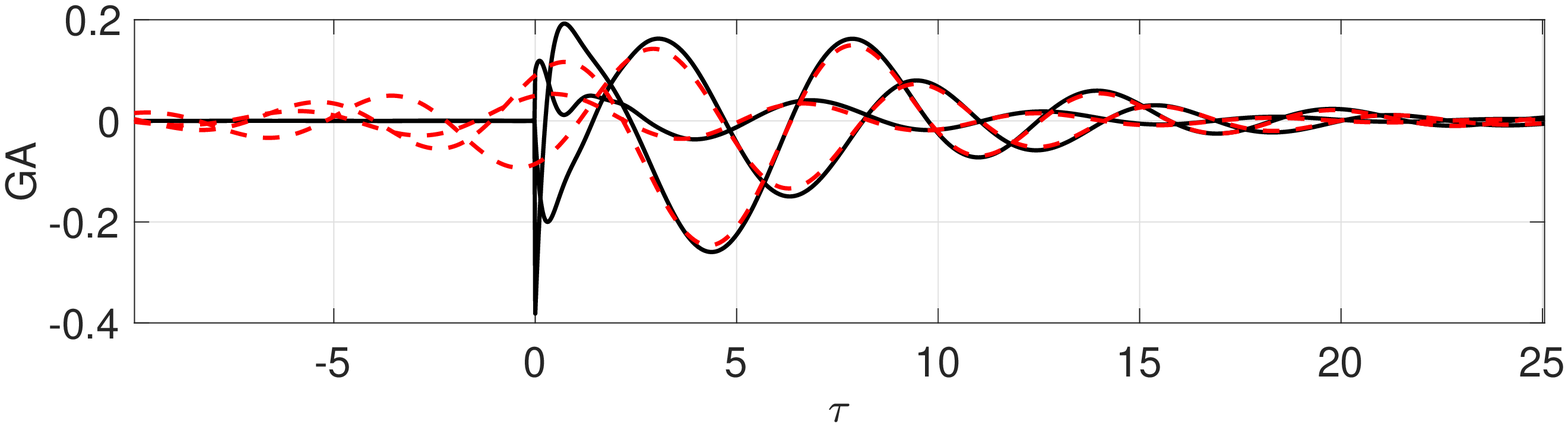}{		\psfrag{GA}{$\GAMMA$}}}
		\caption{Control kernels using an actuator at $ x=-1.5 $ and (a) one sensor, located at $ y=0.5 $, (b) the three sensors indicated in figure \ref{fig:bfsbaseflow}.}
		\label{fig:kernelsmimo}
	\end{figure}

	Figure \ref{fig:MIMO_actPlaceent} shows the expected target energy using the three sensors. A comparison with results obtained using one sensor identifies two trends: (i) lower target PSDs are observed, indicating that the additional sensors are indeed necessary to identify the multiple incoming waves and to accurately estimate the target readings;  (ii) a larger distance between the actuator and the sensors is required for TNC to reproduce the optimal control strategies.  Comparing the kernels when using one and three sensors in figure \ref{fig:kernelsmimo}, it can be seen that non-causal control for three sensors is more ``spread" in $ \tau $, and thus has more signal content for $ \tau<0 $. It is then expected that its truncation leads to a more significant  degradation of the control.

	\subsubsection{Full-rank target control} \label{sec:control_full}
	From figures \ref{fig:SISO_actPlaceent} and \ref{fig:MIMO_actPlaceent}, we conclude that the actuator  should be located shortly downstream of the sensors. With  sensor and actuator placements defined, control kernels for full-rank targets are constructed. This required  that the following additional time-domain solutions be performed: the last equation in the sensor direct-adjoint-direct problem for each of the sensors, and the full direct-adjoint system for each actuator. In total nine additional time-domain solutions were used. 
	
\begin{figure}[t]
		\centering
		\psfragfig[scale=0.5]{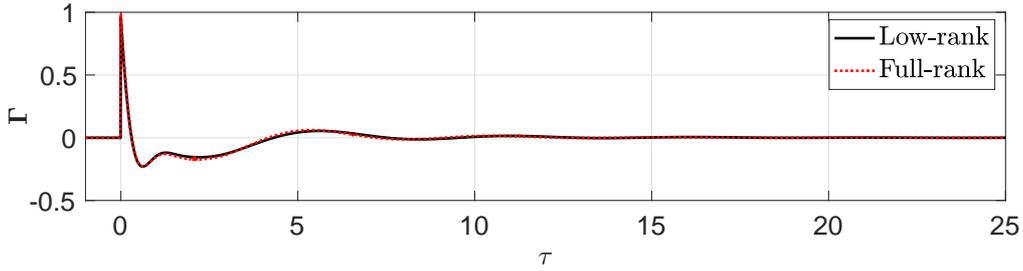}{	\psfrag{GA}{$\GAMMA$}}
		\caption{Comparison between control kernels for low- and full-rank targets.}
		\label{fig:kernellowfull}
	\end{figure}

\begin{figure}
		\centering
		\includegraphics[scale=0.5]{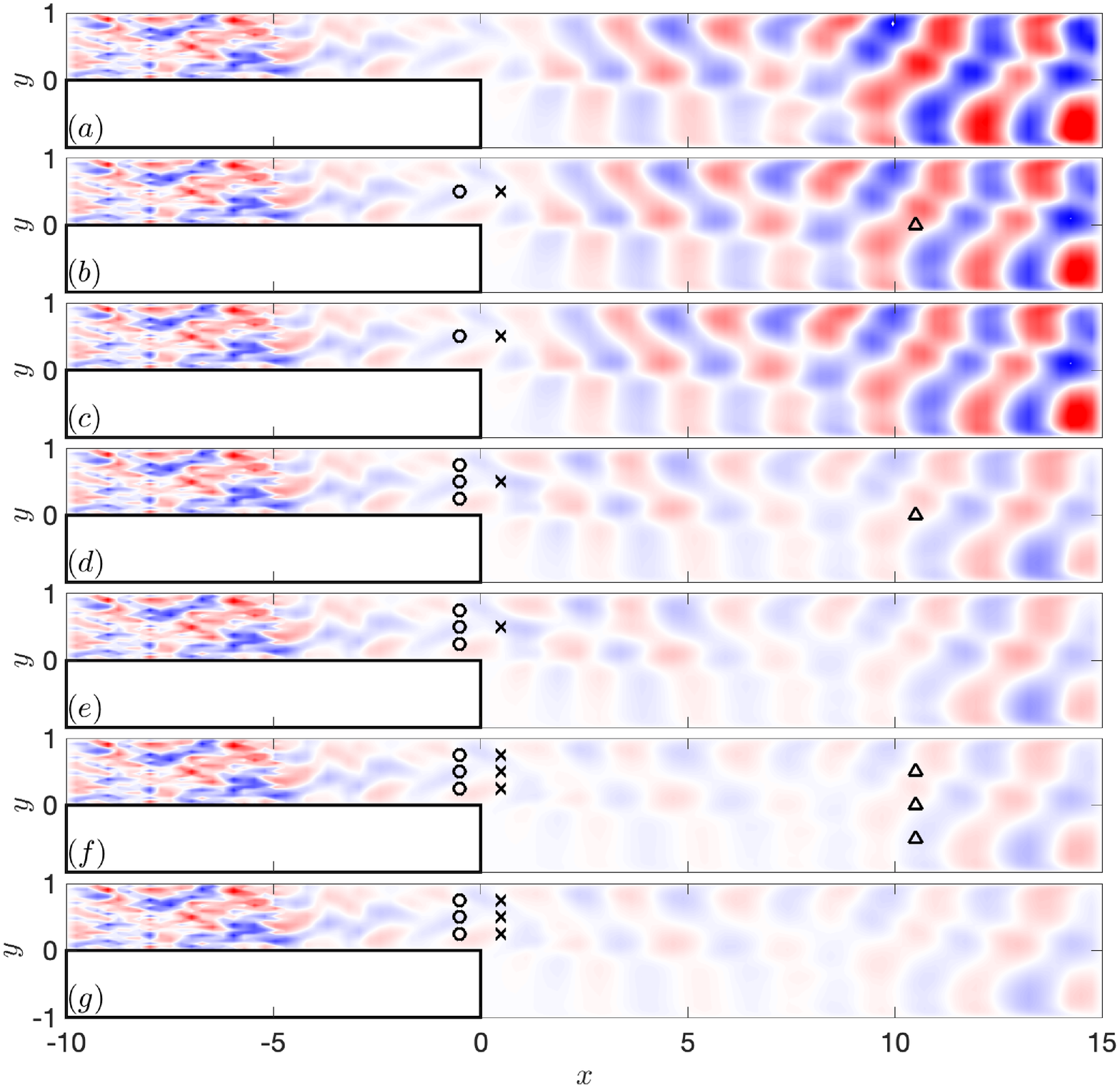}
		
		\includegraphics[scale=0.5]{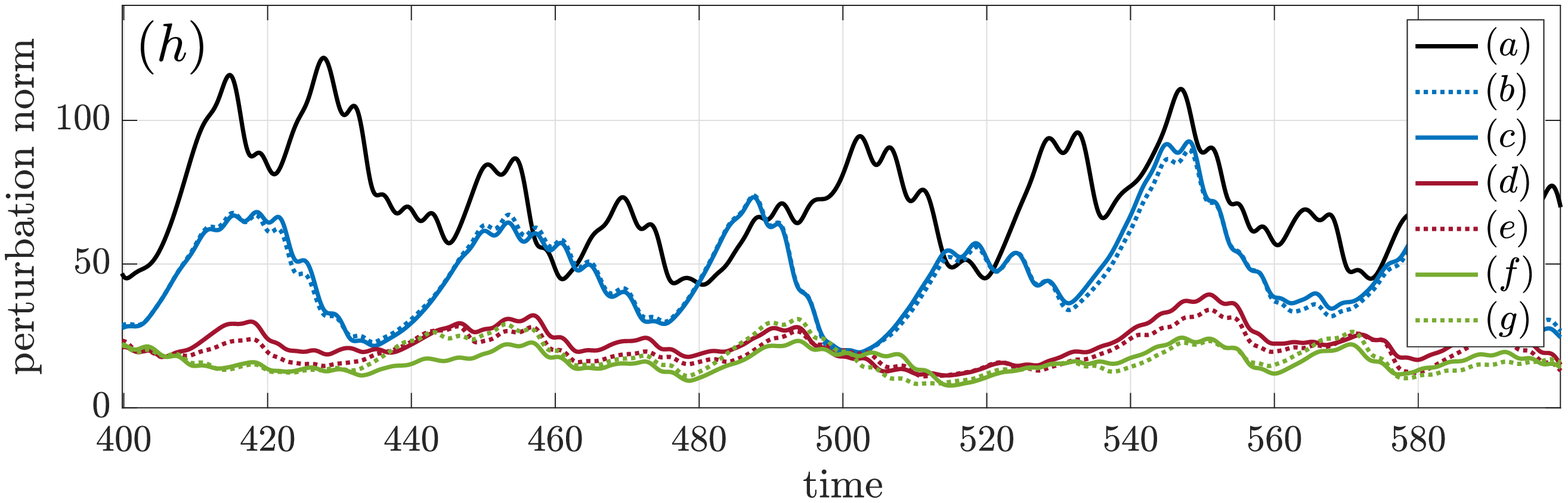}
		
		\caption{Snapshots for the uncontrolled flow (a) and controlled flows using the causal control kernel (b)-(g). The sensors and actuators used in each scenario are indicated by circles and crosses. Low-rank targets are indicated  by triangles on the figures, full rank targets, i.e. $ \Cz= \M{I} $, were used when the figures do not contain triangles. In (h), norms for perturbations for  $ x>0 $, with colours highlighting sensor and actuators used and solid/dotted lines low/full target ranks.  }
		\label{fig:control_snapshots}
	\end{figure}

	Figure \ref{fig:kernellowfull} compares the control kernel for a rank-1 and full-rank $ \Cz $. The close match between the two kernels indicates that downstream of the step the problem  can be well approximated by a rank-1 model, and that the high-rank behaviour of the problem is indeed  restricted to the receptivity of the Kelvin-Helmholtz modes to channel disturbances. This scenario is further supported by results presented in figure \ref{fig:control_snapshots}, where flow snapshots and time-series for the perturbation norm are shown for several configurations of the sensor, actuators, and targets. Control designed for full-rank targets and multiple actuators do lead to perturbation energy reduction, but the biggest improvement is provided by the use of multiple sensors.
	
\subsection{Adjoint-less and empirical application}\label{sec:adjointless_bfs}

Here we illustrate  a data-driven application of the proposed method. The necessary data can be obtained from experimental setups, from numerical solutions of the non-linear problem, or from the direct linearized problem disturbed by stochastic forcing. We focus here on the latter scenario, with the use of a single sensor and target. The generalization to more complex configurations, with more sensors, actuators, and targets, is  straightforward. From sensor and target time series, the  auto-correlation and cross-correlations are computed in the time domain.  As previously described in \S~\ref{sec:estimation_and_control}, \eqref{eq:yzCSDs}, the terms $ \hat{\GG} $ and $ \hat{\Gg} $ are equivalent to the corresponding CSDs.

\begin{figure}[t]
	\centering
	\psfragfig[scale=0.5]{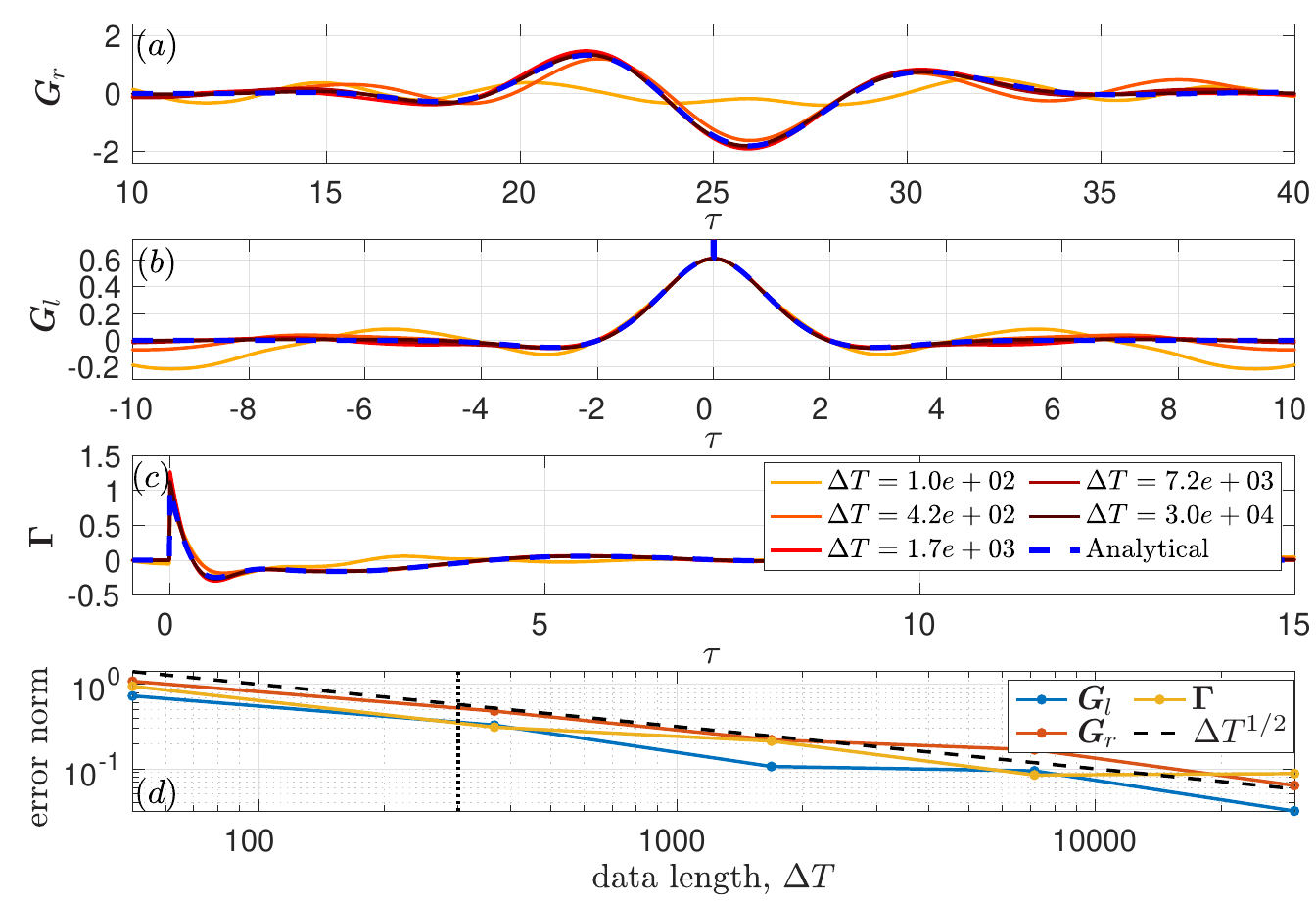}{
	\psfrag{GGl}{$\GG$}
	\psfrag{Gr}{$\Gg$}
	\psfrag{GA}{$\GAMMA$}
	\psfrag{DeltaTh}	{$\Delta T^{1/2}$}
}
	\caption{Comparison between $ \GG $, $ \Gg $, $\GAMMA_c $ and errors between the ``empirical" functions and the ones obtained using the time march method proposed . Results are presented for several time lengths used when estimating the auto- and cross-correlations. Results with the time march method require time integration of approximately 300 time units, indicated by the dotted line in (d).  A white sensor noise was added to $ \hat{\GG} $ to regularize its factorization.}
	\label{fig:sensor-cross-correlation}
	\end{figure}

Figure \ref{fig:sensor-cross-correlation} compares estimates of $ \GG $ and $ \Gg $,  obtained ``empirically",  from the numerical experiment with different data lengths,  and ``analytically", obtained with time marching of direct and adjoint equations.  The resulting  control kernels are also compared. Convergence trends are shown using the error, defined as
\begin{equation}\label{key}
	error = \frac{||  \tilde {\GG }-\GG|| }{||\GG||},
\end{equation}
where the  $L^2$ norm is used. The terms $ \tilde {\GG } $  and $ {\GG } $ indicate, respectively, the empirical and analytical terms, with similar expressions for the other terms.  The trends suggest convergence rates scaling with $ \Delta T^{-1/2} $, where $ \Delta T $ is the  time-series length used to estimate $ \GG $ and $ \Gg $.

Although significantly more expensive, and typically less accurate than results obtained with the use of adjoint solvers, the use of numerical experiments does extend the applicability of the method to virtually any scenario and solver, and also to derive control laws directly from experiments, which typically can be carried out to obtain long time series for better convergence of the required correlations.
Note that if experimental data is used, sensor noise is already present in the data. However, if noise levels are small, adding a small noise may be required to better condition the factorization of $\hat{\GG}$.

	\subsection{Estimation and control of the non-linear system}
	
	 Control laws using two different configurations for controlling the non-linear problem are investigated. As shown in \S~\ref{sec:control_mimo}  and \S~\ref{sec:control_full}, the control kernels for full- and low-rank targets provide similar results, and we thus focus on the latter. We consider two scenarios, one using one sensor, one actuator, and one target, and another using three of each. \modification{Although including statistics of the nonlinear terms within $\hat{\F}$ in the nonlinear case would likely improve the results \citep{amaral2021resolventbased},  we chose not to pursue this so that the control law remains fixed as the forcing amplitudes increase, simplifying comparisons.} 
	
	A series of non-linear simulations were performed using spatially and temporally white forcing in the upstream region indicated in figure \ref{fig:bfsbaseflow} with different forcing amplitudes defined by the RMS value $\epsilon$, i.e., $\Mh{F} = \epsilon^2 \I$. 
	For reference, $\epsilon=10^{-2}$ leads to perturbations  that reach 10-20\% of the baseflow velocity at the targets, and thus is well into the non-linear regime.
		 As the control and estimation kernels  are constructed to minimize the expected value of their respective cost functionals, a comparison should be based on an ensemble  of simulations. Here we assume ergodicity of the system, and the ensemble averaging is replaced by time averaging. \modification{This assumption is validated using multiple runs for some configurations. } 
	The results for the linear problem presented in this section used an outflow condition on the right-most domain, in order to properly compare the linear and non-linear systems. The estimation and control laws should be based on a linear system that represents the non-linear problem; however, for this problem, the impact of the different boundary conditions used in the linear and non-linear systems is negligible, as reported next.
	
		\begin{figure}[t]
		\centering
		\includegraphics[scale=0.5]{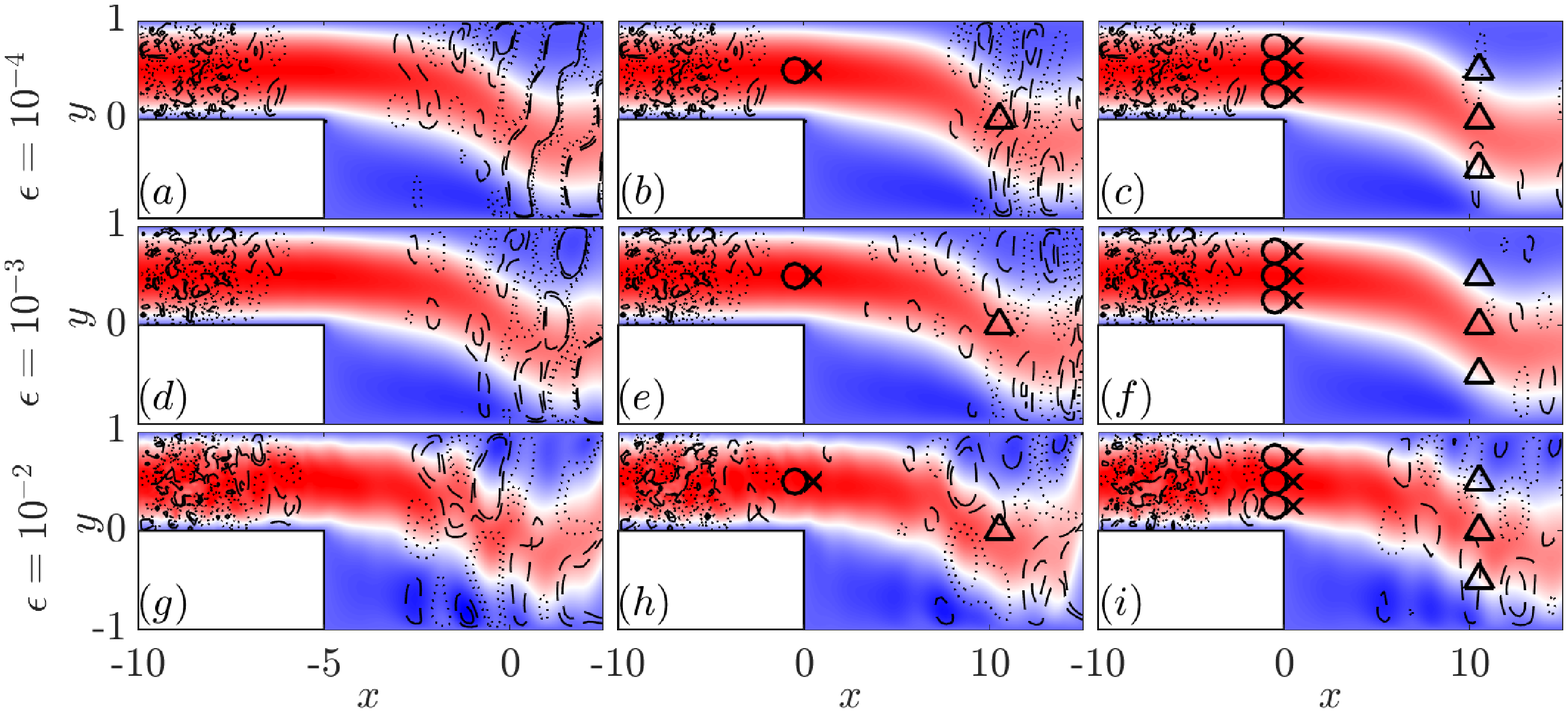}
		
		\includegraphics[scale=0.5]{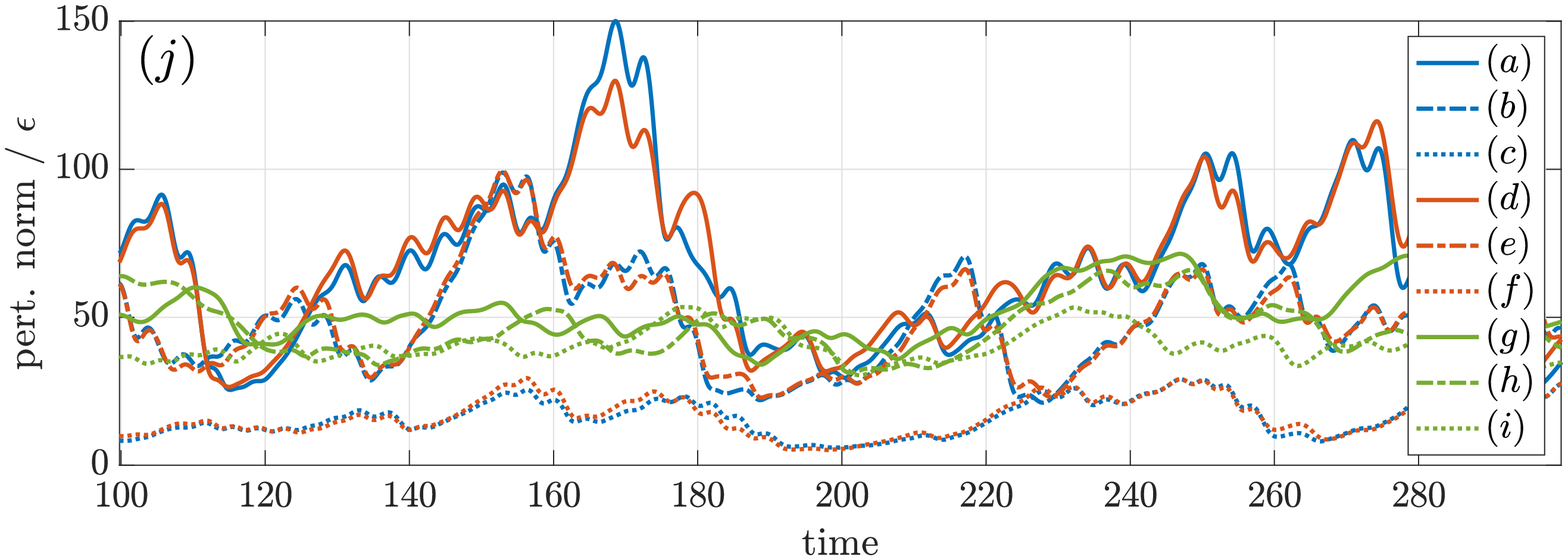}
		
		\caption{(a-i) Snapshots of the non-linear solutions for controlled and uncontrolled flows. Streamwise velocity is shown with the same colour scale as figure \ref{fig:bfsbaseflow}. Dotted/ dashed contour lines indicate velocity excess/deficit of $ \pm 5/\epsilon$ and $ \pm 10/\epsilon $ from the baseflow.  Circles, crosses, and triangles indicate sensors, actuators, and targets used. (j) Evolution of the perturbation norms, normalized by $\epsilon$.}
		\label{fig:non-linearsnapshots_junoh}
	\end{figure}

	\begin{figure}[t]
	\centering
	\includegraphics[scale=0.5]{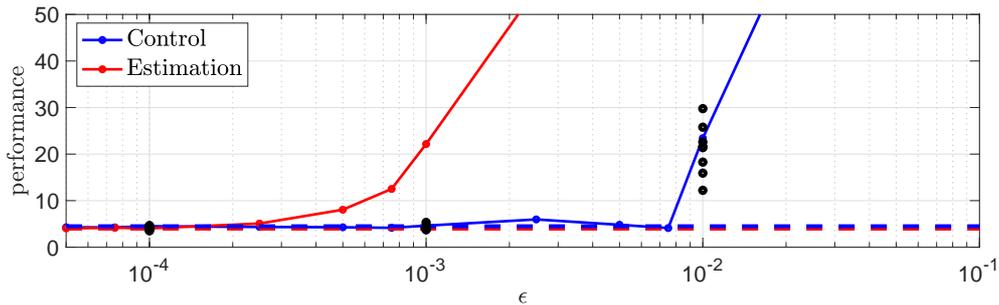}
	
	\caption{Control and estimation performances for the non-linear (solid lines) and linearized (dashed) problems as a function of forcing amplitude. Black dots corresponds to extra runs, with different random seeds for generating the external forcing. Results for the configuration shown in figure \ref{fig:non-linearsnapshots_junoh}(i).}
	\label{fig:non-linearerrors_junoh}
\end{figure}
\begin{figure}[t]
	\centering

	\subfloat[$\epsilon =10^{-4}$]	{		\psfragfig[scale=0.5]{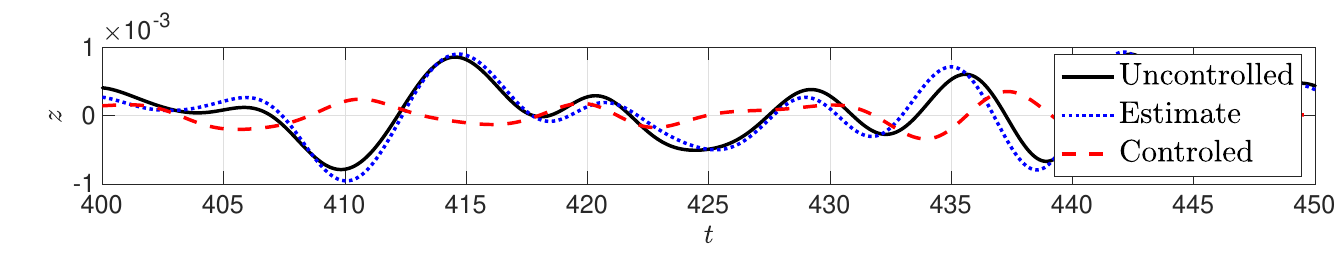}{
			\psfrag{z}{$\z^2$}
		}
	}
	
	\subfloat[$\epsilon =10^{-2}$]	{		\psfragfig[scale=0.5]{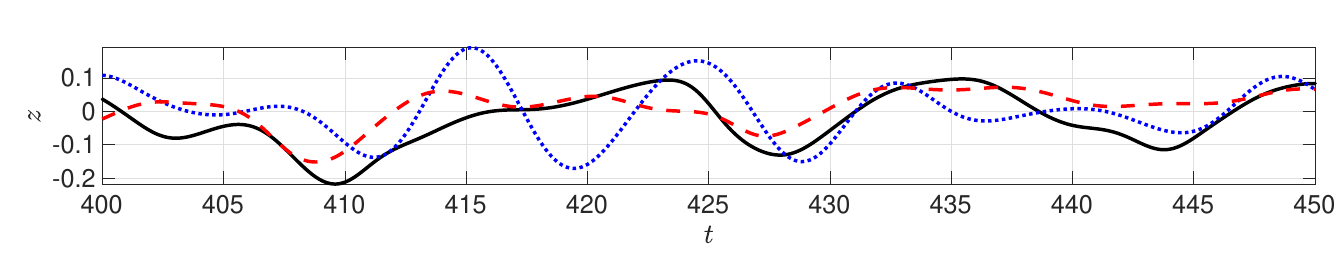}{
			\psfrag{z}{$\z^2$}
		}
	}
	

	\caption{Readings of the target located at $y=0$ for the controlled problem, it's estimate,  and for uncontrolled problem.
		The controlled problem corresponds to the configuration shown in figure \ref{fig:non-linearsnapshots_junoh}(i). 
		\label{fig:targets_timeseries}
	}
\end{figure}

	For small forcing amplitudes, the dynamics are dominated by linear mechanisms, and thus it is  expected that the performance of the estimation and  control laws will be the same as demonstrated in the previous sections. Focusing initially on the control problem, representative snapshots of the flow and time evolution of perturbation norms for controlled and uncontrolled  non-linear systems  are shown in 
	figure \ref{fig:non-linearsnapshots_junoh}. For the lower external forcing RMS,  control of the non-linear system is similar to the control of the linearized problem, but for larger amplitudes it degrades. The perturbation norm, when normalized by the forcing RMS, reduces for larger forcing amplitudes due to non-linear saturation. 
	Estimation and control performances are measured as the non-estimated/controlled target energy fraction, 
	\begin{align} \label{eq:perfMeasure}
		\mathcal{E}_{est}  & = \dfrac{  \sum_i \int  (\z_i(t)-\z_{i,est}(t))^2 dt  } {\sum_i \int  (\z_i(t))^2 dt} ,&
		\mathcal{E}_{con} & = \dfrac{  \sum_i \int  (\z_{i,con}(t))^2 dt  } {\sum_i \int  (\z_i(t))^2 dt}.
	\end{align}
	Figure \ref{fig:non-linearerrors_junoh} shows $\mathcal{E}_{est,con}$ as a function of $\epsilon$. Performances for  the linear and  the non-linear systems are equivalent when small forcing are used, but for the latter it degrades for larger forcing amplitudes. The trend is also observed in figure \ref{fig:targets_timeseries}, where time-series samples of a target for the uncontrolled problem, its causal estimation, and for the controlled problem are shown.
	 \modification{Comparing different runs, indicated by the black dots in figure \ref{fig:non-linearerrors_junoh},  it can be observed that the spread is low for the low forcing amplitudes and increases for higher amplitudes, for which the control performance is degraded. }

	Control and estimation performances for the linear problems are equivalent, indicating that all estimated perturbations are effectively controlled. A similar result is observed for the non-linear problem disturbed by small external forcing.
	Estimation deteriorates for $\epsilon \approx 10^{-3}$,  while control remains effective up to  ${\epsilon \approx 10^{-2}}$.  As the  Kelvin-Helmholtz mode amplifies upstream disturbances, small but finite disturbances are amplified and can exhibit significant non-linear dynamics. Such non-linear effects, which are not accounted for by the linear approach used, degrade estimation performance. For the controlled system, these perturbations are cancelled before they are amplified, and thus the linear assumption is valid for a larger range of external forcing amplitudes for the controlled problem. 
	
	The degraded performance of the approaches presented here when applied to the non-linear system disturbed with high-amplitude forcing is due to the saturation of the uncontrolled flow and to the  violation of the assumption that the disturbances evolve linearly. From figure \ref{fig:non-linearsnapshots_junoh}(j), comparing the normalized perturbation norm for the scenario with $ \epsilon=10^{-2} $ to those with lower values of $ \epsilon $, it can be seen that not only the perturbation norm for the controlled problem is larger, but also the norm of the uncontrolled problem is smaller. Both of these factors impact $ \mathcal{E}_{con} $ as in \eqref{eq:perfMeasure}.
	 The non-linear interactions can be treated as additional external forcing terms in the linearized equation \citep{mckeon2010criticallayer,karban2020ambiguity}. In this context, the degraded performance can be related to inadequacies of the forcing model.  In particular, the non-linear interactions are not, in general, zero-mean when the linearization is performed around a baseflow, thus violating one of the assumptions used to construct the control laws. An alternative is to use a linearization around the mean flow, for which the non-linear terms are, by definition, zero mean. In previous works \citep{martini2020resolventbased,amaral2021resolventbased} it was shown that a more representative  modelling of these forcing terms can lead to significant improvements in estimation, and thus potentially to improvements in control also, for larger disturbances.  Control formulations using a linearization around the mean flow may follow the methods in \cite{hogberg2003relaminarization} or \cite{leclercq2019linear}, with various mean flows used in succesive linearisations of the system as control gains are progressively increased.	Integrating these approaches into the control law is beyond the scope of this study, and will be considered in future work.

		
%
%

	\section{Conclusions} \label{sec:conclusions}
	Causal resolvent-based estimation and control methods based on the Wiener-Hopf framework have been presented. \modification{The approach is an extension of the non-casual resolvent-based estimation methods developed by \cite{towne2020resolvent} and \cite{martini2020resolventbased} to causal estimation and control, and is obtained combining  three different tools: the Wiener-regulator framework~\citep{martinelli2009feedback}, matrix-free methods to obtain the action of the resolvent operator~\citep{martini2020resolventbased,martini2021efficient}, and   numerical methods to solve matrix Wiener-Hopf problems~\citep{daniele2007fredholm}. The resulting method
	is directly applicable to large systems without model reduction or simplified forcing assumptions, requiring only low-rank sensor and actuator setups, which is the case in any practical configuration. } 
	Computational costs are orders-of-magnitude lower than previous approaches for full-rank systems \citep{semeraro2013riccatiless}. If low-rank forcing/targets are used, inexpensive exploration of virtually any sensor/actuator configuration using only  data post-processing can be obtained, allowing the optimization of sensor and actuator placements. Control of systems with high-rank forcing and targets is obtained.  The ability to deal with high-rank targets avoids a possible bias of the control law towards the specific location of a given low-rank target; instead, fluctuations across the domain of interest may be minimized.
	
	Using an open-source implementation of the proposed method,  control of the flow over a backward-facing step is investigated. The flow is disturbed by high-rank forcing, making this test case considerably more challenging than a previous study that focused on rank-1 forcing \citep{herve2012physics}.  Downstream of the step the flow is low rank,  dominated by a Kelvin-Helmholtz instability wave, which is reflected in the fact that a control strategy is only slightly altered if targets are rank-1, rank-3, or full rank, and if one or multiple actuators are used.  Considerable gains were obtained using multiple sensors, which is explained by the presence of several modes from the upstream Poiseuille-like flow and that excite the downstream Kelvin-Helmholtz waves.
	
	Obtaining optimal controllers without model reduction circumvents possible performance losses of ROM-derived controllers when applied to the full system  \citep[page 349]{aastrom2010feedback}. Moreover, the present method handles naturally complex, coloured forcing in space and time, as the forcing CSD, $ \Mh{F} $, can be arbitrarily specified for each frequency. This is likely crucial for the control of turbulent flows, as the behaviour of coherent structures depends strongly on how these are forced \citep{chevalier2006state,towne2018spectral,martini2020resolventbased,nogueira2021forcing,morra2021colour}, but the difficulties involved in estimating and using coloured forcing models for control have hindered their use in previous applications. The method presented here, and forcing estimation methods presented in previous work \citep{martini2020resolventbased}, bridge these difficulties, allowing affordable full-rank, full-coloured controllers to be used. 
	
	This work also sheds light on the wave-cancellation behaviour of optimal control of shear flows \citep{sasaki2018wavecancelling}, which is equivalent to the truncated-non-causal (TNC) control presented here. The approach is optimal whenever the non-causal control kernels do not rely on future information, i.e. while not designed to be causal, they are  causal. This is the case for flows dominated by downstream-travelling modes, such as jets and boundary layers, provided that there is proper spacing between sensors, actuators, and targets. Reducing the distance between these, which may enhance control,  generates non-causal components in the control kernel, and deteriorates the TNC approach, particularly if several sensors/actuators are used. This effect can be considerably reduced with the optimal causal control strategy developed here.

	\section*{Acknowledgements}
	Eduardo Martini acknowledges financial support by CAPES grant 88881.190271/2018-01. Junoh Jung and Aaron Towne were supported by the Air Force Office of Scientific Research (AFOSR) grant \#FA9550-20-1-0214. 	André V. G. Cavalieri was supported by CNPq grant 313225/2020-6.

	\appendix

	\section{Wiener-Hopf problems}\label{app:WH}
	
	Analysing linear differential equations in the Fourier domain has the advantage of decoupling different frequencies for which solutions can be obtained independently. This property has been exploited in a previous study \citep{martini2020resolventbased} to achieve optimal non-causal estimation, i.e. estimation based on both past and future sensor readings.  Imposing causality, however, couples different frequencies, which must then be solved for simultaneously. The resulting equations frequently lead to Wiener-Hopf problems. 
	We here provide a brief introduction to this class of problems and a numerical method to solve them. 
	
	\subsection{Introduction}
	Wiener-Hopf equations appear when solving problems with restrictions applied on half domains. One example is the inversion of the half-convolution, given by
	\begin{align} \label{eq:WHeq_time}
		\int_0^\infty \HH(t-\tau)\M{W}_+(\tau)d\tau = 	\Hh(t) ,\;\;\;    t>0.
	\end{align}
	with $\M{W}_+(\tau) \in  \mathbb{C}^{n_a\times n_z}$ considered as the unknown matrix function to be determined.  The terms $\HH(t) \in  \mathbb{C}^{n_a\times n_a}$ and $\Hh(t) \in  \mathbb{C}^{n_a\times n_z}$ are known matrix functions. In \S~\ref{sec:estimation_and_control}, $ n_a $ and $ n_z $ corresponded, respectively, to the number of  actuators and targets used for control, $\HH$ and $\Hh$ are related to the actuator transfer functions, and $ \M{W}_+  \in  \mathbb{C}^{n_a\times n_y}$ to the optimal full-knowledge  control. Solving \eqref{eq:WHeq_time} allows for the construction of full-knowledge control laws, as will be demonstrated later.
	
	The frequency-domain representation of \eqref{eq:WHeq_time} is obtained by first constructing an equation valid for $ t<0 $. Extending \eqref{eq:WHeq_time} for negative times will in general break the equality of the equation. An \textit{a priori} unknown term $\M{W}_- \in  \mathbb{C}^{n_a\times n_z}$ is thus added to preserve the equality, as 
	\begin{align} \label{eq:WHeq_time_negT}
		\int_0^\infty \HH(t-\tau)\M{W}_+(\tau)d\tau = \M{W}_-(t)	+ \Hh(t) ,\;\;\;    t<0.
	\end{align}
	The integral can be extended to $ -\infty $ by requiring $ \M{W}_+(t<0)=0 $. Similarly, requiring $ \M{W}_-(t>0)=0 $ allows \eqref{eq:WHeq_time} and \eqref{eq:WHeq_time_negT} to be  added, leading to
	\begin{align} \label{eq:WHeq_time_comp}
		\int_{-\infty}^\infty \HH(t-\tau)\M{W}_+(\tau)d\tau = \M{W}_-(t)	+ \Hh(t),
	\end{align}
	which can be expressed in the frequency domain as
	\begin{align} \label{eq:WHeq}
		\hat{\HH}(\omega)\Mh{W}_+(\omega) = 	\Mh{W}_-(\omega)  + \hat \Hh(\omega),
	\end{align}
	where 
	\begin{align}
		\hat{\Hh}(\omega) = \int_{-\infty}^\infty {\Hh}(t) \e^{\ii \omega t} dt,
	\end{align}
	with similar expressions for the other variables. Although this is a single equation for two variables ($ \M{W}_- $ and $ \M{W}_+ $), the restriction that these variables be zero on different temporal half-domains ensures that the problem is well-posed.
	
	Provided $ \M{W}_+ $ is bounded for $ t \to \infty $, the requirement that $ \M{W}_+(t<0) =0$ is equivalent to restricting $ \hat{\M{W}}_+(\omega) $ to be regular in the upper half of the complex plane. This equivalence can be observed from its inverse-Fourier transform,
	\begin{align}
		\M{W}_+(t) =\frac{1}{2\pi} \int_{-\infty}^\infty \hat{\M{W}} _+(\omega) \e^{-\ii \omega t} d\omega,
	\end{align}
	which can be computed for $ t<0 $ by closing the contour around the upper-half plane and using the residue theorem.  As  $ \e^{-\ii \omega t}\to 0 $ for $ |\omega| \to \infty $ if  $ t<0 $ and $ \Im(\omega)>0 $, the integral on the upper contour closure is zero and, as neither $ \hat{\M{W}}_+(\omega) $ nor the exponential function have poles in the top-half plane, the integral is null for all $ t<0 $.	A similar argument holds for $\M{W}_-$ and  $ t>0 $,  with the contour being closed from below.   
	Henceforth, plus and minus subscripts are used to label functions that are regular in the upper- and lower halves of the complex frequency plane, respectively. Both frequency- and time-domain representations of  functions will be used interchangeably for  convenience or clarity. 	
	
	Other related Wiener-Hopf problems read
	\begin{align} \label{eq:WHeq_est}
		\hat{\M{W}}_+(\omega)\hat{\GG}(\omega) = 	\hat{\M{W}}_-(\omega)  + \hat{\Gg}(\omega),
	\end{align}
	and
	\begin{align} \label{eq:WHeq_con}
		\hat{\HH}(\omega)\hat{\M{W}}_+(\omega)\hat{\GG}(\omega) = 	\hat{\M{W}}_-(\omega)  + \hat{\Hh}(\omega)\hat{\Gg}(\omega).
	\end{align}
	where  $ \hat{\GG}  \in  \mathbb{C}^{n_y\times n_y} $ and $ \hat{\Gg}  \in  \mathbb{C}^{n_z\times n_y}$.
	
	\subsection{Solving Wiener-Hopf problems} 
		To obtain optimal causal estimation and partial-knowledge control, the Wiener-Hopf problems  \eqref{eq:WFcausal} and \eqref{eq:gammaplus} need to be solved. In what follows, formal solutions for these equations are presented. These solutions are based on the factorization of the kernels into plus and minus components. As analytical factorization are known only for special cases,  a  numerical method to factorize matrix functions, tailored for functions are known only numerically, is presented.

		Before proceeding, we define the two types of factorizations that will be used: additive and multiplicative.  Multiplicative factorization of a  matrix function $ \Mh{D} $ reads 
		\begin{align}\label{eq:multiFac}
			\Mh{D}(\omega) =  \Mh{D}_-(\omega)  \Mh{D}_+(\omega), 
		\end{align}  
		while an additive factorization reads
		\begin{align}\label{eq:addFac}
			\Mh{D}(\omega) =  \left(\Mh{D}(\omega)\right)_ -+ \left(\Mh{D}(\omega)\right)_+,
		\end{align}
		where all factors have the same size as the original matrix, and multiplicative factorizations are only defined for square matrices.  
		To differentiate these two types of factorizations, multiplicative factors will have the subscripts applied directly to them, as in \eqref{eq:multiFac}, and  additive factors will be presented with the subscripts applied outside the parenthesis, as in \eqref{eq:addFac}.

		These factorizations are not unique.  
		For a multiplicative factorization as in \eqref{eq:multiFac}, a valid factorization is constructed as $ {\hat{\HH}}_- \M{J} $ and  $ \M{J}^{-1} {\HH}_+ $,  for any constant and invertible matrix  $ \M{J}$.  Likewise, new additive factorizations are obtained by respectively  adding and subtracting a constant to the plus and minus factors.
		
		Any  multiplicative factorization can be used to solve Wiener-Hopf problems with the methods presented in this work, and thus we do not impose any extra condition to make it unique. However, we restrict additive factorizations to
		to standard factorizations~\citep{noble1959methods,daniele2014wiener}, that is
		\begin{equation}
					\left(\Mh{D}(\omega) \right)_\pm \to 0, \text{ for } \omega\to \pm\infty.
		\end{equation}
		
		\modification{ Note that the multiplicative factorization is also known as spectral factorization~\citep{claerbout1976fundamentals} in the signal processing community and is frequently expressed in terms of the Z, instead of the Fourier,  transform. Typical methods to obtain this factorization are the root method, which provides an analytical factorization if the poles of the kernel are known, and the Levinson algorithm, which is based on recursion to solve a de-convolution problem. Any of these methods can in principle be used for the solution of the Wiener-Hopf problems presented here. In this work, we use the strategy proposed by \cite{daniele2007fredholm}.}
		 
	\subsubsection{Formal solution}
	To obtain a solution for \eqref{eq:WHeq}, a multiplicative factorization of the kernel $ \hat{\HH} $, 
	\begin{align}\label{eq:WHfact_mul}
		\hat{\HH}(\omega) =  \hat{\HH}_-(\omega)  \hat{\HH}_+(\omega),
	\end{align}  
	is used. After manipulation, \eqref{eq:WHeq} becomes
	\begin{align}\label{eq:WHeq_dev}
		\hat{\HH}_+(\omega)\hat{\M{W}}_+(\omega) = \hat{\HH}_-^{-1}(\omega) \hat{\M{W}}_-(\omega)  + \hat{\HH}_-^{-1}(\omega) \hat{\Hh}(\omega).
	\end{align}
	Using an additive factorization of $ \hat{\HH}_-^{-1}(\omega) \hat{\Hh}(\omega) $, \eqref{eq:WHeq_dev} is  re-written as
	\begin{align} \label{eq:WH_L}
		\hat{\HH}_+(\omega)\hat{\M{W}}_+(\omega) - \left(\hat{\HH}_-^{-1}(\omega) \hat{\Hh}(\omega)\right)_+ = 
		\hat{\HH}_-^{-1}(\omega) \hat{\M{W}}_-(\omega) + \left(\hat{\HH}_-^{-1}(\omega) \hat{\Hh}(\omega)\right)_- .
	\end{align}
	
	Equation \eqref{eq:WH_L} is in Wiener-Hopf form, with only plus (minus) functions on the left(right)-hand side. Thus, the left- and right-hand sides are analytical functions in the lower and upper complex half planes for $\omega$, respectively.  Solution of the Wiener-Hopf equation amounts to stating that the left- and right-hand sides are analytical continuations of each other, which allows to define a single function of $\omega$ that is analytical everywhere. 
	
	That each side of the equation contains only plus or minus terms suggests that each side can be solved independently, as, loosely speaking, each side is an equation for for positive/negative times only. To formalize this idea, we make use of the following assumptions,
	\begin{enumerate}
		\item $ \hat{\HH}(\omega)  $ is bounded and positive definite,
		\item $ \hat{\HH}(\omega)  $ has no poles on the real line,
		\item   $ \hat{\Hh}(\pm \infty)\to 0 $,
	\end{enumerate}
	and define
	\begin{equation}\label{key}
		\mathcal L(\omega)  =	\hat{\HH}_+(\omega)\hat{\M{W}}_+(\omega) - \left(\hat{\HH}_-^{-1}(\omega) \hat{\Hh}(\omega)\right)_+ = 
		\hat{\HH}_-^{-1}(\omega) \hat{\M{W}}_-(\omega) + \left(\hat{\HH}_-^{-1}(\omega) \hat{\Hh}(\omega)\right)_- .
	\end{equation}


	Assumption (i) guarantees that $ \hat{\HH}_\pm $ is invertible, and that $ \hat{\HH}_-^{-1} $ does not create any poles in the  right-hand side of \eqref{eq:WH_L}.
	Assumption (ii) guarantees that   \eqref{eq:WH_L} is valid on a strip around the real axis, $ {-\epsilon<\Im(\omega)<\epsilon} $. 
	As the left-/right-hand-side of \eqref{eq:WH_L} are the analytical continuation of this strip in the upper/lower-half plane, these two functions and $\mathcal{L}(\omega)$ are regular everywhere. Since they are  bounded, by Liouville's theorem,  they are also constant.  	
	Finally, assumption (iii) and the use of a standard additive factorization in \eqref{eq:WH_L}, guarantees that the left-hand side of \eqref{eq:WH_L}  goes to zero for $ \omega\to\infty $, and thus that $ \mathcal{L}(\omega) =0$.
	 
	The solution of \eqref{eq:WHeq}, is obtained as 
	\begin{align}
		\label{eq:WHcausalsol_H}
		\hat{\M{W}}_+(\omega) =& \hat{\HH}^{-1}_+(\omega) \left(\hat{\HH}_-^{-1}(\omega) \hat{\Hh}(\omega)\right)_+, \\
		\hat{\M{W}}_-(\omega) =& \hat{\HH}_-\,\,(\omega) \left(\hat{\HH}_-^{-1}(\omega) \hat{\Hh}(\omega)\right)_-.
	\end{align}
	
	In this work, the assumptions (i)-(iii) are satisfied by construction. The kernels are constructed from a Hermitian quadratic form of the resolvent operator, to which a constant and Hermitian positive-definite matrix is added, thus  guaranteeing assumption~(i).
	The restriction to stable systems guarantees that the resolvent operator has no poles in the real line, and thus neither  do the kernels, guaranteeing assumption (ii).  Finally, as the term $ \hat{\HH} $ is a linear function of the resolvent operator, and since $ \R \propto 1/\omega $ for 
	 $ \omega\to\infty $, assumption (iii) is also guaranteed.
	
	Two other Wiener-Hopf problems, \eqref{eq:WHeq_est} and  \eqref{eq:WHeq_con}, are used in this study. The first, which appears when solving for the full-knowledge control kernel in Appendix \ref{sec:WHcontrol}, reads 
	\begin{align} \label{eq:WHeq_2}
		\hat{\M{W}}_+(\omega)\hat{\GG}(\omega) = 	\hat{\M{W}}_-(\omega)  + \hat{\Gg}(\omega), 
	\end{align}
	where $ \hat{\GG} \in  \mathbb{C}^{n_y\times n_y}$,  $ \hat{\Gg} \in  \mathbb{C}^{n_z\times n_y}$, and $ \Mh{W}_\pm \in  \mathbb{C}^{n_z\times n_y}$ are matrix functions. Making similar assumptions for $ \hat{\GG}$ and $\hat{\Gg} $ as the ones made for $ \hat{\HH}$ and $\hat{\Hh} $,  solutions are obtained as
	\begin{align} 
		\label{eq:WNsolutioncausal_G}
		\hat{\M{W}}_+(\omega) = \left( 	\hat{\Gg}(\omega) \hat{\GG}_-^{-1}(\omega) \right)_+ \hat{\GG}_+^{-1}(\omega),
	\end{align}
	where $ \hat{\GG}(\omega) $ has a multiplicative factorisation with different convention from $ \hat{\HH} (\omega)$, given by
	\begin{align}
		\hat{\GG}(\omega) = \hat{\GG}_+(\omega) \hat{\GG}_-(\omega).
	\end{align}
	
	The second problem, which appears when solving for the partial knowledge optimal control kernel in \S~\ref{sec:partialKownControl},  reads
	\begin{align} \label{eq:WHeq_3}
		\hat{\HH}(\omega)\hat{\M{W}}_+(\omega)\hat{\GG}(\omega) = 	\hat{\M{W}}_-(\omega)  + \hat{\Hh}(\omega)\hat{\Gg}(\omega),
	\end{align}
	where now $ \Mh{W}_\pm \in  \mathbb{C}^{n_a\times n_y}$. With the same assumption as before, solutions are given by
	\begin{align} 
		\label{eq:WNsolutioncausal_HG}
		\hat{\M{W}}_+(\omega) = \hat{\HH}_+^{-1}(\omega)\left( 	\hat{\HH}_-^{-1}(\omega) \hat{\Hh}(\omega)\hat{\Gg}(\omega) \hat{\GG}_-^{-1}(\omega) \right)_+ \hat{\GG}_+^{-1}(\omega).
	\end{align}

	\subsubsection{Numerical  Wiener-Hopf factorizations}\label{sec:danielesMethod}
		An analytical expression for additive factorization reads~\citep{noble1959methods}
		\begin{equation}\label{eq:additiveFac}
			\left(\Mh{W}(\omega)\right)_\pm  = \frac{\pm1}{2\pi\ii}\int \frac{\Mh{W}(\omega')}{\omega - \omega'}d\omega' ,
		\end{equation}
		with integration contours idented below or above the pole at $\omega$ for $+$ and $-$ functions, respectively.
		However, when the factors are desired numerically, additive factorizations can be easily obtained using Fourier transforms. 
		Applying an inverse transform, the time-domain representation of the function is obtained. This representation is then split into its plus (minus) component by multiplication with a Heaviside-step function, i.e., setting to zero all values for $ t>0 $($ t<0 $) . A Fourier transform is then used to recover the frequency-domain representation. 
		The function thus obtained constitutes a standard factorization: if the original function is smooth, i.e. its spectral content goes to zero for high enough frequencies, so will the factors calculated with the procedure just described.

	Multiplicative factorization for scalar problems can be reduced to an additive factorization using a logarithm function to convert multiplication into  addition~\citep{noble1959methods,peake2004unsteady}, with the factorization reading
	\begin{equation}\label{key}
		\hat{\HH}_\pm(\omega)  = \exp\left({\frac{\pm1}{2\pi\ii}\int \frac{\ln(\hat{\HH}(\omega'))}{\omega - \omega'}d\omega'} \right).
	\end{equation}
 	This  procedure, however, requires that the quantities commute, which is not generally the case when $ \hat{\HH} $ is a matrix. 
	Analytical Wiener-Hopf factorizations are only known for special classes of matrices \citep{daniele1978factorization}, and an analytical method for the factorization of general matrices is still unknown.   
	
	In this work, we use a method similar to the one described by \cite{daniele2007fredholm} to obtain multiplicative matrix factorizations for kernels that are known numerically, rather than analytically.
	
	The multiplicative factorization, satisfying \eqref{eq:WHfact_mul},  can be obtained from $ n_a $ independent solutions of 
	\begin{equation}\label{eq:homWH}
		\hat{\HH}(\omega) \hat{\V{w}}_{i,+}(\omega) = \hat{\V{w}}_{i,-}(\omega),
	\end{equation}	
 as
	\begin{align}
		\hat{\HH}_- (\omega) &= \Matrix{\hat{\V{w}}_{1,-}(\omega)\,\hat{\V{w}}_{2,-}(\omega)\, \dots\, \hat{\V{w}}_{n,-} (\omega)}, \label{eq:danieleH+}\\
		\hat{\HH}_+(\omega)  &= \Matrix{\hat{\V{w}}_{1,+}(\omega)\,\hat{\V{w}}_{2,+}(\omega)\, \dots\, \hat{\V{w}}_{n,+}(\omega)}^{-1} \label{eq:danieleH-},
	\end{align}
	where $ n_a $ is the size of the square matrix $ \hat{\HH} $.
		That is, matrices $\hat{\HH}_-$ and $\hat{\HH}_+^{-1}$   have vectors $\Vh{w}_{i,\mp}$ as columns, respectively.
	
	To obtain solutions of \eqref{eq:homWH}, we divide it by $ \omega-\omega_0 $, with $ \Im(\omega_0)<0 $, and  integrate along a  line that crosses the real axis and closes around the lower-half plane. Defining $ \Vh{x}_i = \Vh{w}_{i,+}/ (\omega-\omega_0)  $,
	\eqref{eq:homWH} becomes a Fredholm integral equation of the second kind \citep{daniele2007fredholm},
	\begin{align} \label{eq:fredholm}
		\Vh{x}_{i}(\omega) +\frac{1}{2\pi\ii} \int_{-\infty}^{\infty}\frac{ \hat{\HH}^{-1}(\omega) \hat{\HH}(u)-1}{u-\omega}\Vh{x}_{i}(u) du  =  \hat{\HH}^{-1}(\omega)\frac{\Vh{w}_{i,-}(\omega_0)}{\omega-\omega_0}. 
	\end{align}
	
	Note that \eqref{eq:fredholm} has only one unknown, $ \Vh{x} $,  while \eqref{eq:homWH} has two, $ \hat{\V{w}}_{i,-} $ and $ \hat{\V{w}}_{i,+} $. 	The integration of the unknown term $\hat{\V{w}}_{i,-}(\omega)$ is carried out with the residue theorem, leading to  $ \Vh{w}_{i,-}(\omega_0) $, which  is constant and can be arbitrarily specified.   Choosing it as the canonical basis ($ \Vh{w}_{i,-}(\omega_0) =\V{e}_i $) is an obvious choice. The parameter $\omega_0$ can be arbitrarily chosen, although different values can change convergence requirements for the numerical solution of the equation. As discussed by \cite{daniele2007fredholm}, $\omega_0$ introduces an apparent singularity in the equation that, while not impacting the analytical solutions, can lead to numerical instabilities for approximate, numerical, solutions. Values close to the real axis lead to a right-hand side  that has sharp variations, and thus requires finer frequency discretization to be resolved, whereas values excessively far from the real axis cause the left-hand side to have significant values on a larger domain, thus requiring the discretization of a larger frequency range. \cite{daniele2007fredholm} suggest choosing $\omega_0$ such that it corresponds to singularities of the physical problem under study, but in the context of the problem studied here the choice is not obvious. It is thus necessary to check convergence using  different values   of $\omega_0$ and/or different frequency discretizations.
	
	\cite{daniele2007fredholm}  discretized  \eqref{eq:fredholm} to construct a matrix representing its right-hand side. The numerical solution was obtained by solving the resulting linear problem. Deformation of the integration path into the complex $ \omega $ plane was used to improve the convergence rate of the solutions whenever the kernel had poles close to the real line. Similarly, \cite{atkinson2008algorithm} used different integration weights and collocations points to deal with such singularities.  Throughout this work, we  focus on kernels that are obtained numerically, and thus only available on the real frequency line. Deformation of the integration path is thus unfeasible, and convergence is obtained by refining the frequency discretization. 
	
	To solve \eqref{eq:fredholm}, a linear problem with size $ n_a^2 n_\omega $  has to be solved, where $ n_\omega $ is the number of frequency points used, and $ n_a $ the size of the square matrix $ \hat{\HH} $.  This approach becomes unpractical for  the frequency discretization required for convergence of the results in this study. Instead,  we rewrite  \eqref{eq:fredholm} as
	\begin{equation}\label{eq:fredhmol_Hilbert}
		\Vh{x}_{i}(\omega) 
		+\frac{1}{2\ii} \mathcal{H} ( \Vh{x}_{i}) (\omega)
		-\frac{1}{2\ii} \hat{\HH}^{-1}(\omega) \mathcal{H} ( \hat{\HH}\Vh{x}_{i})(\omega) 
		=\hat{\HH}^{-1}(\omega) \frac{\Vh{w}_{i,-}(\omega_0)}{\omega-\omega_0},
	\end{equation}
	where 
	\begin{equation}\label{}
		\mathcal{H}(\Vh{x}) = P.V. \frac{1}{\pi}\int_{-\infty}^{\infty} \frac{1}{\omega-u} \Vh{x}(u) du,
	\end{equation}
	is the Hilbert transform of $ \Vh{x}(\omega) $.	Hilbert transforms can be efficiently computed numerically using fast-Fourier transforms \citep{todoran2008discrete}, and thus the left-hand side  of \eqref{eq:fredhmol_Hilbert} can be obtained without the construction of the matrix that represents it. The problem is thus well suited for solutions via iterative methods, such as GMRES, used here.  
	
	As mentioned by \cite{zhou2009novel}, using Fourier transforms to compute Hilbert transforms can lead to significant errors at the extremities of the signal, due to the implicit assumption of periodicity. The signals need thus to be zero-padded to avoid such errors. Due to the slow decay of the term $1/(\omega-u)$, large paddings can be necessary.  Padding of 20 times the time signal has been used throughout this study. 
	
	Using $ n_a $ linearly independent solutions, $ \hat{\HH}_\pm(\omega) $ can be obtained from 
	\begin{align}
		\Vh{w}_{+,i}(\omega) =&\Vh{x}_i (\omega)(\omega -\omega_0), \\  \Vh{w}_{i,-}(\omega) =&   \hat{\HH}(\omega)  \Vh{w}_{i,+}(\omega)  ,
	\end{align}
	as in \eqref{eq:danieleH+}-\eqref{eq:danieleH-}.
	
	A factorization with the order of plus and minus functions exchanged, i.e., $ \hat{\GG}  = \hat{\GG}_+ \hat{\GG}_- $, can be obtained via the same method using an auxiliary matrix $ {\hat{\GG}'  = \hat{\GG}^* } $, where~$ ^* $ represent complex conjugation. From a factorization of $  \hat{\GG}'$, the desired factorization is obtained as  $ \hat{\GG}_+  = \hat{\GG}_-'^{*}  $ and $ \hat{\GG}_- = \hat{\GG}_+'^{*}  $.

		\subsubsection{Convergence}

	\begin{figure}[t]
	\centering
	\subfloat[Error scaling with sampling rate.]{\includegraphics[scale=0.5]{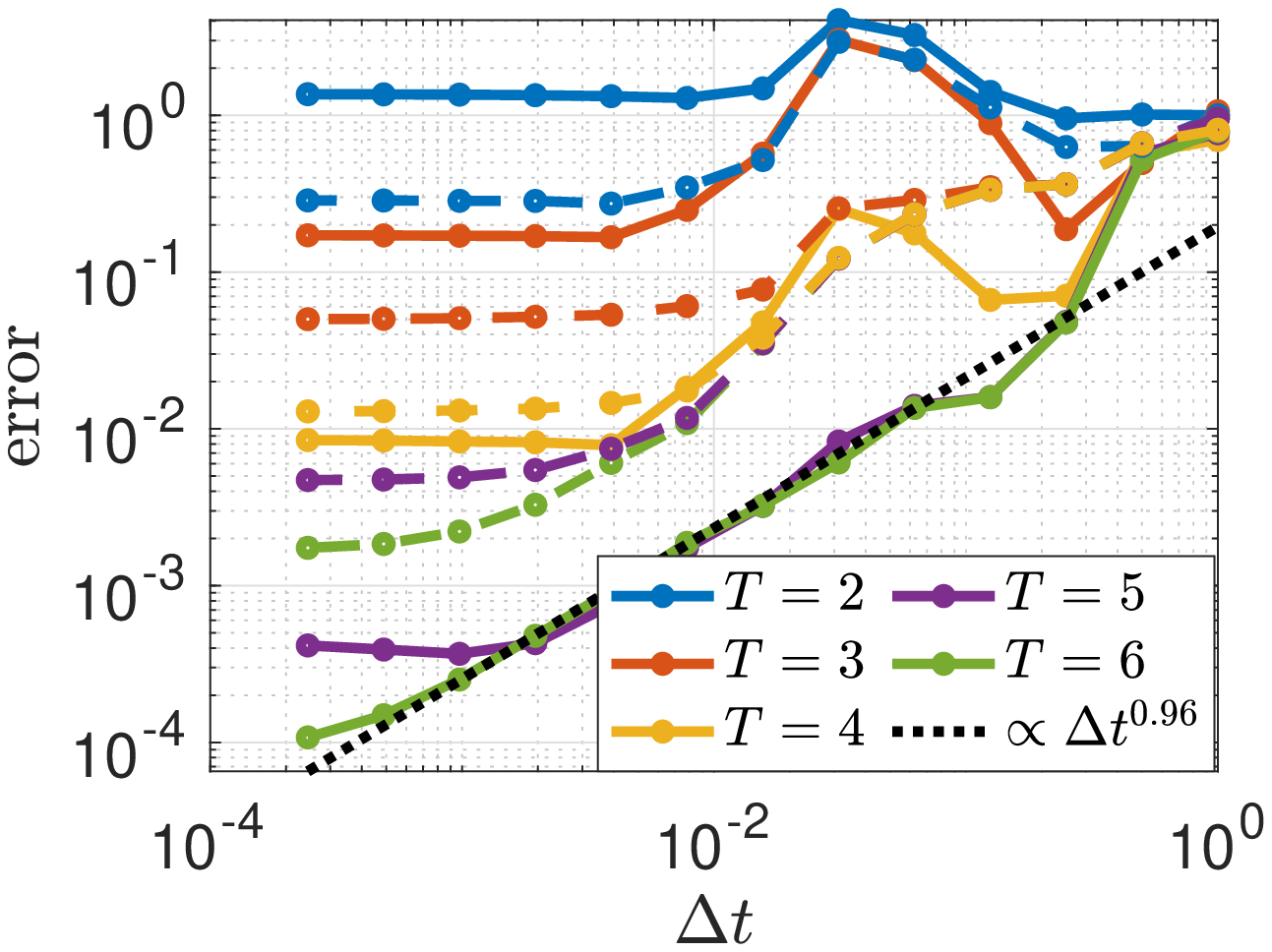}}
	\subfloat[Error scaling with $T$ for the largest sampling rate.]{\includegraphics[scale=0.5]{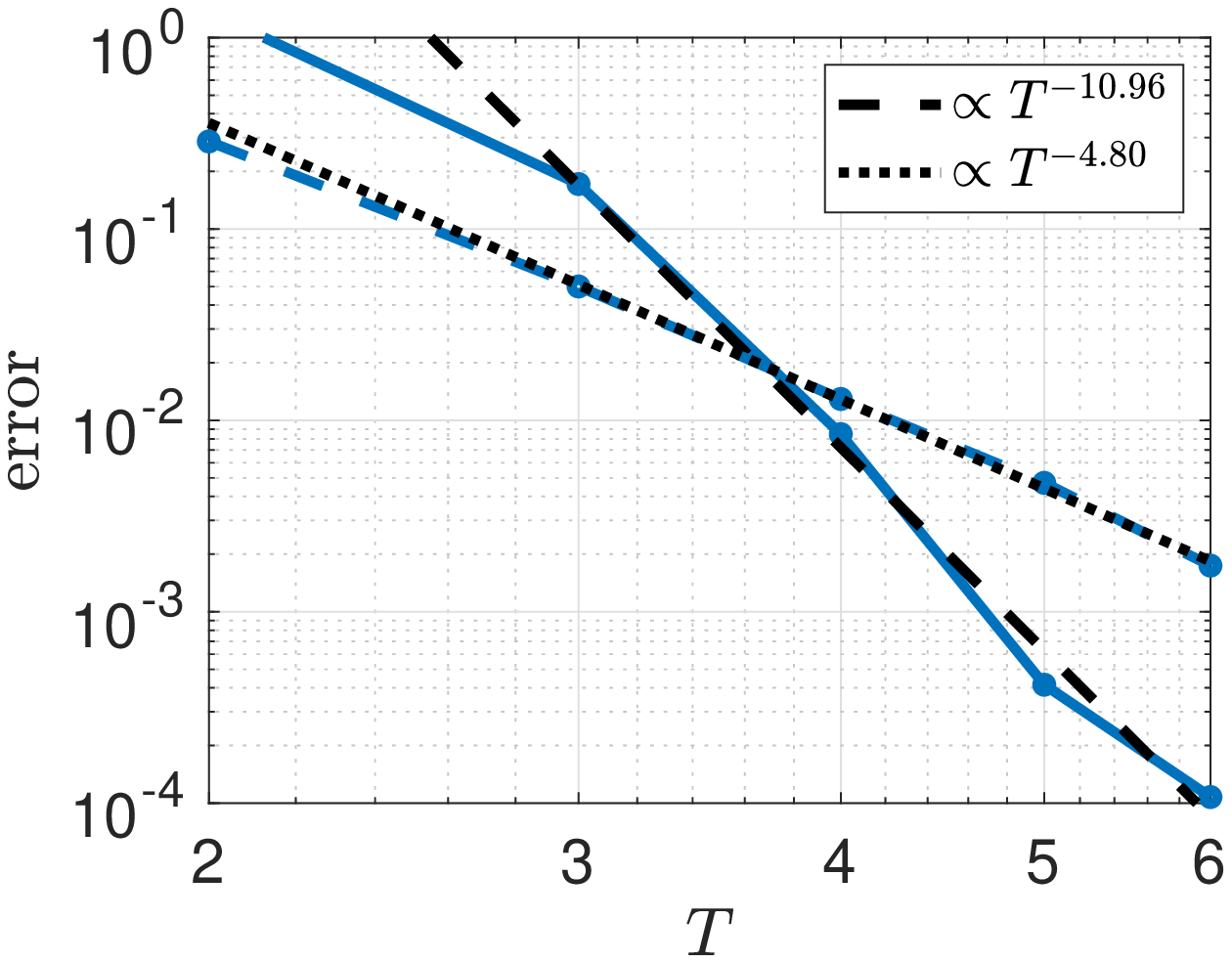}}
	
	\caption{Trends for errors and cost of the Wiener-Hopf factorization for a sensor located at $ X=5 $ and a target at $ x=35 $. The solid(dashed) line correspond to an actuator located at $ x=15(7) $. }
	\label{fig:WH_scalling}
\end{figure}

\begin{figure}[t]
	\centering
	\subfloat[Error scaling with sampling rate.]{\includegraphics[scale=0.5]{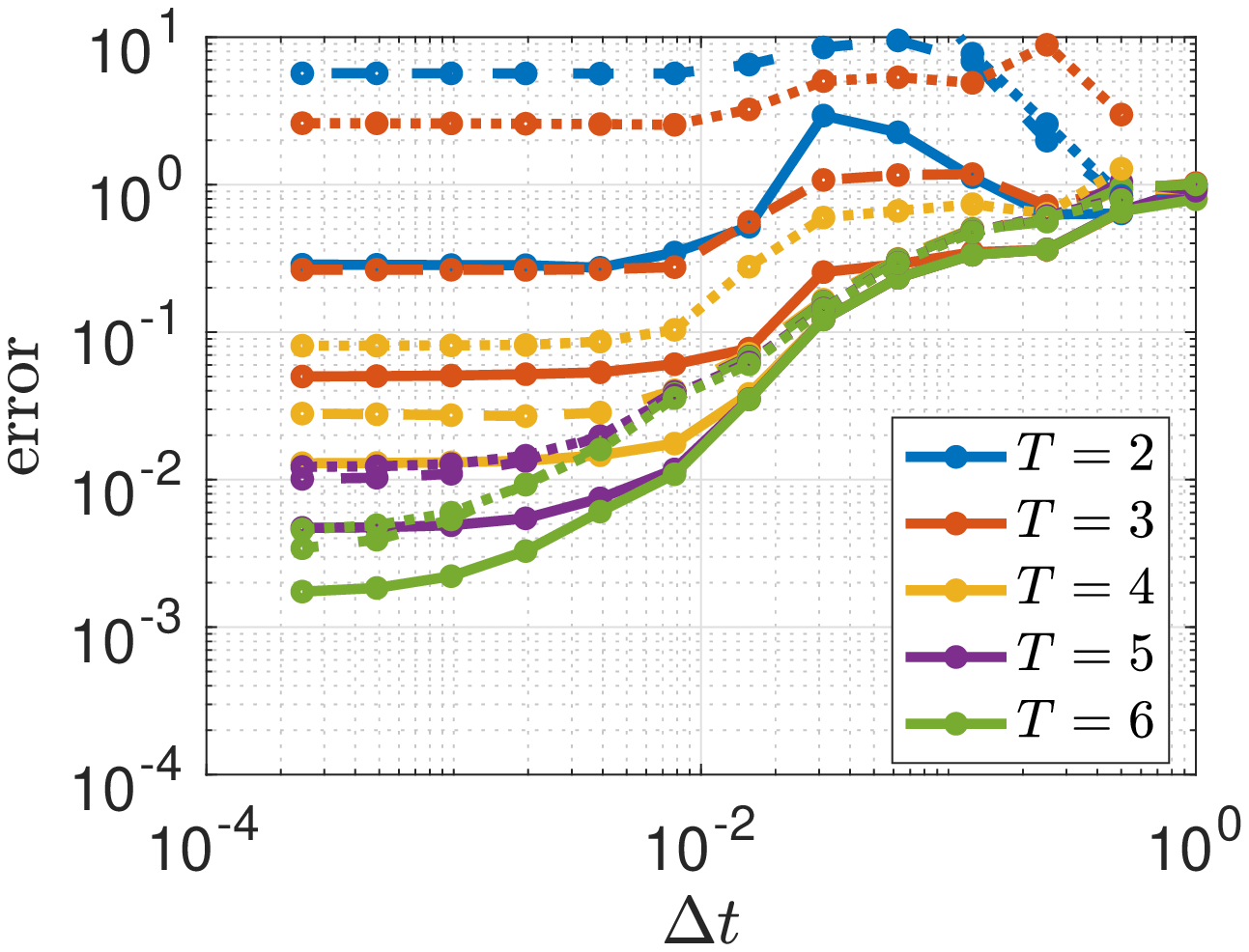}}
	\subfloat[Factorization wall-time.]{\includegraphics[scale=0.5]{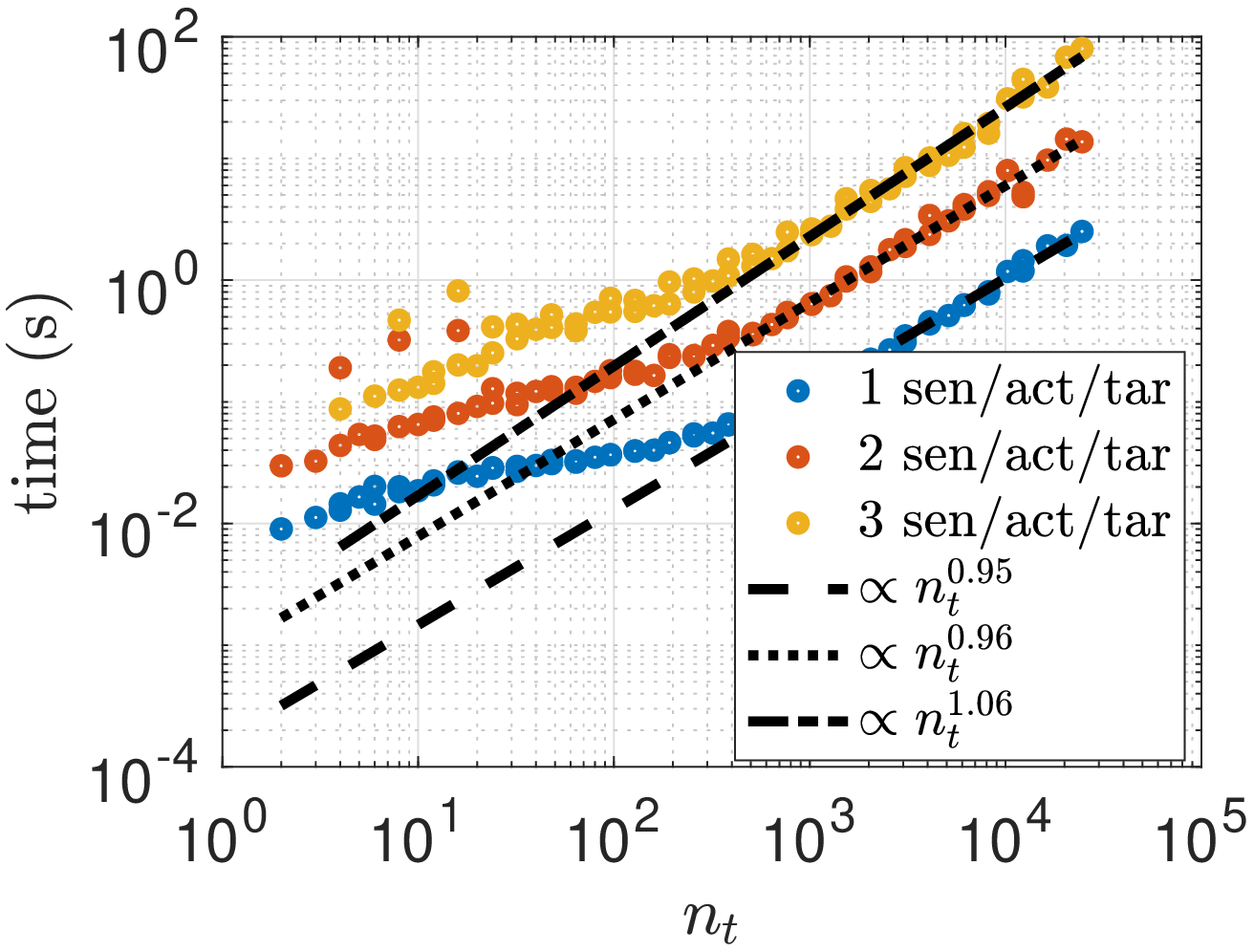}}
	
	\caption{Same as figure \ref{fig:WH_scalling} with sensors are located at $ x=5,10,15 $, actuators at $ x=7,12,17 $ and targets at $ z=30,35,37 $. On the left, solid, dashed, and solid lines corresponds to the configuration where only the first, the first and second, and all sensors, actuators and targets are used. The wall time to perform the factorizations is shown on the right.}
	\label{fig:WH_scalling_MIMO}
\end{figure}
	
		To access the convergence of the method, we compare the control kernels obtained for the Ginzburg-Landau system described in \S~\ref{sec:comparison_on_GL}. A normalized error is defined as
		\begin{equation}\label{key}
			\sqrt{	\frac{ \int_0^\infty | \GAMMA'_c(\tau)	- \GAMMA'_{lqg}(\tau)| ^2	d\tau }{\int_0^\infty |  \GAMMA'_{lqg}(\tau)| d\tau}},
		\end{equation}
		where $ \GAMMA'_c $ is the kernel computed using the Wiener-Hopf approach and $ \GAMMA'_{lqg} $ is obtained as in \eqref{eq:galmmalqg}. A time interval $ [-T,T ] $ was discretized with points spaced by  $ \Delta t $, corresponding to a sampling frequency $ \omega_s= 2\pi\Delta t$ and a frequency resolution of $ \Delta\omega = 2\pi/T $. 
		
		 The normalized error is shown in figure \ref{fig:WH_scalling}, for different values of $ T $ and $ \Delta t $. The factorization scales linearly with the number of points in the frequency discretization. The linear convergence of the kernel with $ \Delta t $ is a consequence of its discontinuity at $ \tau=0 $: this is the convergence rate of Fourier series for discontinuous functions.  Note that when actuators are close to the sensor, the discontinuity is stronger, leading to the larger errors seen when the actuator is at $ x=7 $. The converge trend is nevertheless unaffected. Convergence with respect to the domain size $  T $ is very fast, and thus has a small impact on the overall cost. 
		 
		 The effect of having multiple sensors and actuators on the factorization is explored in figure \ref{fig:WH_scalling_MIMO}. Adding sensor/actuators leads to an increase in the required $ T $ and $ \Delta t$ for a given accuracy, but does not significantly affects the convergence trends. 
		  It is also seen that the cost scales linearly with the number of points used for time/frequency discretization. The increase with $ n_a $ and $ n_y $ is due to the need to perform matrix multiplication, which scales with the square of its size. In all scenarios, good accuracy is obtained within a few minutes on a standard notebook. Note also that this cost does not scale with the size of the system, and thus remain roughly the same for any system. When applied to complex flows,  factorization is thus orders of magnitude less-costly than the time-domain solutions of the direct and  adjoint problems described in \S~\ref{sec:timemarching}.

	\section{The full-knowledge control problem} \label{sec:WHcontrol}
	Complementing the optimal estimation (\S~\ref{sec:estimation}) and  partial-knowledge control (\S~\ref{sec:partialKownControl}), we here present the derivation of the optimal full-knowledge control.

	Analogous to the procedure used in \S~\ref{sec:estimation}, optimal control is obtained by minimizing a cost functional given by
	\begin{align}\label{eq:controlCostFun}
		\begin{aligned}
			J &= \int_{-\infty}^{\infty} \left( \z ^\dagger (t)  \z(t)  + \V{a} ^\dagger(t)  \M{P} \V{a}(t) \right) dt 
			= \int_{-\infty}^{\infty} \left( \hat{\z} ^\dagger(\omega)   \hat{\z} (\omega) + \Vh{a} ^\dagger (\omega)  \M{P} \Vh{a}(\omega) \right) d\omega.
		\end{aligned}
	\end{align}

	Full system knowledge control implies the system state for the current time is known, from which $ \z_1 $ is known can be computed. If external forcing is present, $ \z_1 $ need to be updated at each time instant.  An actuation, $ \V{a}_{nc}(t) $, that minimizes the cost functional can be obtained in terms of $ \z_1 $ alone.
	Expanding terms  in  \eqref{eq:controlCostFun} gives
	\begin{align}
		\begin{aligned}
			J =& \int_{-\infty}^{\infty}  \left( (\hat{\z}_1(t) + \hat{\z}_2(t)  ) ^\dagger  (\hat{\z}_1(t) + \hat{\z}_2(t)  )  + \Vh{a}_{nc} ^\dagger(t)  \M{P} \Vh{a}_{nc}(t)  \right) d\omega
			\\ =& 
			\int_{-\infty}^{\infty}  \left( (\hat{\z}_1(\omega) + \Raz(\omega) \Vh{a}_{nc}(\omega)  ) ^\dagger   (\hat{\z}_1(\omega) + \Raz \Vh{a}_{nc}(\omega)  )  + \Vh{a}_{nc} ^\dagger(\omega)  \M{P} \Vh{a}_{nc}(\omega)  \right) d\omega,
		\end{aligned}
	\end{align}
	and differentiation with respect to $ \Vh{a}_{nc}^\dagger(\omega) $ leads to 
	\begin{align} \label{eq:non-causalLQR}
		\hat{\HH}(\omega) \Vh{a}_{nc}(\omega) = \hat{\Hh}(\omega) \hat{\z}_1(\omega) , 
	\end{align}	
	where
	\begin{align} \label{eq:Hh_def}
		\hat{\HH}(\omega) =& \Raz^\dagger(\omega)  \Raz(\omega)  +\M{P} , &
		\hat{\Hh}(\omega) =& -\Raz^\dagger(\omega) .
	\end{align}
	
	As in the estimation problem, causality, i.e. $ \V{a}_c(t<0)=0 $, can be enforced with Lagrange multipliers. Taking the derivative of  the modified  cost functional given by
	\begin{align}\label{eq:controlCostFunCausal}
		J' = \int_{-\infty}^{\infty} \left( \hat{\z}  ^\dagger (t) \hat{\z}(t)   + \V{a}_c ^\dagger (t) \M{P} \V{a}_c(t)   + \LAMBDA_-(t)\V{a}_c(t) + \LAMBDA^\dagger_-(t)\V{a}_c^\dagger(t)  \right) dt,	
	\end{align}
	with respect to $ \V{a}_c^\dagger $ yields the Wiener-Hopf problem
	\begin{align}\label{eq:WHlqr}
		\hat{\HH}(\omega)  \Vh{a}_c(\omega) + \hat{\LAMBDA}_-(\omega)=  \hat{\Hh}(\omega) \hat{\z}(\omega),
	\end{align}	
	which has the structure of \eqref{eq:WHeq} and solution given by \eqref{eq:WHcausalsol_H}.

	\bibliographystyle{unsrt}
	\bibliography{References.bib}

\end{document}

%% file: Figures/System.eps_tex
\begingroup%
  \makeatletter%
  \providecommand\color[2][]{%
    \errmessage{(Inkscape) Color is used for the text in Inkscape, but the package 'color.sty' is not loaded}%
    \renewcommand\color[2][]{}%
  }%
  \providecommand\transparent[1]{%
    \errmessage{(Inkscape) Transparency is used (non-zero) for the text in Inkscape, but the package 'transparent.sty' is not loaded}%
    \renewcommand\transparent[1]{}%
  }%
  \providecommand\rotatebox[2]{#2}%
  \newcommand*\fsize{\dimexpr\f@size pt\relax}%
  \newcommand*\lineheight[1]{\fontsize{\fsize}{#1\fsize}\selectfont}%
  \ifx\svgwidth\undefined%
    \setlength{\unitlength}{639.40988159bp}%
    \ifx\svgscale\undefined%
      \relax%
    \else%
      \setlength{\unitlength}{\unitlength * \real{\svgscale}}%
    \fi%
  \else%
    \setlength{\unitlength}{\svgwidth}%
  \fi%
  \global\let\svgwidth\undefined%
  \global\let\svgscale\undefined%
  \makeatother%
  \begin{picture}(1,0.32901706)%
    \lineheight{1}%
    \setlength\tabcolsep{0pt}%
    \put(0,0){\includegraphics[width=\unitlength]{System.eps}}%
    \put(0.13718882,0.27969172){\makebox(0,0)[lt]{\lineheight{1.25}\smash{\begin{tabular}[t]{l}$\Bf$\end{tabular}}}}%
    \put(0.13525356,0.17959968){\makebox(0,0)[lt]{\lineheight{1.25}\smash{\begin{tabular}[t]{l}$\Ba$\end{tabular}}}}%
    \put(0.42571367,0.22899279){\makebox(0,0)[lt]{\lineheight{1.25}\smash{\begin{tabular}[t]{l}$\R$\end{tabular}}}}%
    \put(0.54835915,0.13590186){\makebox(0,0)[lt]{\lineheight{1.25}\smash{\begin{tabular}[t]{l}$\Cy$\end{tabular}}}}%
    \put(0.79169369,0.13613636){\makebox(0,0)[lt]{\lineheight{1.25}\smash{\begin{tabular}[t]{l}$\Cz$\end{tabular}}}}%
  \end{picture}%
\endgroup%

%% file: Figures/SeparatedSystem.eps_tex
\begingroup%
  \makeatletter%
  \providecommand\color[2][]{%
    \errmessage{(Inkscape) Color is used for the text in Inkscape, but the package 'color.sty' is not loaded}%
    \renewcommand\color[2][]{}%
  }%
  \providecommand\transparent[1]{%
    \errmessage{(Inkscape) Transparency is used (non-zero) for the text in Inkscape, but the package 'transparent.sty' is not loaded}%
    \renewcommand\transparent[1]{}%
  }%
  \providecommand\rotatebox[2]{#2}%
  \newcommand*\fsize{\dimexpr\f@size pt\relax}%
  \newcommand*\lineheight[1]{\fontsize{\fsize}{#1\fsize}\selectfont}%
  \ifx\svgwidth\undefined%
    \setlength{\unitlength}{660.15138245bp}%
    \ifx\svgscale\undefined%
      \relax%
    \else%
      \setlength{\unitlength}{\unitlength * \real{\svgscale}}%
    \fi%
  \else%
    \setlength{\unitlength}{\svgwidth}%
  \fi%
  \global\let\svgwidth\undefined%
  \global\let\svgscale\undefined%
  \makeatother%
  \begin{picture}(1,0.35804863)%
    \lineheight{1}%
    \setlength\tabcolsep{0pt}%
    \put(0,0){\includegraphics[width=\unitlength]{SeparatedSystem.eps}}%
    \put(0.21955698,0.28821187){\makebox(0,0)[lt]{\lineheight{1.25}\smash{\begin{tabular}[t]{l}$\Bf$\end{tabular}}}}%
    \put(0.21847685,0.05123018){\makebox(0,0)[lt]{\lineheight{1.25}\smash{\begin{tabular}[t]{l}$\Ba$\end{tabular}}}}%
    \put(0.04328462,0.05261276){\makebox(0,0)[lt]{\lineheight{1.25}\smash{\begin{tabular}[t]{l}$\GAMMA$\end{tabular}}}}%
    \put(0.39478915,0.05456297){\makebox(0,0)[lt]{\lineheight{1.25}\smash{\begin{tabular}[t]{l}$\R$\end{tabular}}}}%
    \put(0.39558551,0.28930815){\makebox(0,0)[lt]{\lineheight{1.25}\smash{\begin{tabular}[t]{l}$\R$\end{tabular}}}}%
    \put(0.64124856,0.05324315){\makebox(0,0)[lt]{\lineheight{1.25}\smash{\begin{tabular}[t]{l}$\Cz$\end{tabular}}}}%
    \put(0.64171606,0.28860939){\makebox(0,0)[lt]{\lineheight{1.25}\smash{\begin{tabular}[t]{l}$\Cz$\end{tabular}}}}%
    \put(0.48434034,0.13247871){\makebox(0,0)[lt]{\lineheight{1.25}\smash{\begin{tabular}[t]{l}$\Cy$\end{tabular}}}}%
    \put(0.48593306,0.22140865){\makebox(0,0)[lt]{\lineheight{1.25}\smash{\begin{tabular}[t]{l}$\Cy$\end{tabular}}}}%
  \end{picture}%
\endgroup%

%% file: Figures/ControlSchemeClosed.eps_tex
\begingroup%
  \makeatletter%
  \providecommand\color[2][]{%
    \errmessage{(Inkscape) Color is used for the text in Inkscape, but the package 'color.sty' is not loaded}%
    \renewcommand\color[2][]{}%
  }%
  \providecommand\transparent[1]{%
    \errmessage{(Inkscape) Transparency is used (non-zero) for the text in Inkscape, but the package 'transparent.sty' is not loaded}%
    \renewcommand\transparent[1]{}%
  }%
  \providecommand\rotatebox[2]{#2}%
  \newcommand*\fsize{\dimexpr\f@size pt\relax}%
  \newcommand*\lineheight[1]{\fontsize{\fsize}{#1\fsize}\selectfont}%
  \ifx\svgwidth\undefined%
    \setlength{\unitlength}{621.14282227bp}%
    \ifx\svgscale\undefined%
      \relax%
    \else%
      \setlength{\unitlength}{\unitlength * \real{\svgscale}}%
    \fi%
  \else%
    \setlength{\unitlength}{\svgwidth}%
  \fi%
  \global\let\svgwidth\undefined%
  \global\let\svgscale\undefined%
  \makeatother%
  \begin{picture}(1,0.49791651)%
    \lineheight{1}%
    \setlength\tabcolsep{0pt}%
    \put(0,0){\includegraphics[width=\unitlength]{ControlSchemeClosed.eps}}%
    \put(0.56272887,0.31925628){\makebox(0,0)[lt]{\lineheight{1.25}\smash{\begin{tabular}[t]{l}$\Cy$\end{tabular}}}}%
    \put(0.79437503,0.31682215){\makebox(0,0)[lt]{\lineheight{1.25}\smash{\begin{tabular}[t]{l}$\Cz$\end{tabular}}}}%
    \put(0.45066236,0.40382572){\makebox(0,0)[lt]{\lineheight{1.25}\smash{\begin{tabular}[t]{l}$\R$\end{tabular}}}}%
    \put(0.15983514,0.45500992){\makebox(0,0)[lt]{\lineheight{1.25}\smash{\begin{tabular}[t]{l}$\Bf$\end{tabular}}}}%
    \put(0.16154762,0.35449839){\makebox(0,0)[lt]{\lineheight{1.25}\smash{\begin{tabular}[t]{l}$\Ba$\end{tabular}}}}%
    \put(0.11413483,0.22314453){\makebox(0,0)[lt]{\lineheight{1.25}\smash{\begin{tabular}[t]{l}$\Ba$\end{tabular}}}}%
    \put(0.23715986,0.22225643){\makebox(0,0)[lt]{\lineheight{1.25}\smash{\begin{tabular}[t]{l}$\R$\end{tabular}}}}%
    \put(0.35465262,0.22225643){\makebox(0,0)[lt]{\lineheight{1.25}\smash{\begin{tabular}[t]{l}$\Cy$\end{tabular}}}}%
    \put(0.27405677,0.08674333){\makebox(0,0)[lt]{\lineheight{1.25}\smash{\begin{tabular}[t]{l}$\GAMMA$\end{tabular}}}}%
    \put(0.08009553,0.02208961){\makebox(0,0)[lt]{\lineheight{1.25}\smash{\begin{tabular}[t]{l}$\GAMMA'$\end{tabular}}}}%
  \end{picture}%
\endgroup%

%% file: Figures/scheme_runs.eps_tex
\begingroup%
  \makeatletter%
  \providecommand\color[2][]{%
    \errmessage{(Inkscape) Color is used for the text in Inkscape, but the package 'color.sty' is not loaded}%
    \renewcommand\color[2][]{}%
  }%
  \providecommand\transparent[1]{%
    \errmessage{(Inkscape) Transparency is used (non-zero) for the text in Inkscape, but the package 'transparent.sty' is not loaded}%
    \renewcommand\transparent[1]{}%
  }%
  \providecommand\rotatebox[2]{#2}%
  \newcommand*\fsize{\dimexpr\f@size pt\relax}%
  \newcommand*\lineheight[1]{\fontsize{\fsize}{#1\fsize}\selectfont}%
  \ifx\svgwidth\undefined%
    \setlength{\unitlength}{771.99993896bp}%
    \ifx\svgscale\undefined%
      \relax%
    \else%
      \setlength{\unitlength}{\unitlength * \real{\svgscale}}%
    \fi%
  \else%
    \setlength{\unitlength}{\svgwidth}%
  \fi%
  \global\let\svgwidth\undefined%
  \global\let\svgscale\undefined%
  \makeatother%
  \begin{picture}(1,0.57772027)%
    \lineheight{1}%
    \setlength\tabcolsep{0pt}%
    \put(0,0){\includegraphics[width=\unitlength]{scheme_runs.eps}}%
    \put(0.06292571,0.48534942){\makebox(0,0)[lt]{\lineheight{1.25}\smash{\begin{tabular}[t]{l}$\Cy$\end{tabular}}}}%
    \put(0.59079473,0.48513852){\makebox(0,0)[lt]{\lineheight{1.25}\smash{\begin{tabular}[t]{l}$\Ba$\end{tabular}}}}%
    \put(0.34719971,0.40407643){\makebox(0,0)[lt]{\lineheight{1.25}\smash{\begin{tabular}[t]{l}$\Rfy^\dagger$\end{tabular}}}}%
    \put(0.34871847,0.34076241){\makebox(0,0)[lt]{\lineheight{1.25}\smash{\begin{tabular}[t]{l}$\Ray^\dagger$\end{tabular}}}}%
    \put(0.34554929,0.24759876){\makebox(0,0)[lt]{\lineheight{1.25}\smash{\begin{tabular}[t]{l}$\Rfy\Rfy^\dagger$\end{tabular}}}}%
    \put(0.3469469,0.19029624){\makebox(0,0)[lt]{\lineheight{1.25}\smash{\begin{tabular}[t]{l}$\Rfz\Rfy^\dagger$\end{tabular}}}}%
    \put(0.34396665,0.07320282){\makebox(0,0)[lt]{\lineheight{1.25}\smash{\begin{tabular}[t]{l}$\Raz^\dagger\Rfz\Rfy^\dagger$\end{tabular}}}}%
    \put(0.86810485,0.40270681){\makebox(0,0)[lt]{\lineheight{1.25}\smash{\begin{tabular}[t]{l}$\Ray$\end{tabular}}}}%
    \put(0.86962364,0.33939285){\makebox(0,0)[lt]{\lineheight{1.25}\smash{\begin{tabular}[t]{l}$\Raz$\end{tabular}}}}%
    \put(0.86676149,0.21514109){\makebox(0,0)[lt]{\lineheight{1.25}\smash{\begin{tabular}[t]{l}$\Raz^\dagger\Raz$\end{tabular}}}}%
    \put(0.8659903,0.06989022){\makebox(0,0)[lt]{\lineheight{1.25}\smash{\begin{tabular}[t]{l}$\Rfy\Raz^\dagger\Raz$\end{tabular}}}}%
  \end{picture}%
\endgroup%